\documentclass[11pt,a4paper,fleqn,oneside,doublespace]{book}
\usepackage{doublespace}
\usepackage{a4,graphicx,psfrag}
\usepackage{latexsym,tabularx,amsmath,subfigure,amssymb,amsfonts,eufrak}
\usepackage[dvips]{epsfig}
\newcommand{\half}{\frac{1}{2}}

\newcommand{\realnumbers}{{\mathbb R}}
\newcommand{\complexnumbers}{{\mathbb C}}

\newcommand{\sige}{\sigma_{\scriptscriptstyle 1}}
\newcommand{\sigt}{\sigma_{\scriptscriptstyle 2}}
\newcommand{\bRN}{\mathbf{R}_{N}}

\newcommand{\bRn}{\mathbf{R}_{n}}
\newcommand{\brn}{\mathbf{r}{_n}}

\newcommand{\br}{\mathbf{r}}
\newcommand{\bR}{\mathbf{R}}
\newcommand{\bracc}{\mathbf{r'}}
\newcommand{\bRacc}{\mathbf{R'}}

\newcommand{\px}{p_{x}}
\newcommand{\py}{p_{y}}
\newcommand{\pz}{p_{z}}
\newcommand{\bolz}{k_{\scriptscriptstyle \rm{B}}}
\newcommand{\bfeta}{\mathbf{\eta}}
\newcommand{\bnu}{\mathbf{\nu}}
\newcommand{\Nc}{N_{\rm{c}}}
\newcommand{\Nw}{N_{\rm{w}}}

\makeatletter
\@addtoreset{equation}{chapter}
\@addtoreset{footnote}{chapter}

\renewcommand{\@makecaption}[2]{
\vspace{10pt}%
\setstretch{1}{\bf #1: \it #2}
}
\makeatother

 \def\department#1{\gdef\@department{#1}} 
 \def\submitdate#1{\gdef\@submitdate{#1}} 
  
 \def\statement{ 
         Thesis presented in partial fulfilment of the requirements 
         for the degree of \\ \uppercase\expandafter{Master of Science} at the  
         University of Stellenbosch. \\ [12pt] 
         Supervisor : Dr. Kristian M\"uller-Nedebock\\ [8pt] 
         } 
 
 \def\titlepage{ 
     \thispagestyle{empty} 
     \vglue .5in\vfil       
     \begin{singlespace} 
     \begin{center} 
         \parindent=0pt 
         {\Huge\bf Polymer Networks at Surfaces}\\ [12pt] 
         By \\ [12pt] 
         {\sc Wendy Leigh Vandoolaeghe}\\ [10cm] 
     \statement 
     April 2003
     \end{center} 
     \end{singlespace} 
    }

\def\declaration{ 
    \hbox{ } 
    \vspace{5cm} 
    \begin{center} 
        {\bf \large\uppercase{Declaration}} \\ [12pt] 
        I, the undersigned, hereby declare that the work contained  
        in this thesis is my own original work and that I have not  
        previously in its entirety or in part submitted it at any 
        university for a degree. 
        \vskip 2cm 
        \leftline{\rule[15pt]{5cm}{.3pt}  
                  \hspace{-5cm} 
                  Signature 
                  \hspace{8cm} 
                  \rule[15pt]{4cm}{.3pt}  
                  \hspace{-4cm} 
                  Date 
                  \hfil}          
    \end{center} 
    \clearpage
} 
 \def\abstract{ 
    \hbox{ } 
    \vspace{1cm} 
    \begin{center} 
        {\bf \large\uppercase{Abstract}} \\ [12pt] 
        \parbox[h]{14cm}
{In this thesis the formation and properties of a polymer gel on and at a
surface are investigated. The gel under investigation is defined as a
three-dimensional network of macromolecules that form permanent links
with one another and also with confining planar
surfaces. The
precise location of the crosslinks on the wall or on another
macromolecule is not known prior to linking, and will differ from
sample to sample. However, once the crosslinks are formed, they are
assumed to be permanent. This random linking is the
source of the disorder in the system, over which a quenched average has to be
taken.  An existing model \cite{DeamEdwards} of
network formation, with polymer-polymer crosslinks, is  extended to incorporate a surface
and polymer-surface crosslinks. Within the framework of replica
theory, statistical averages and physical properties of the system are calculated by means of a variational
approach. Macroscopic information, in terms of
the free energy of
deformation, is obtained by using two \emph{different} potentials to simulate
the crosslinks mathematically.} 
    \end{center} 
    \clearpage
} 

 \def\thanks{ 
    \hbox{ } 
    \vspace{1cm} 
    \begin{center} 
        {\bf \large\uppercase{Acknowledgements}} \\ [12pt] 
        \parbox[h]{14cm}{
The realization of this thesis would not have been possible without
financial assistance  from Stellenbosch University,
the National Research Foundation (NRF) and the Harry Crossley bursary.
The financial assistance of the Department of Labour (DoL)
          towards this research is hereby acknowledged. Opinions
          expressed and conclusions derived at, are those of the
          author and are not necessarily to be attributed to the DoL.

\vspace*{1cm}

          }
          \end{center}
          \clearpage
}

\renewcommand\listfigurename{List of Figures} 
\renewcommand\contentsname{Table of Contents}

\department{Physics}
\title{\Huge\bf Polymer Networks at Surfaces}
\date{\today}
\author{\sc Wendy Leigh Vandoolaeghe}

\begin{document}
  \titlepage

  \frontmatter      
   \declaration
  
    \addcontentsline{toc}{chapter}{Abstract --- Opsomming}
   \abstract
        
   \addcontentsline{toc}{chapter}{Acknowledgements}
   \thanks
     
  \addcontentsline{toc}{chapter}{\contentsname}
   \tableofcontents  
  
  \mainmatter       
  
  
  \renewcommand{\chaptermark}[1]{\markright{\sc
    \chaptername~\thechapter~---~#1}}


\chapter{Introduction}\label{chap:intro}
Polymers in restricted geometries exhibit very specific and
interesting properties. The widespread interest in polymers in
confined media, is proof of the importance of understanding these properties,
which play a vital role in many industrial applications, but
also in nature\footnote{Polymers in confined geometries is a
  multidisciplinary field, attracting research interest from scientists in chemistry, material science, biology and experimental,
  simulation, and theoretical physics. In 1998 the European Associated
  Laboratory (of the Institut
  Charles Sadron, Strasbourg and the Max Planck Institute 
  for Polymer Research, Mainz) was created for the purpose of
  studying polymers in confined media.}.
Fundamental theoretical physics research 
provides the basis for investigating and predicting  properties of
these systems. Problems related to
this field are diverse, and include the adsorption behaviour of gels,
surface coatings, membranes in nano pores and biopolymers
in restricted geometries.

In particular, the formation of polymer \emph{networks} at surfaces, is crucial for a number
of applications, where  surfaces have to be  protected  against forms
of mechanical
or chemical stress, such as abrasion and corrosion
\cite{Paul}. Surface-attached
networks also play an important role in several
biomedical concepts, for example, to provide biocompatible, but stable
coatings on implant surfaces \cite{Ratner}. Most theoretical
treatments of these types of systems have been on the level of scaling
theory, and analytical treatments have been lacking. Recently,
Allegra and \mbox{Raos \cite{Allegra}} investigated a confined polymer
network, but modelled the effect of confining walls on a network by an harmonic potential
and completely ignored the possibility of wall attachments. 

The aim of this thesis is to gain a better understanding,  of a
polymer network (or gel)
that has formed at a  confining surface. In particular, we 
investigate the simplest case where the confining geometry is two parallel planar
surfaces. However, before we embark on the statistical physics theory treatment of this
problem, some relevant polymer background should be given.  

\section{What is a polymer?}
\label{sec:polymer}
A polymer is a substance\footnote{Although the terms \emph{polymer} 
  and  \emph{macromolecule} are used interchangeably, the former strictly
  refers to any type of polymeric material (rubbers, biopolymers,
  fibres, glassy and crystalline polymers) of which the macromolecule is the
  essential, common building block \cite{TreloarPolSci}.}  composed of
macromolecules, which have long
sequences of one or more types of atoms or groups of atoms linked to
each other by primary, usually covalent, bonds. Macromolecules are
formed by the process of polymerization, that is, 
by linking together many, say $N$,  monomer units through chemical reactions. The
long chain nature of macromolecules is responsible for the 
characteristic properties of polymers and sets them apart from other materials.

\subsection{Flexibility}
Flexible polymers have a large number of internal degrees of
freedom. The typical primary structure of macromolecules is a
linear chain (the backbone) of atoms connected by chemical bonds  and some \emph{pendant} atoms or groups to satisfy the
remaining valencies. By rotation about the single bonds in the
backbone the molecule changes its shape or conformation. Since there
are many of
these bonds, a wide spectrum of conformations is
available to a macromolecule\footnote{The degree of
  polymerization, $N$ is usually more than a hundred. For example,
  polymerization of $N\sim 10^{4}$ ethylene
  ($\mathrm{CH_{2}}=\!\!=\mathrm{CH_{2}}$) monomers will yield a giant
  polyethylene 
  ($[\!\!\!\text{---}\mathrm{CH_{2}}\text{---}\!\!\!]_{N}$)
macromolecule. Imagine now that this molecule only had three possible
bond rotations; then the total number of shapes it may assume will be
$\sim 3^{N}$. Furthermore, the conformation is continuously changing due to thermal
motion. 
A DNA molecule has $N\approx 10^{9}$ links. Detailed
analysis of these configurations is futile! In order to investigate the properties of a system of molecules, it is
beneficial to consider a macroscopic system. In the case of
macromolecules, even a \emph{single} molecule is a macroscopic system with
infinitely many possible conformations. Consequently, we can calculate
relevant thermodynamic quantities of even single polymer chains  by means
of statistical mechanics.}. The rotation of
chemical bonds may be hindered by
bulky pendant groups, so that some of the conformations become
unfavourable. Sometimes the interaction between neighbouring
groups leads to preferred sequences of bond orientations, which emerge
as helical or folded sections in the molecules. Thus, a polymer is termed flexible if thermal motion is
strong compared to the energy barriers associated with backbone
rotation. 

\subsection{Ideality}
\label{sec:ideal}
The simplest measure of the length of a polymer chain is the
contourlength $L$. This is the length of the stretched-out molecule, that
is, for a chain of $N$ bonds of length $\ell$ the contourlength is
$N\ell$. However, this length does not give a realistic measure of the size
of the polymer chain, which in a molten state or in a dilute solution
is \emph{coiled} up.

The conformational properties of long flexible chains can be described
by the universal random walk model first introduced by Kuhn \cite{Kuhn}. A chain
is considered as a sequence of $N$ randomly orientated bonds, each of
length $\ell$. If the bonds are completely independent of each other, the
conformation of the polymer chain resembles the trajectory of a diffusing particle under
the action of a random force, for which the solution is well
known \cite{Flory, Wiegel}. If $\bR$ is the end-to-end vector of the linear
macromolecule, the mean square displacement is given by 
\begin{equation}
 \langle R^{2} \rangle = \ell^2 N
\end{equation}
Therefore, the characteristic size\footnote{For non-linear, branched or star-polymers the
  radius of gyration $R_{g}$, which is the root mean square distance
  of the segments from the centre of mass, is the approriate quantity, and is given by
$R_{g}^{2}=\frac{1}{6}\ell^{2}N$.} $R$ of the polymer is proportional
to $N^{\half}$. The probability distribution of an ideal
chain endpoint to be at a distance $R$ from the initial point  is given by
the Gaussian probability function $P (R, N)\propto
\exp\left(-\frac{3R^{2}}{2 N\ell^{2}}\right)$ [also see Section \ref{sec:freegausschain}]. 

For ideal chains, the finite volume
of the segments and solvency effects are completely ignored. In
reality, segments cannot overlap --- called volume exclusion --- which
leads to chain expansion. The statistics of real,
non-intersecting chains are described by self-avoiding walks instead
of random walks \cite{DeGennesScaling}. 

However, polymers can adopt \emph{ideal} dimensions in
solutions in a so-called $\Theta$-solvent. In a good solvent, a chain
expands from its unperturbed, ideal dimensions to maximize the number of
\emph{segment-solvent} contacts and the coil is said to be swollen. In a poor
solvent, the chains will contract to minimize interactions with the
solvent. However, competing with this effect is the tendency for chains
to expand to reduce unfavourable \emph{segment-segment} interactions (which
is the excluded volume effect). If these two interactions are in
balance, the polymer molecule will adopt unperturbed dimensions, and
the 
solvent is said to be a $\Theta$-solvent \cite{DeGennesScaling}. 
In short, if the concentration of molecules is high enough to be
classified as a dense melt, the molecules will be forced to
interpenetrate, so that (ideal) \emph{screened} excluded-volume
statistics may be assumed.

\subsection{The Entropic Spring}
\label{sec:entropicspring}
The entropy of a macromolecule is described by the formula
\begin{equation}
  \label{eq:entropy}
  S=\bolz\ln\Omega\,,
\end{equation}
where $\bolz$ is the Boltzmann constant and $\Omega$ is
the number of possible conformations. When the chain is deformed, the
entropy change $\Delta S$ is given by,
$\Delta S = \bolz \ln \Omega_{\lambda}/\Omega_{0}$, where the
subscripts $0$ and $\lambda$ refer to the initial condition of no
strain and the condition of the oriented state under stress,
respectively. 

Near an impenetrable surface, the geometric restriction leads to a
\emph{lower} conformational entropy of the polymer.

\subsection{Elasticity}
\label{sec:elas}
Rubberlike elasticity is the consequence of molecular
arrangements --- in other words --- chain flexibility. This is quite different from the elasticity of ordinary
solids such as pure metals and crystals, where the elasticity or
the resistance to deformation under external force, arises from the
distortion of the intermolecular potential fields. For macromolecules
under strain, the intermolecular potential energy $U$ remains nearly
constant with or without strain \cite{Treloar}. The elastic driving force is
therefore entirely from the tendency for macromolecules to randomize in
order to attain the maximum entropy, and minimum free energy. For
this reason, a macromolecule is often called the \emph{entropic spring}. The entropy
spring becomes stiffer at higher \mbox{temperature $T$}, since the tendency to
randomize becomes stronger with more vigorous segmental Brownian
motions like those of molecules in a liquid. This is  in contrast with
the behaviour of most crystalline solids in which the potential energy is
weakened and the stiffness is diminished at higher temperatures as a
consequence of thermal expansion, which moves atoms further apart.

The above discussion on polymer elasticity can be summarized in terms of a
thermodynamic equation of state for the stress $f$, given by
\begin{equation}
\label{eq:fundstress}
f \propto \left(\frac{dF}{d\varepsilon}\right)_{T} =
\underbrace{\left(\frac{dU}{d\varepsilon}\right)_{T}}_{\cong 0}-T \left(\frac{d
    S}{d\varepsilon}\right)_{T}\,,
\end{equation}
where $F$ is the Helmholtz free energy and $\varepsilon$ is the length
of the strained sample. For an ordinary solid, like a diamond, the reverse of
\eqref{eq:fundstress} is true, that is,  $T\left(\frac{d
    S}{d\varepsilon}\right)_{T}$ is zero. 

\subsection{Polymer Gels}
\label{sec:polnetworks}
Gels are macroscopic network polymers that have three-dimensional structures in
which each chain is connected to all the others by a sequence of
junction points, called crosslinks. 

The crosslinks, \emph{together} with the condition of chain flexibility, are
responsible for rubber elasticity\footnote{This is the simplest view of a
material that is expected to contain  network inhomogeneities
\cite{Panyukov} and other
defects, like trapped entanglements, which will alter its elastic
behaviour.}. The classical theory of rubber
elasticity \cite{Kuhn} predicts that the free energy, $\mathcal{F}$ of deformation
is proportional to the sum of squares of the principal extension
ratios:
\begin{equation}
  \label{eq:kuhnf}
  \mathcal{F} = \frac{\Nc \bolz T}{V} \sum_{i=x,y,z}\,\lambda_{i}^{2}\,,
\end{equation}
where $\Nc /V$ is the crosslink density of the network sample. A major
discrepancy of this model is that it assumes an \emph{affine}
deformation: if a macroscopic rubber sample is deformed by
{\boldmath$\lambda$}, then the end-to-end vector $\bR$ of any subchain between two
junction points will be equal to {\boldmath$\lambda\cdot\bR$}, after
deformation. The affinity assumption  implies that the crosslinks are
spatially fixed, and do not fluctuate. In the James and Guth model
\cite{JamesGuth}, the crosslinks are essentially unrestricted, and the resultant
free energy in \eqref{eq:kuhnf} is altered by a factor of
$\half$. In 1975 a pioneering network model was introduced by Deam and Edwards
\cite{DeamEdwards}, which models the effect of the network on a given
chain by a harmonic localizing potential. In order to calculate the free
energy of deformation, they resorted to the replica method from spin
glass theory. The resultant Deam and Edwards free energy
\eqref{eq:DeamEdwardsfreeEnergy}, again shows the same
strain-dependency than \eqref{eq:kuhnf}, but with a different front
factor and more terms depending on the crosslink density. In these
\emph{phantom} models, excluded volume interactions are ignored, and the polymer chains can pass through each other.

In reality,
polymers displaying rubber elastic behaviour, will deviate from
phantom models like
\eqref{eq:kuhnf} mainly due to the presence of chain entanglements
\cite{BallEdwardsDoi}. A detailed review of the role of entanglements
and attempts to understand these topological constraints, like the
slip-link and tube
models,  can be found in \cite{VilgisEdwards, VilgisKhol}.

Although network models are numerous\footnote{A recent review, and
  experimental comparison between different  models of unconfined
  networks is given by \cite{Urayama}.}, the demand for theoretical
treatments of \emph{surface-attached}, \emph{confined} networks, has
remained unfulfilled.

\section{Outline of chapters}
The thesis is structured as follows. In Chaper 1 we present a single,
free macromolecule that is subsequently confined in box and plate
systems. Within this simple single chain-system we introduce the concept of a
\emph{stitch} and deterministic bulk and wall crosslinkages. 

Chapter 2 deals with  quenched disorder, due to random
crosslinking, in the context of confinement and a simple
stitch-network model.

In Chapter 3 the basic statistical crosslinking model for a confined,
surface-attached model is presented. We extend the Deam and Edwards \cite{DeamEdwards} phantom
model, to include two parallel confining walls and the possibility of
surface attachments or \emph{wall-links}. This model treats the
localizing effect of the crosslinks on the system, by a harmonic
potential with a strength that is isotropic and
strain-independent. In Chapter 4 we improve upon this assumption, by
employing an inhomogeneous localization potential. For simplicity, this
is done in the framework of a \emph{brush}-network. In both models, we
start by formulating the problem in the language of 
statistical mechanics and constructing 
the partition function. Thereafter, the free energy of deformation and
the stress-strain equation are calculated. 

In this thesis we adopt the philosophy of first finding a tractable
theory, under reasonable assumptions, before attempting more realistic
situations. The first simplification, used throughout this work, is
the assumption of \emph{ideal}, Gaussian chains. The omission of 
excluded-volume effects (in contrast with the neglect of trapped entanglements) from the theory, is still a laboratory-attainable
assumption to make. These flexible, intersecting chains are
illustrated throughout the text by kinky, spaghetti-like lines, emphasizing the
fact that we are dealing with theoretic,  simplified entities. The second assumption, is that of a sufficiently
crosslinked network. This underlines the fact that we do not work near
the so-called
\emph{sol-gel} (second-order) phase transition region \cite{GoldbartII}. 

In the last chapter we briefly conclude the work and give an outline of
possible improvements and future developments.
\begin{figure}[!h]
\begin{center}
  {\psfrag{(A)}[][l]{(a)}\psfrag{(B)}[][r]{(b)}\includegraphics[width=0.65\textwidth]{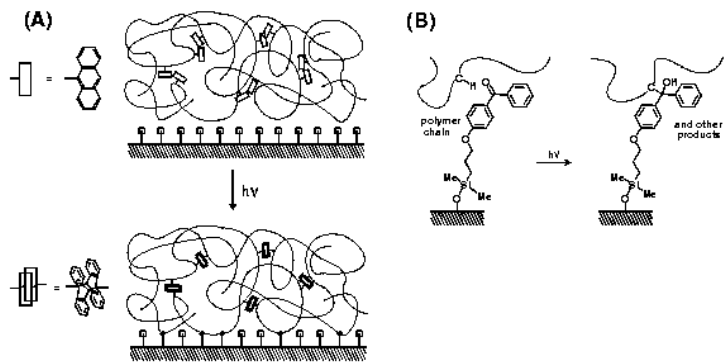}}
\caption[Schematic description of instantaneous crosslinking and
surface attachment.]{\label{fig:chem1}(A) A schematic description 
  of the preparation of a surface-attached network by simultaneous
  photochemical crosslinking and surface attachment.(B) The
  photochemical reaction used for surface attachment. \emph{[Picture taken
  from \cite{Prucker}.]}}
    \end{center}
\end{figure}



\chapter{The Single Confined Chain}\label{chap:singlechain}
Before we can probe the physics of a  polymer network formed at a
surface, we have to formulate the problem in terms of a specific
mathematical model.  In this chapter we present the fundamental
concepts that we shall employ to construct such a model. A vital
approach throughout is to use suitable Green's functions
to investigate the statistical properties of the system. Using the
Green's function approach, we firstly introduce a free Gaussian
chain\footnote{We do not attempt to give a complete introduction to the
  mathematics of ideal chains, random flight polymers and the random walk
  analogy, as this will (and already does) expend many polymer theory textbooks
  \cite{DoiEdwards, Wiegel, DeGennesScaling, Doi}. A comprehensive
  study of the Brownian chain and associated probability law is given
  in \cite{DeCloiz}.}, and then place it in a confining environment. 
The behaviour of a macromolecule is determined by the
number of different conformations it can take. By placing the polymer in a
restricted geometry, only certain specific conformations are
selected. After confining the chain, we link its two
endpoints onto the confining surface. These surface links define the
simplest case of \emph{wall-links}, and restrict the confined polymer to an even greater extent. Lastly, we introduce a single
polymer-polymer link, a so-called \emph{bulk-link}, under very specific
conditions. 

The work in this chapter together with the prefatory treatment of disorder in
Chapter \ref{chap:disorder} will serve as a platform for the full calculations.


\section{A free Gaussian chain}
\label{sec:freegausschain}
\begin{figure}[h!]
\begin{center}
  \includegraphics[width=0.75\textwidth]{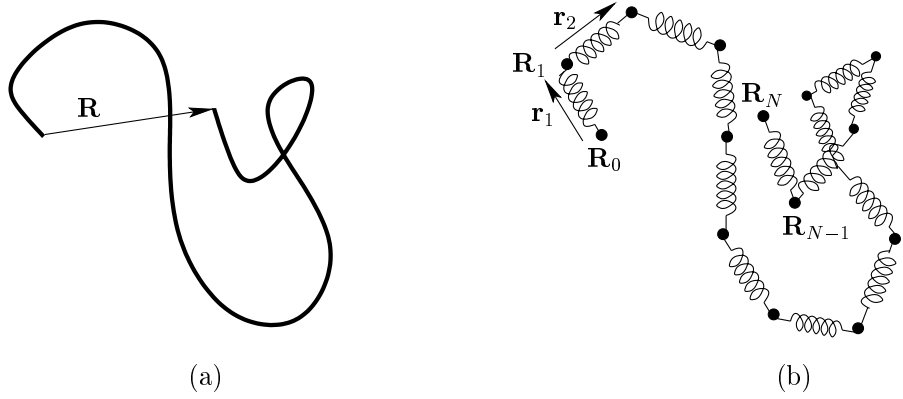}
\caption[(a) An ideal, long flexible phantom chain, and (b) its Gaussian model counterpart.]{(a) An ideal, long flexible phantom chain, and (b) its Gaussian model counterpart. The shape of the polymer can be represented by the set of position vectors of the beads 
\mbox{$S=\{\bR_{0}, \bR_{1},\ldots, \bR_{N}\}$}, or by the set of bond
vectors $\{\br_{1}, \br_{2},\ldots, \br_{N}\}$, where
$\brn=\bRn-\bR_{n-1}$. The end-to-end vector $\bR=\bRN-\bR_{0}$ characterizes the
size of the chain.}
\end{center}
\label{fig:beadspring}
\end{figure}

One way of representing an ideal macromolecule is via the standard
Gaussian or \emph{bead-spring} model, Figure \ref{fig:beadspring}
(b). For the standard Gaussian model, the chain conformation is
specified by the set $S=\{\bR_{0}, \bR_{1},\ldots,
\bR_{N}\}$ of $N$ \emph{beads}, which can be thought of as $N$
repeating monomer units of the chain.  
The conformations of an ideal macromolecule coincide with the random walk path of a Brownian particle. Since Brownian motion is a Markovian process, the ideal chain also belongs to the class of Markov chains. Let $\psi(\bRn, \bR_{n-1})$ represent the linear memory that describes the 
bonds between a pair of link neighbours. The memory of the chain
direction is lost over a distance comparable to its persistence length\footnote{The persistence length of a long, flexible, ideal chain
  is related to the effective Kuhn segment length $\ell$ of a freely jointed
  chain of $N=L/\ell$ segments, such that $\langle\bR^{2}\rangle=N\ell^2$. In the
  context of the Gaussian chain model, the Kuhn step length is defined
  by $\langle(\bRn-\bR_{n-1})^{2}\rangle=\ell^2$, and denoted by the term (RMS)
  \emph{link length}.} $\approx \ell$. The probability of a given polymer shape $S$ is thus given by 
$\Psi[S] =  \prod_{n=1}^{N}\,\psi(\bR_n,\bR_{n-1})$, and represents the connectivity of the ideal 
macromolecule. For the bead-spring model, the bond lengths have the
Gaussian distribution, such that
$\psi(\bRn, \bR_{n-1})$ is defined as follows:
\begin{equation}
        \label{eq:beadspring}
        \psi(\bRn,\bR_{n-1}) = \left(\frac{3}{2 \pi \ell^{2}}\right)^{\frac{3}{2}}\, \exp\left\{-\frac{3}{2 \ell^{2}}
        (\bRn-\bR_{n-1})^{2} \right\}\,.
\end{equation}
The partition function is computed by integrating the distribution
$\Psi[S]$ over all possible conformations [8]. 

When the chain is very long, the monomer index $n$ may be regarded as
a continuous variable. In this continuous representation, 
the discrete bond vector $\brn=\bRn-\bR_{n-1}$ is replaced by the
functional derivative $\partial \bRn/\partial n$ \cite{DoiEdwards}. If the endpoints of the
macromolecule are fixed, at say positions $\br$ and $\br'$ in space, the partition function is the Wiener path integral
\begin{eqnarray}
 \label{eq:greenidealchain}
        G(\br,\bracc,N)&=&
        \mathcal{N}\int_{\bR(0)=\bracc }^{ \bR(N)=\br }[\,
        \mathcal{D}\bR\,]\, \exp\Big\{-
        \frac{3}{2\ell^{2}} \int_{0}^{N}\bigg(\frac{\partial
            \bRn}{\partial n}\bigg)^{2}\,dn\Big\} ,
\end{eqnarray}
which is a Green's function and a solution of the following diffusion type equation:
\begin{equation}
   \label{eq:diffeqb}
   \Big[\, \frac{\partial}{\partial
  N}-\frac{\ell^{2}}{6}\frac{\partial^{2}}{\partial\br^{2}}\,\Big]\,G(\br,\bracc;N)=\delta(\br-\bracc)\,\delta(N).
\end{equation}
The solution of \eqref{eq:greenidealchain} or \eqref{eq:diffeqb} is the Gaussian distribution, given by
\begin{equation}
        \label{eq:gaussb}
        G(\br,\br';N)= \left(
          \frac{3}{2 \pi \ell^{2}N} \right)^{3/2} \exp \Big\{ -\frac{3}{2
            \ell^{2}} \frac{(\br-\br')^{2}}{N }\Big\} .
\end{equation}
If all interactions are neglected, a free polymer chain is
characterized by an average
end-to-end length of \mbox{$R=\langle \mathbf{R}^{2}\rangle=\ell N^{\frac{1}{2}}$.}
The number of configurations of a free, ideal chain between points
$\br'$ and $\br$ and chain contourlength $L$ between those points
thus corresponds to the statistical weight of a random walk starting at
position $\br'$ and ending at $\br$, in $N=L/\ell$ steps, where $\ell$ is the Kuhn step-length of a polymer chain.
\section{A Single Chain in a Box}
\label{sec:singlechaininbox}
Consider a phantom molecule confined in a box of
volume \mbox{$V=h^{2}h_{z}$}, with length $h$ in the $x$ and $y$
directions, and $h_{z}$ in the $z$-direction. The statistical weight
of the confined polymer chain which starts at $\bracc$ and ends at $\br$ in
$N$ steps is given by the Green's function G($\bf{r}$ ,$\bf{r'}$ ;$N$),
written as a Wiener path integral \cite{DoiEdwards} :
\begin{eqnarray}
 \label{eq:greena}
        G(\br,\bracc;\,N)&=&
        \mathcal{N}\int_{\bR(0)=\bracc }^{ \bR(N)=\br }[\,
        \mathcal{D}_{\rm{box}} \bR\,]\, \exp\Big\{-
        \frac{3}{2\ell^{2}} \int_{0}^{N}\bigg(\frac{\partial
            \bRn}{\partial n}\bigg)^{2}\,dn\Big\} ,
\end{eqnarray}
where the normalization $\mathcal{N}$
refers to the number of configurations, \eqref{eq:gaussb} of a
completely free polymer chain, starting at $\bracc$ and ending at
$\br$ in $N$ steps, with Kuhn step length $\ell$. The notation $\mathcal{D}_{box}\bR$ means evaluating the path integral only in the 
allowable box-region. In terms of an attractive potential $A$, the Green's
function \eqref{eq:greena} can be written in terms of unconfined integration \cite{FreedEdwards},
\begin{eqnarray} 
 \label{eq:greenb}
         G(\br,\bracc\,;N)&=&
        \mathcal{N}\int_{\bR(0)=\bracc }^{\bR(N)=\br } [\,
        \mathcal{D}\bR\,]\, \exp\Big\{-
        \frac{3}{2\ell^{2}} \int_{0}^{N}\dot{\bR}_n^{2}\,dn + \int_{0}^{N}\,A(\bRn)\,dn\Big\}, 
\end{eqnarray}  
where
\begin{equation}
  \label{eq:pota}
  A(\mathbf{R},n)\equiv \ln\Theta(\bRn)\,, \qquad\textrm{with}\qquad  \Theta(\bR)= \left\{ \begin{array}{ll} 1 & \textrm{if
             $\bR$ in $V$}\\
           0 & \textrm{otherwise} \end{array} \right.
\end{equation}
represents an infinite potential wall. The solution of
\eqref{eq:greena} is equivalent to solving 
the inhomogeneous
differential equation 
\begin{equation}
  \label{eq:diffeqa}
  \Big[\, \frac{\partial}{\partial
  N}-\frac{\ell^{2}}{6}\frac{\partial^{2}}{\partial\br^{2}}-A(\br,N)\,\Big]\,G(\br,\bracc;N)=\delta(\br-\bracc)\delta(N),
\end{equation}
which, with the substitution of the potential $A$ in
\eqref{eq:pota}, simplifies to a diffusion equation of the form given
by equation \eqref{eq:diffeqb}, together with the boundary condition
that $G(\br,\bracc;N)=0$ at the confining surface.  The solution $G$ is constructed from eigenfunctions that vanish on and
outside the boundaries of the box\footnote{The solution was obtained
  by using an eigenfunction expansion method \cite{Morse}, but can
  also be obtained by using the method of images \cite{Carslaw}.}.  The coordinates are separable 
\begin{equation}
  \label{eq:greenc}
  G(\br,\bracc;N) = G_{x}(x,x';N)\,G_{y}(y,y';N)\,G_{z}(z,z';N)\, ,
\end{equation}
and since the polymer in a box is a symmetric problem, all the
parts, for example $G_{x}$, have the same form:
\begin{equation}
  \label{eq:onegreen}
  G_{x}(x,x';N)=\frac{2}{h}\sum_{\px = 1}^{\infty}\sin\frac{\px\pi
    x}{h}\,\sin\frac{\px\pi x'}{h}\,
  \exp\left(-\frac{\ell^{2}}{6}\frac{\pi^{2}}{h^{2}}\,\px^{2}\,N\right).
\end{equation}
Equation \eqref{eq:greenc} is the probability that one endpoint
$(n=N)$ of a
box-confined polymer is $\br$, provided that the other endpoint
$(n=0)$ is
at $\br'$.
The partition function of all possible conformations, for a polymer
with \emph{endpoints free}, is given by 
\begin{eqnarray}
  \label{eq:partitionconfineda}
  \mathcal{Z} &=& \int\,d\br\,\int\,d\br'\,G\,(\br,\br';N)\\
   \label{eq:partitionconfinedb} 
   &=&\left(\frac{2}{\pi}\right)^{6}\,
   \sum_{p_x,\,p_y,\,p_z=1,3,\ldots}^{\infty}\,\left(\,\frac{8\, h^{2}\, h_{z}}{p_{x}^{2}\,p_{y}^{2}\,p_{z}^{2}}\,
   \right)\,\exp\left\{-\frac{\ell^{2} \pi^{2}N}{6}\Big[\,\frac{(p_{x}^{2}+p_{y}^{2})}{h^{2}}+\frac{p_{z}^{2}}{h_z^2} \,\Big] \right\}.
\end{eqnarray}
The endpoints of the confined molecule in
\eqref{eq:partitionconfinedb} are not \emph{fixed} at positions somewhere in the box or at the boundaries, as
will be the case in the next section.
\section{A polymer \emph{stitch}}
\label{sec:stitch}
\begin{figure}[htb]
\begin{center}
  \includegraphics[width=0.5\textwidth]{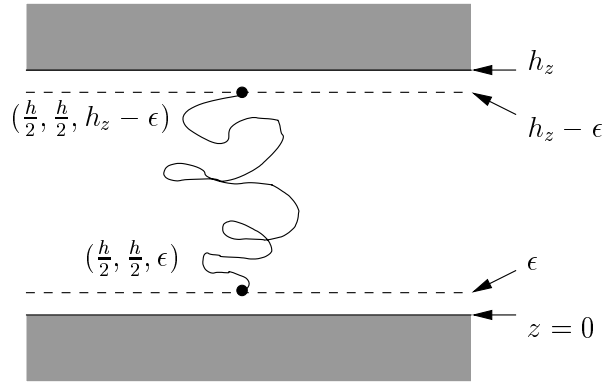}
\caption[A single polymer chain stitch.]{\label{fig:onestitch}A single polymer chain ''stitch'' with end-points held  at
  two fixed points}
\end{center}
\end{figure}

Next we add two wall-links to the system by linking the ends of the
polymer to the top and bottom planes of the box, creating  a polymer
\emph{stitch}, as illustrated in Figure~\ref{fig:onestitch}. 
We are interested in the partition function of a single polymer chain
\emph{stitch} in a box with volume $V=h^2 h_z$. To this end, we fix the end
points of the chain a small distance $\epsilon$ from the floor and lid of the box\footnote{Instead of a
  polymer that is cross-linked \emph{exactly on} the wall, we have
  localized the wall cross-link a distance $\epsilon$, of the order of
  the link length $\ell$, away from the walls, otherwise the partition function for the polymer vanishes as
  should be expected.}, permanently, at 
\begin{equation}
    \bracc = (\frac{h}{2},\frac{h}{2},\epsilon) \qquad
    \text{and}\quad \br = (\frac{h}{2},\frac{h}{2},h_{z}-\epsilon),
    \label{eq.coora}
\end{equation}
By substituting these coordinates in equation \eqref{eq:greenc} and
\eqref{eq:onegreen}, we obtain the partition function for a
\emph{stitch} confined in a box:
\begin{eqnarray}
  \label{eq:part1}
  \mathcal{Z}(\br,\bracc;N)&=&
  Z_{x,y}(N,h)\,\,Z_{z}(N,h_z)\,,
\end{eqnarray}
with its three Cartesian parts written as follows:
\begin{eqnarray}
  \label{eq:part2}
  Z_{x,y}(N,h)&=& \Big(\frac{2}{h}\Big)^{2}\,\sum_{\px,\py
    =1,3,\ldots}^{\infty}\,\exp\Big\{-\frac{\ell^{2}}{6}\frac{\pi^{2}}{h^{2}}\,(\px^{2}+\py^{2})\,N\Big\},\\
  \label{eq:part3}
  Z_{z}(N,h_z)&=&\frac{2}{h_z}\, \sum_{\pz = 1}^{\infty}\, 
     \sin\big(\frac{\pi \pz\epsilon}{h_{z}}\big)\,\sin\big(\frac{\pi\pz[h_{z}-\epsilon]}{h_{z}}\big)\,
    e^{-\frac{\ell^{2}}{6}\frac{\pi^{2}}{h_{z}^{2}}\,\pz^{2}\,N}.
\end{eqnarray}
It is at this point necessary to look at the relationships that exist
between the box dimensions, $h_z$ and $h$ and the length scale of the
polymer $\sqrt{N}\ell$.

\subsection{Three limiting cases}
\label{sec:limitingcases}
\begin{figure}[h]
\begin{center}
  \includegraphics[width=0.85\textwidth]{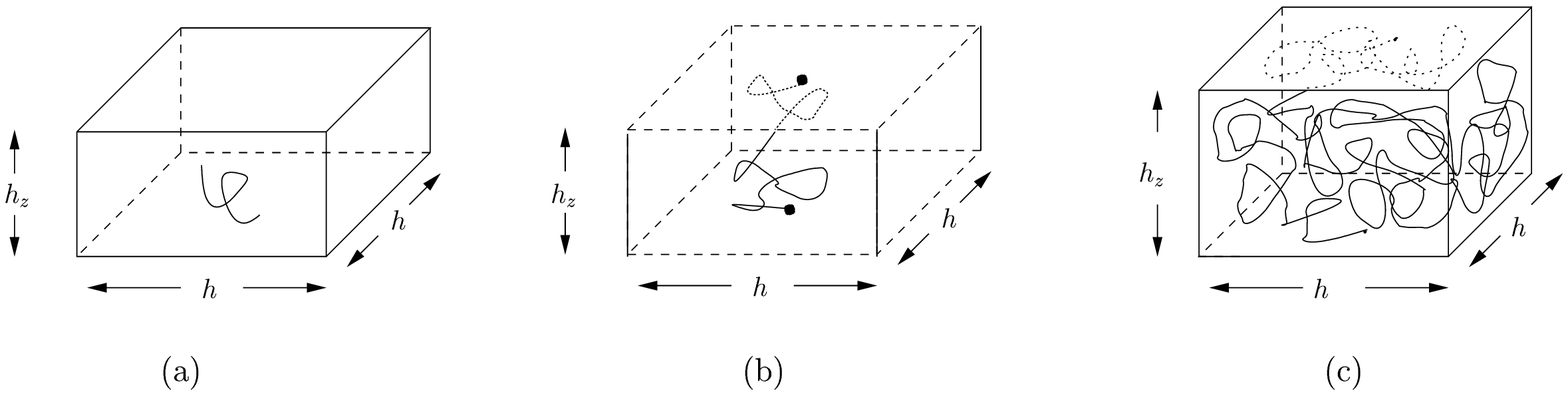}
\caption[Illustrating the relationship between the size of the polymer
and the confinement dimensions.]{\label{fig:3boxstitch}Relationship between the size of the polymer and confinement
  dimensions: (a) A small polymer in a big box, \eqref{eq:limitf0};
  (b) Polymer big enough to be a stitch, \eqref{eq:limitf1}; (c) A
  large polymer, densely occupying the box, \eqref{eq:limitf2}.}
\end{center}
\end{figure}

The relationship between the confining box dimension, $h$ and $h_{z}$,
and the size of the chain $\sqrt{N}\ell$ determines the type of
physical situation we are facing. It will also dictate in which of
these limits future calculations will be done. There exist three distinct limits:
\begin{eqnarray}
\label{eq:limitf0}
  \textrm{(i)}\qquad \sqrt{N}\ell\, \ll& h_{z}  & \ll \,h\, \\
\label{eq:limitf1}
  \textrm{(ii)}\quad \qquad  h_{z}\,  \ll& \sqrt{N}\ell & \ll \,h\\
\label{eq:limitf2}
 \textrm{(iii)}\quad \qquad  h_{z}\,   \ll& \sqrt{N}\ell & \text{ and}
 \quad h\,   \ll\: \sqrt{N}\ell\,,
\end{eqnarray}depicted graphically in Figure~\ref{fig:3boxstitch}.
In the first case (i) \eqref{eq:limitf0}, we have a small polymer in a large
confining environment, which intuitively means that it will be very difficult to fix the endpoints a distance
$h_z$ from each other. This difficulty is portrayed by the
sum argument in \eqref{eq:part3}, which in this case oscillates
rapidly around zero. Physically, (i) is analogous to a
\emph{short} random walk in a \emph{large} cavity
\cite{GrosbergKhok}. Since the polymer is too small to fix at the top
and bottom faces of the confining box, we investigate the original
partition function \eqref{eq:partitionconfineda}, for a polymer with
\emph{free} ends. Since the characteristic chain size $R$ is small relative to the box
dimensions, the sums in
the partition function \eqref{eq:partitionconfinedb} will be dominated by $\frac{1}{p^{2}}$ terms
\cite{DoiEdwards}. In this limit\footnote{If $\sqrt{N}\ell \ll
  h_{i}$, the partition function $\mathcal{Z} \simeq h^{2}\,h_{z}
  \equiv V$, since $\exp\{-\frac{1}{6}
  \pi^{2}\,p_{i}^{2}(\,\frac{\sqrt{N}\ell}{h_{i}}\,)^{2}\}\,\simeq 1$,
  $i=\{x,y,z\}$. The remaining sum is then of the
  form $\sum_{p=1,3,\dots}^{\infty}\,1/p^{2}$, which converges to $\frac{\pi^{2}}{8}$.}, the free energy $F$ is like that of a
perfect gas, with corresponding pressure $P$, given by:
\begin{equation}
  \label{eq:freeideal}
  F \approx -\bolz T\,\ln\,V\,,\qquad\text{and}\qquad   P =
  \frac{\partial F}{\partial V} \Rightarrow P\,V = \bolz \,T\,.
\end{equation}
In the second relation (ii) \eqref{eq:limitf1} the stretched
out chain is small compared with $h$, but long enough to reach the
endpoint cross-link locations in the $\hat{z}$ direction. The sum in
$Z(N,h_z)$ in \eqref{eq:part3} is then entirely dominated by the
$p_{z}=1$ term, so that the $h_z$-dependent part is given by:
\begin{equation}
  \label{eq:zlimit1}
   Z_{z}(N,h_z,\lambda)\cong\frac{2}{ h_z}\, 
     \sin\big[\frac{\pi\epsilon}{h_{z}}\big]\,\sin\big[\frac{\pi( h_{z}-\epsilon)}{ h_{z}}\big]\,
    e^{-\frac{\ell^{2}}{6}\frac{\pi^{2}}{h_{z}^{2}}N}.
\end{equation}
On the other hand, a continuous Gaussian approximation compares well with the sum over $p_{x}$ and 
$p_{y}$ in \eqref{eq:part2}, that is:
\begin{eqnarray}
   \label{eq:gausslimit1}
   Z_{x,y}(N,h) &\cong&
   \left(\frac{1}{2}\right)^{2}\,\left(\frac{2}{h}\right)^{2}\,\int\limits_0^\infty\int\limits_0^{\infty}\,
   e\,^{\{-\frac{N\ell^{2}\pi^{2}(\rho_{x}^{2}+\rho_{y}^{2})}{6h^{2}}\}}\,d\rho_{x} \,d\rho_{y}\\
   \label{eq:gausslimit1ans}  
   &=& \frac{3}{2\pi\ell^{2}N},
 \end{eqnarray}
which is the same as not restricting the $\hat{x}$ and $\hat{y}$
dimensions \emph{at all}, and describing each of these components of the chain by a
free chain Green's function \eqref{eq:gaussb}. In this limit of
sufficiently large
$h$, and infinitesimal $\epsilon$, the problem
is equivalent to a chain restricted between two parallel plane surfaces with an
\emph{$h$-independent} partition function,
\begin{equation}
   \label{eq:part7}
   \mathcal{Z}(h_{z})\simeq\frac{\pi\epsilon^{2}}{3N\ell^2
     h_z^{3}}\,\exp\left(\,-\frac{N\ell^2\pi^2}{6 h_z^2}\right)\, \qquad \qquad \quad \qquad
   \Longleftrightarrow \,\textrm{limit (ii)} \quad   \left[h_{z}\ll \sqrt{N}\ell  \ll h\right]\,.
\end{equation}
In the third  relation (iii) \eqref{eq:limitf2} the polymer contourlength
is large compared with both $h$ and $h_z$. Consequently, the
dominating contribution to the sums in \eqref{eq:part2} and
\eqref{eq:part3}, is given by the $p_{x}=p_{y}=p_{z}=1$ term, leading to a
partition function (in the limit of small $\epsilon$),
\begin{equation}
  \label{eq:part9}
  \mathcal{Z}\,(h,h_{z}) \simeq
   \frac{8\pi^{2}\epsilon^{2}}{h^{2}h_z^{3}}\,
   \exp\left\{-\frac{N\ell^2\pi^2}{6}\Big(\frac{2}{h^2}+\frac{1}{h_z^2}\Big)\right\}\, \quad \Longleftrightarrow \,\textrm{limit (iii)} 
    \quad \left[h_{z}   \ll  h  \ll\sqrt{N}\ell \right]
\,,
\end{equation}
which is dependent on all the parameters of the confinement. In this
instance the single polymer chain will tend to \emph{fill} up all the
available space. 
\section{\emph{Stitch} Strains}
\label{sec:stitchelas}
The mechanical properties of a polymer \emph{stitch} may be
investigated by looking at what happens during strain. It is possible
to characterize the stress acting at a point on a substance in terms
of three \emph{principal} stresses or extension ratios, acting along mutually perpendicular
principal axes \cite{Treloar, Young}. The extension ratio $\lambda_i$ is defined as
the deformed length in direction $\hat{i}$ divided by
the original length of the sample \cite{TreloarPolSci}. 

In principle there are many possible deformations that can be employed
to verify the predictive ability of theoretical elastomer models
describing stress-strain behaviour. However, in practice it is
uniaxial deformation that is most often studied, due to its
experimental simplicity\footnote{However, only general \emph{bi}-axial strains cover all
accessible pure homogeneous deformations for an incompressible
material, and is the preferred method of testing real network
theories that account for topological features like trapped entanglements \cite{Urayama}.}.

An important characteristic of macromolecules in the gel state is that
any type of applied stress will most readily influence the gel's shape,
without appreciably changing its volume \mbox{\cite{TreloarGee,Treloar}}. This observation has led to
the terminology \emph{incompressible} or \emph{indilatable} to describe
the mechanical behaviour of gels. In most cases we shall thus be concerned with an \emph{isovolumetric},
uniaxial deformation \eqref{eq:isomatrix}, where an elongation (compression) by a factor
$\lambda$ along the $\hat{z}$ axis results in compression (dilation)
by a factor $1/\sqrt{\lambda}$ along the other axes. The tensor
$\mathbf{\Lambda}$ expresses the isovolumetric, uniaxial
(macroscopic) deformation of a point
$\br \to \br'=\mathbf{\Lambda}\cdot \br$, 
\begin{equation}
  \label{eq:isomatrix}
  \mathbf{\Lambda} =
  \left( \begin{array}{ccc}
      \lambda^{-\frac{1}{2}} & 0 & 0 \\
      0 &  \lambda^{-\frac{1}{2}} & 0 \\
      0 & 0 & \lambda
      \end{array} \right), \qquad \lambda>0\,.
\end{equation}
The fundamental property of polymer gels, namely the constancy of
volume during deformation, makes it possible to define the complete
state of strain in terms of a single parameter $\lambda$.  
\subsection{Case 1: Deforming the plate system}
\label{sec:stitchmedium}
\begin{figure}[h]
\begin{center}
  \includegraphics[width=0.9\textwidth]{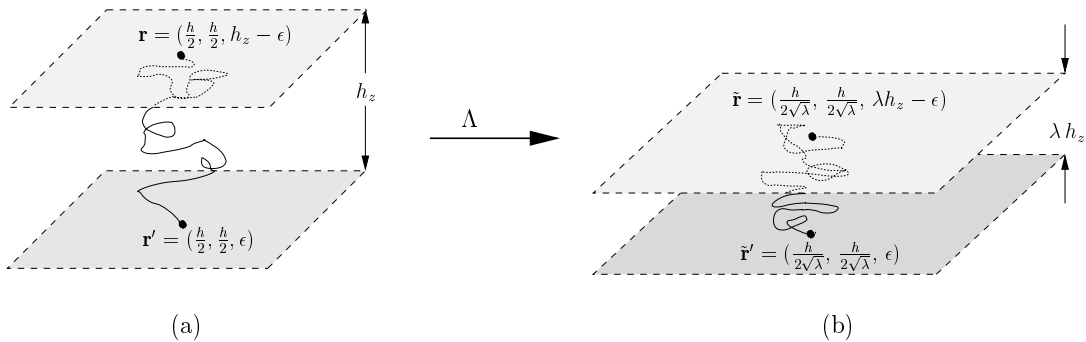}
\caption[Polymer stitch in the unstrained state, and after a uniaxial compression]{\label{fig:deformnormalstitch}Polymer stitch in the (a) the unstrained state; (b) strained
  (unidirectional compression) state, confined between two long
  parallel plates, corresponding to limit (ii).}
   \end{center}
\end{figure}

Firstly, we investigate the effect of strain, defined by $\mathbf{\Lambda}$, on a
macromolecule linked at fixed positions to two parallel plates
(Figure~\ref{fig:deformnormalstitch}).  This situation corresponds to
limit (ii) with $h \to \infty$.
In this case it is appropriate to use the
partition function $\mathcal{Z}$, already derived in
\eqref{eq:part7}, to calculate the free energy of
the plate-system as follows:
\begin{eqnarray}
  \label{eq:vrysimpelelonga}
  F &=&-\bolz T\,\ln\,\mathcal{Z}(h_{z}) \\
   \label{eq:vrysimpelelongb} &=&
  -\bolz T\,\ln \,\left[\frac{\pi\epsilon^{2}}{3N\ell^{2}h_{z}^{3}}
  \right]+\frac{\bolz T N \pi^{2}\ell^{2}}{6 h_{z}^{2}}\,.
\end{eqnarray}
If we perform a simple elongation ($\lambda>1$) or compression ($\lambda<1$), the change in free energy is given
by: 
\begin{equation}
  \label{eq:changefsimpel}
  \bigtriangleup\,F =  \bolz T\,\ln \,\lambda +\frac{\bolz T N \pi^{2}\ell^{2}}{6 
    h_{z}^{2}}\,\left(\frac{1}{\lambda^{2}}-1 \right)\,,
\end{equation}
which has similar characteristics to that of De Gennes' scaling theory
calculation \cite{DeGennesScaling} for the
energy required to squeeze an ideal chain (of unperturbed size
$R_{0}$) trapped in a tube or cavity of
diameter $h_z$:
\begin{equation}
  \label{eq:Degennes}
  \bigtriangleup\,F_{\rm{deGennes}} \cong \bolz T \,\frac{R_{0}^{2}}{h_{z}^{2}}\,\qquad,
  R_{0}= N^{\frac{1}{2}}\,\ell.
\end{equation}
The free energy of the form in \eqref{eq:Degennes} holds generally for
any type of ideal chain confinement \cite{GrosbergKhok}. It originates from the
decrease in the number of available conformations --- or lowering of the
entropy --- when the polymer is
restricted. The only force
acting on the system during strain is the tensile force in the
direction of extension (or compression) $\lambda$. 
\begin{figure}[t]
\psfrag{f}[][l]{$f$}\psfrag{lam}[][l]{$\lambda$}\psfrag{Confined}[][c]{\emph{Confined}}\psfrag{Free}[][r]{\emph{Free}}
{\centering\includegraphics[width=0.45\textwidth]{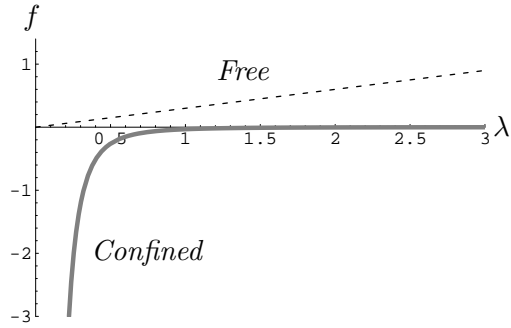}
\caption[Stress-strain plot for a polymer stitch confined
between two parallel planar surfaces.]{Plot of the stress $f$ \eqref{eq:tensileforce} versus the
  deformation ratio $\lambda$ for a plate system, in limit (ii). The
  solid line is for a stitch confined between the plates; the dashed
  line corresponds to a free chain (with fixed endpoints). Since the
  free energy was evaluated in the limit (ii), only small elongations,
  $\lambda\,h_{z} \ll \sqrt{N}\ell$
  are allowed. \label{fig:plotstres1}}}
\end{figure}
The magnitude of the force per unit cross-sectional area measured in the unstrained
state, depicted in Figure \ref{fig:plotstres1}, is given by 
\begin{equation}
  \label{eq:tensileforce}
  f = \frac{1}{V}\,\frac{\partial (\bigtriangleup F)}{\partial\lambda}
  = \frac{\bolz T}{V}\left(\frac{1}{\lambda}-\frac{N \pi^{2}\ell^{2}}{3h_{z}^{2}}\,\frac{1}{\lambda^{3}}\right)\,.
\end{equation}
An alternative  way to model a \emph{stitch} 
is to simulate the effect of the wall confinement by a simple harmonic potential
\cite{Allegra}. If the potential is $U
=\frac{q^{2}\ell^2}{6}\int_{0}^{N}\,\bR^{2}dn$, the wall-confinement is
enforced by choosing the localization parameter to be of the form $q
\simeq h_{z}^{-\frac{1}{2}}$. The partition function in this case is
analogous to the propagator\footnote{Similar calculations that employ a
  Green's function solution for a harmonic localization of the
  crosslinks, is given in more detail in Chapter \ref{chap:realnetwork}.} of a particle in an harmonic potential,
with initial and terminal points of its trajectory fixed on one of the
plates \cite{Feynman}. The elastic free energy calculated from the
harmonic approximation is consistent with \eqref{eq:changefsimpel}, and
confirms the form of the stress-strain graph, Figure
\ref{fig:plotstres1}.

When an \emph{unconfined} ideal chain \eqref{eq:gaussb} is deformed from an end-to-end
length $h_{z}$ to $\lambda\,h_{z}$, its end links will undergo an
external stretching force of magnitude $f = \bolz T\,\frac{3 \lambda
  h_{z}^{2}}{N\ell^2}$ \cite{DeGennesScaling,GrosbergKhok}. 
The important difference between a confined and
unconfined polymer \emph{stitch}, is illustrated by the stress-strain
relationship in Figure \ref{fig:plotstres1}.

\subsection{Case 2: Deforming the box system}
\label{sec:deformlongstitch}
\begin{figure}[!h]
\begin{center}
  \includegraphics[width=0.9\textwidth]{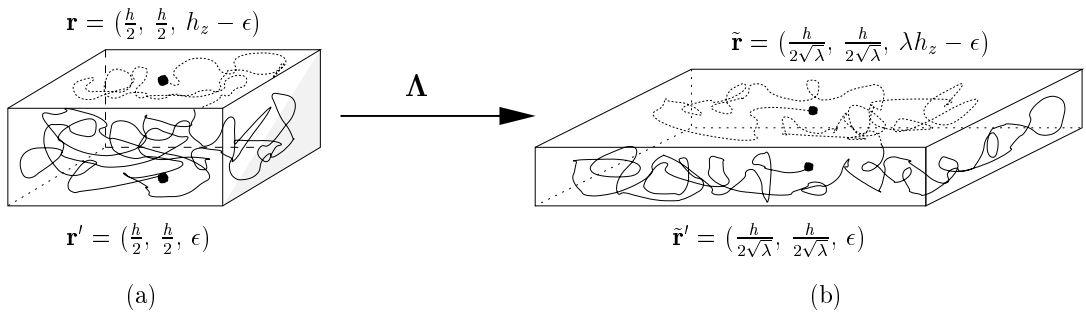}
\caption[Polymer stitch in a box in the unstrained state, and after a uniaxial
compression.]{\label{fig:deformstitch}Polymer stitch in the (a) the unstrained state; (b) strained
  (unidirectional compression) state, in the limit (iii).}
   \end{center}
\end{figure}

\begin{figure}[!h]
\psfrag{f}[][l]{$f$}\psfrag{l}[][l]{$\lambda$}\psfrag{Compression}[][c]{\emph{Compression}}\psfrag{Elongation}[][l]
{\emph{Elongation}}
{\centering\includegraphics[width=0.46\textwidth]{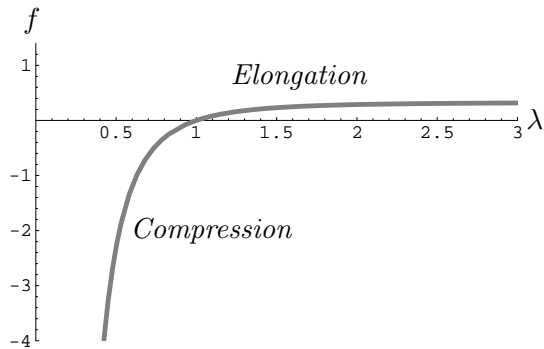}
\caption[Plot of stress versus strain for a long polymer confined in a
box.]{Plot of the stress ($f$)~\eqref{eq:tensileforce1} versus the
  deformation ratio ($\lambda$), in limit (iii) when \quad \mbox{$h \ll
  \sqrt{N}\ell$}, for a long polymer confined
  in a box.} \label{fig:plotstres2}}
\end{figure}

Secondly, we investigate the effect of strain, defined by $\mathbf{\Lambda}$, on a
macromolecule in the limit (iii), $\sqrt{N}\ell \gg h$,
as shown in Figure \ref{fig:deformstitch}. In this limit it is appropriate
to perform calculations using the partition function already
derived in \eqref{eq:part9}. 
\begin{flushleft}
It is found that the work done to strain the system, 
\begin{eqnarray}
  \label{eq:straingrootpol1}
   \bigtriangleup\,F =  \frac{1}{6}\,\bolz\,T \,N\ell^{2}\pi^{2}\left(\frac{2
       \lambda-1}{h^{2}}+\big(\frac{1}{\lambda^{2}}-1\big)\frac{1}{h_{z}^{2}}\right)\,,
\end{eqnarray}
is both $h$ and $h_{z}$ dependent. The only force
acting on the system during strain is the tensile force in the
direction of extension (or compression) $\lambda$. The stress-strain
relationship is given by 
\begin{equation}
  \label{eq:tensileforce1}
  f = \frac{1}{V}\frac{\partial(\bigtriangleup F)}{\partial\lambda}
  =\frac{ \bolz T N\ell^{2}\pi^{2}}{3\,V}\left(\frac{1}{h^{2}}-\frac{1}{\lambda^{3}\,h_{z}^{3}}\right)\,
\end{equation}
and is only valid for small elongations $\lambda\,h_{i}\ll
\sqrt{N}\ell$. Both elongation ($\lambda>1$) and compression
($\lambda<1$) is represented by a
single curve on the next page (Figure~\ref{fig:plotstres2}), illustrating the
theoretical relationship between force (stress) and $\lambda$ when
relation (iii) holds. 
\end{flushleft}
\section{One bulk-link}
\label{sec:onebulklink}
\begin{figure}[h]
\begin{center}
  \includegraphics[width=0.8\textwidth]{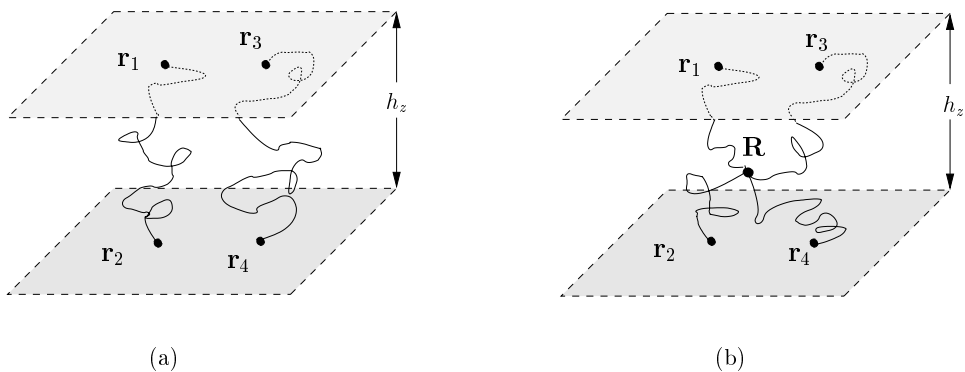}
\caption[Introducing a junction point in a two-stitch polymer system.]{\label{fig:twostitch}Two polymer stitches, (a) without a junction point; (b) with
  a junction point at $\bR=(X,\,Y,\,Z)$. The wall links are situated
  at permanent points, $\br_{1}=(x_{1},y_{1},h_{z}-\epsilon)$,
  $\br_{2}=(x_{2},y_{2},\epsilon)$,
  $\br_{3}=(x_{3},y_{3},h_{z}-\epsilon)$ and  $\br_{4}=(x_{4},y_{4},\epsilon)$.}
  \end{center}
\end{figure}

Consider two, identical macromolecule \emph{stitches}, each of length
$L=N\ell$, confined between two parallel
plates, in Figure~\ref{fig:twostitch}(a).
Next, the chains are joined together permanently
at \mbox{position $\bR$,} such that both chains meet at half of their respective
contourlengths, $\frac{1}{2}N$. We obtain
a system consisting of four wall-links and one bulk-link, as shown in
Figure~\ref{fig:twostitch} (b).
The partition function for the system is thus the  statistical weight
of one large star polymer\footnote{A regular $n$-star polymer is a macromolecule containing a single branch point from which $n$ linear
chains (identical with respect to constitution and degree of
polymerization) emanate \cite{GrosbergKhok}. The size of such a macromolecule is described
by the mean square of the radius of gyration:
${R_{g}}_{4\,\rm{arms}}^{2}=\frac{5}{4}\,N\ell^2/6$.} of total
length $2 L$. The statistical weight is determined by averaging over all
possible positions of the junction point $\bR$ for the four
half-chains \cite{DoiEdwards}, that is:
\begin{equation}
  \label{eq:partitionstich1}
  \mathcal{Z} = \int\,d \bR\, G\,\big(\br_{1},\bR;\frac{N}{2}\big)\,G\,\big(\bR, \br_{2};\frac{N}{2}\big)\,G\,\big(\br_{3},\bR;\frac{N}{2}\big)\,G\,\big(\bR,\br_{4};\frac{N}{2}\big)\,,
\end{equation}
with the points $\br_{1}$, $\br_{2}$, $\br_{3}$ and $\br_{4}$ at fixed positions
on the plates, shown in Figure \ref{fig:twostitch} (b).
The Green's function $ G\,\big(\br_{1},\bR;\frac{N}{2}\big)$ represents the
statistical weight of a chain portion  which starts at $\br_{1}$ and
ends at $\bR$ in $\frac{N}{2}$ steps. Since the chains are only
confined in the $\hat{z}$ dimension, each Green's function $G$ is the
product of two free chain contributions, \eqref{eq:gaussb} and one confined
statistical weight contribution \eqref{eq:onegreen}, for example: 
\begin{eqnarray}
  \label{eq:greensex}
  G\,(\br_{1},\bR;\frac{N}{2}) &=&  G_{x}\,(x_{1},X;\frac{N}{2})\, G_{y}\,(y_{1},Y;\frac{N}{2})\, G_{z}\,(z_{1},Z;\frac{N}{2}) \\
  \label{eq:greensexb}
  &=& \frac{3}{\pi \ell^{2}N}
  \,\exp \left\{-\frac{3}{N\ell^{2}}\left(\,(X-x_{1})^{2}+(Y-y_{1})^{2}\, \right) \right\}
  \nonumber \\ & & {} \times
  \frac{2}{h_{z}}\,\sum_{p=1}^{\infty}\,\sin\frac{\pi p
    Z}{h_{z}}\,\sin\frac{\pi p
    (h_{z}-\epsilon)}{h_{z}}\,\,e^{-\frac{\ell^{2}\pi^{2}p^{2}}{12 h_{z}^{2}}N }\,.
\end{eqnarray}
If this one-link system is strained by simple elongation $\mathbf{\Lambda}$, as
defined in \eqref{eq:isomatrix}, the change in free energy is given by
\begin{eqnarray}
  \label{eq:onelinkfreeE}
  \bigtriangleup F &=& -\bolz T\,\ln
  \frac{\mathcal{Z}_\lambda}{\mathcal{Z}_{1}}\,,
\end{eqnarray} 
where $Z_{\lambda}$ denotes the strained and $Z_{1}$ the unstrained
states of the system, defined in \eqref{eq:partitionstich1}. In the limit (ii), $\sqrt{N}\ell
> h_{z} \gg \epsilon$, equation \eqref{eq:onelinkfreeE} reduces to:
\begin{eqnarray} 
 \label{eq:onelinkfreeEb}
 \bigtriangleup F &\simeq&
 \bolz T\,\left\{\,7\ln\lambda
   +\frac{\ell\pi^{2}N}{3\,h_{z}^{2}}\,\Big(\frac{1}{\lambda^{2}}-1\Big)\right.
   \nonumber \\
   & & {  }\qquad \left. + { }\frac{3}{N\ell^{2}}
   \underbrace{\left[\Big(\sum_{i=1}^{4}\Big(\,x_{i}^{2}+y_{i}^{2}\,\Big)-\frac{1}{4}\,\Big(\sum
     x_{i}\Big)^{2}-\frac{1}{4}\,\Big(\sum y_{i}\Big)^{2}
      \right]}_{\equiv\, d^{2}}\,\Big(\frac{1}{\,\lambda}-1\,\Big)\right\}\,,
\end{eqnarray}
with $d^{2}$ depending on the chosen, albeit fixed, relationship of
the wall linkages. The constant $d^{2}$ is translation invariant and ought not play a dominant role, since the
problem is symmetric in the $\hat{x}$ and $\hat{y}$
coordinates\footnote{The deformation factor $(1/\lambda-1)$ after the constant $d^{2}$ is zero if the plates are simply
  stretched in the $\hat{z}$ coordinate by a factor $\lambda$. In
  Equation \eqref{eq:onelinkfreeEb} $d^{2}$ is present since an (optional)
  \emph{isovolumetric} deformation was enforced.}. 
\begin{figure}[!h]
\begin{center}
  \includegraphics[width=0.4\textwidth]{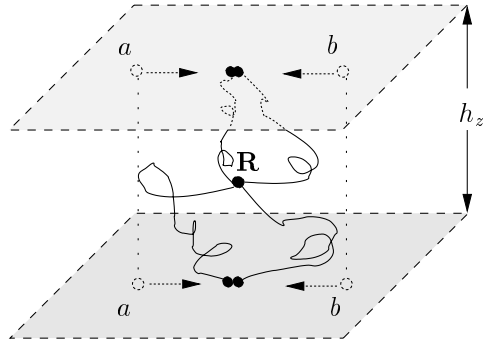}
\caption[Maximum entropy and the single bulk-link.]{\label{fig:dzero}Two polymer stitches with
  a junction point at $\bR=(X,\,Y,\,Z)$. The top and bottom wall links are situated
  at permanent positions with equal $x$ and $y$ coordinates, such that
  the left links have coordinates $(x_a,y_a,\epsilon)$ and
  $(x_a,y_a,h_{z}-\epsilon)$, and the right has coordinates
  $(x_{b},y_b,\epsilon)$ and $(x_b,y_b,h_z-\epsilon)$, for top and
  bottom plates, respectively. The arrows show movement of the fixed links to 
  a common position that correspond to maximum entropy.}   
    \end{center}
\end{figure}
However, for the
above expression to be valid in the limit \eqref{eq:limitf1},
one must limit the distance between wall links for
this two-stitch system. This can most easily be seen, when the number
of coordinate constants $x_{i}$ and $y_{i}$ is reduced, such that $x_{1}=x_{2}\equiv x_{a}$ and
$x_{3}=x_{4}\equiv x_{b}$. In this case 
$d^{2}=(x_{a}-x_{b})^{2}+(y_{a}-y_{b})^{2}$, and gives a measure of the distance
between two wall-linkages on each plate. The entropy of the system
will thus increase as the distance between points $\br_{a}$ and
$\br_{b}$ decreases (Figure \ref{fig:dzero}). Conversely, when $d$, or the distance between wall-links in
the $x$-$y$ plane, increases, the chain portions are more stretched
out and fewer configurations are available to the polymer. 

When the stress $f$ is calculated in the
manner of \eqref{eq:tensileforce}, we obtain a stress-strain
relationship similar to \eqref{eq:tensileforce}:
\begin{equation}
  \label{eq:tensileforcebulklink}
  f_{\rm{link}} = \frac{\bolz T}{V}\left(\frac{7}{\lambda}-\frac{2 N
      \pi^{2}\ell^{2}}{3h_{z}^{2}}\,\frac{1}{\lambda^{3}}-\frac{3\,d}{N\ell^{2}}
    \,\frac{1}{\lambda^{2}}\right)\,.
\end{equation}
The coefficient of the confinement term in \eqref{eq:tensileforcebulklink}, $\frac{2 N
      \pi^{2}\ell^{2}}{3h_{z}^{2}}$ has increased by a factor two
    compared to \eqref{eq:tensileforce}, which portrays the relative
    difficulty in deforming a system consisting of two chains with an added bulk linkage. Also, if we
    let $d^{2}\equiv 0$, we recover the case of only two
    wall-links. Since $d^{2}\ll \sqrt{N \ell}$, the second term in
    \eqref{eq:tensileforcebulklink} will dominate, such that the
    stress $f_{\rm{link}}$ will not differ much from $f$ for the simple stitch
    system (Figure \ref{fig:plotstres1}). This is illustrated
    by a similar stress-strain plot of
    \eqref{eq:tensileforcebulklink}, shown in comparison with the classic stress-strain curve
    for an unconfined ``network'' of only one bulk-link
    (and no wall-links) \cite{Treloar}.
\begin{figure}[!h]
\psfrag{f}[][l]{$f_{\rm{link}}$}\psfrag{lam}[][c]{$\,\lambda$}\psfrag{Classic}[][l]{\emph{Classic
  rubber elasticity}\,\,$\searrow$}\psfrag{Det}[][r]{$\longleftarrow$\emph{Confined single bulk-link}}
{\centering\includegraphics[width=0.46\textwidth]{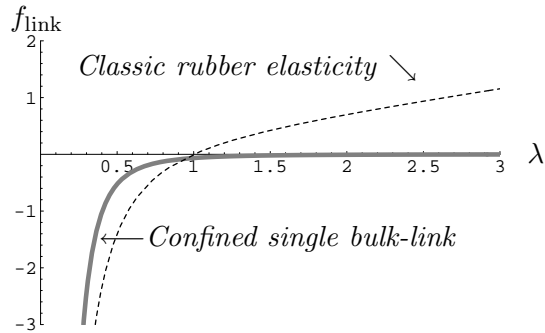}
\caption[Stress-strain plot for the confined, single bulk-link plate system.]{Plot of the stress ($f_{\rm{link}}$) \eqref{eq:tensileforcebulklink} versus the
  deformation ratio ($\lambda$) for the confined, single bulk-link
  plate system (solid line),
  in limit (ii). The dashed line is the stress-strain curve
  corresponding to the classic rubber elasticity model, with \mbox{$f \propto\lambda - 1/\lambda^{2}$}, for only one
  bulk link. \label{fig:plotstres3}}}
\end{figure}
\section{Contemplating stitches}
\label{sec:concluding}
In this chapter we briefly introduced the basic concepts which play a
part in the formation of a network at a surface: the Gaussian chain,
confinement, homogeneous strain, wall-links and lastly a
\emph{non-random}, single bulk
link. 

Confinement
reduces the number of possible chain shapes,
and decreases the entropy. In a confined chain, the role of
the characteristic chain dimension $R_{0}$ is reversed, relative to its
free chain counterpart. This leads to a chain which does not obey the Hooke
``law'' for stretching, so typical of an ideal chain. 

Since the
Green's function of confinement included intractable sums, we were
forced to work in certain limits, most notably, $h_z \ll
\sqrt{N}\ell$. Consequently the stitches were not allowed to stretch
too much. This restraining condition also applies to an unconfined
phantom chain. Its 
force-extension relation is subject to the
same limitations as the Gaussian distribution function from which it is 
derived. For end-to-end extensions approaching the total contourlength
of the chain, the Gaussian approximation becomes progressively
inaccurate, and will fail if $\lambda_{\rm{max}} \sim \sqrt{N}$.

It might be a reasonable, first approximation
to simulate the real network (discussed in the Introduction) by a
\emph{network} of stitches: two parallel planar surfaces stitched together,
with the possibility of link-formation in the bulk region between the plates. 
If we envisage such a network, restricted between two
walls, we should work in the limit \eqref{eq:limitf1} since the
\emph{stitch}-system has only
confinement in one coordinate (as opposed to a box-confinement). Each chain must also be large enough
so that it can be fixed to opposite walls with ease. These links are formed an
infinitesimal distance $\epsilon$ from the walls. It was shown, for
example in Section \ref{sec:stitchelas}, 
that $\epsilon$ does not influence the macroscopic properties of the
system.

During network formation, crosslinkages are formed at random. However, in
this chapter we placed them at completely determistic positions. This descrepancy will be attended to in the next chapter.

\chapter{The Stitch Network}\label{chap:disorder}
\emph{How does one model a system with
  permanent, but random, constraints?}

\begin{figure}[h!]
\centering 
\epsfig{figure=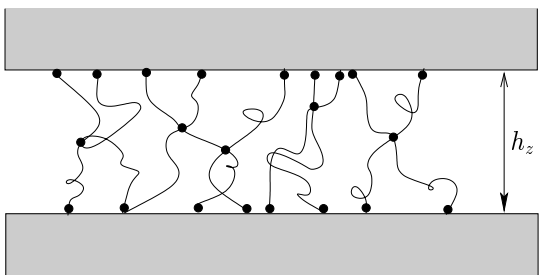,width=0.45\textwidth}
\caption[A confined polymer network of stitches.]{A polymer network of stitches, confined between two
  plates of spacing $h_{z}$.\label{fig:stitchnetworka}}
\end{figure}

In this thesis we are investigating the properties of a polymer gel,
formed at a surface. The gel consists of long macromolecules that
form cross links with the surface, in addition to the polymer-polymer
links. We specifically consider a gel confined between two parallel plane surfaces. This symmetric problem is related to two plates, which are
\emph{stitched} together by very long chain threads as shown in  Figure
\ref{fig:stitchnetworka}. In the previous chapter we isolated one long thread, and fixed its
terminal ends to the top and bottom confining plates. This was called a
\emph{stitch}. We included a \emph{mock} bulk link that had its  exact
location on the chains,
chosen  beforehand. In a real network, the precise location of the
linkages is \emph{not} known beforehand, and will tend to vary from
one sample to the next. Any realistic network model should thus take
account of this randomness.

In the present chapter we shall apply the replica method to the stitch
network model to
show how one can describe and handle random cross-linking mathematically.
\section{A random bulk link}
\label{sec:section1}
The first crucial step is to rectify the assumption of a \emph{given} bulk link (Section
\ref{sec:onebulklink}) by making it random. Let $s$ measure the arc length, $s\in [0,L]$,
along a Gaussian chain of contourlength\footnote{In terms of the
notation previously used, $s=n\ell$, such that $L=N\ell$, where $\ell$
is the average link length introduced in Section \ref{sec:singlechaininbox}.} $L$. We simplify the toy model of
stitches by having specific, but arbitrarily placed  wall links. 

They are always chosen to be at
the terminal points, corresponding to $s=0$ and $s=L$, of the polymer
chain. 
We choose to divide the stitch network into a macroscopic
number of, say $M$, subsystems, each having one random bulk link. The
total free energy of the stitch network system would then consist of the sum of the free energies of the
$M$ subsystems, plus a contribution that comes from the interactions of
the subsystems.  For a system of phantom chains, we shall ignore any
interactions. 

\begin{figure}[htb]
\centering 
\epsfig{figure=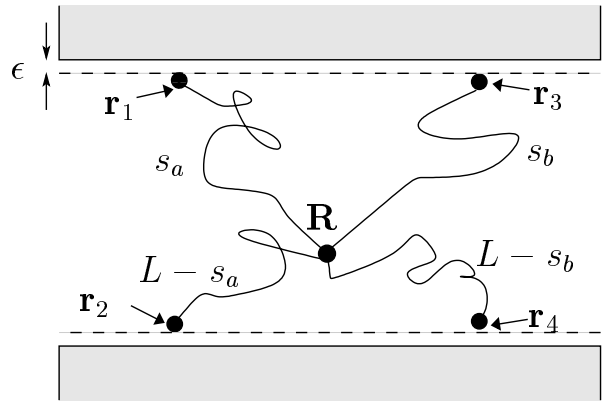,width=0.5\textwidth}
\caption[Isolating two stitches with a common linkage.]{A portion of the stitch network, as seen in isolation, consisting of two stitches
  linked at spatial position $\bR$.\label{fig:stitchnetworkb}}
\end{figure}

We continue by isolating a subsystem, a portion
of the stitch network, as follows. Consider 
two stitches, each of length $L=N\ell$, confined between two walls of
spacing $h_z$, illustrated in Figure \ref{fig:stitchnetworkb}.
 
The wall links are formed an
infinitesimal distance $\epsilon$ from the walls\footnote{The parameter
  $\epsilon$ does not influence the physical properties of the
  system. For example, the stress-strain relationship is entirely independent of
$\epsilon$, as investigated in section \ref{sec:stitchelas}.}, at 
spatial positions $\br_{1}$,  $\br_{2}$, $\br_{3}$ and  $\br_{4}$,
which do \emph{not} vary from sample to sample.   The
bulk link is formed at a \emph{random} arc length location $s_a$ and
$s_b$ on the two chains. This random linking is the source
of the disorder in the system. 
Intuitively, any relevant observable of the stitch-system  should depend on
some general averaged characteristics of the random crosslinking. 

The main goal of this chapter is to \emph{introduce} general methods  for
dealing systematically with the statistical mechanics of a random
system. For the simple case of one bulk link, it might not seem worthwhile to employ strategies like the
replica method (Section \ref{sec:replicas}). This is especially the
case when the crosslinkages are distributed uniformly.  

In the sections that follow  we adapt the basic
Green's function approach, developed in Chapter \ref{chap:singlechain},
to a system with disorder.
\pagebreak
\section{Quenched disorder}
\label{sec:avedisorder}
In Chapter \ref{chap:singlechain} we used standard statistical averaging to compute the free
energy of the system at a \emph{given} cross linking position. For a
real network, the linkage positions $\{\mathbf{\mathcal{S}}\}$ on the chains are unknown, but we will assume
that they are variables which vary randomly from sample to sample -- called quenched variables --
distributed according to a probability distribution
$P(\mathbf{\mathcal{S}})$. Statistical physics averages that are
generally employed, should be altered for quenched variables. 
\subsection{Green's functions}
\label{sec:greensub}
The partition function, or statistical weight, of the system with its degrees of freedom frozen at
a particular set $\mathbf{\mathcal{S}}=\{s_{a},s_{b}\}$, and with its junction point at a
specific spatial position $\bR$, is given by 
\begin{equation}
  \label{eq:Zdisorder1}
  \mathcal{Z}\,(\bR,\,\mathbf{\mathcal{S}}) =
  G\,\big(\br_{1},\bR;s_{a}\big)\,G\,\big(\bR,
  \br_{2};L-s_{a}\big)\,G\,\big(\br_{3},\bR;s_{b}\big)\,
  G\,\big(\bR,\br_{4};L-s_{b}\big)\,.
\end{equation}
In Equation \eqref{eq:Zdisorder1} we  employed the composition property of the Green's function to
divide the total contour length ($=2L$) into four smaller chain
intervals. This is done analogous to \mbox{Section \ref{sec:onebulklink}},
but with one crucial difference, namely that the interval lengths are
not pinned down at $L/2$.  
The Green's function \eqref{eq:greensexb} of one chain section, with unknown radius length
$s_{a}$, is thus given by
\begin{eqnarray}
  \label{eq:greensdisex}
  G\,(\br_{1},\bR;s_{a}) &=&  G_{x}\,(x_{1},X;s_{a})\, G_{y}\,(y_{1},Y;s_{a})\, G_{z}\,(z_{1},Z;s_{a}) \\
  \label{eq:greensdisexb}
  &=& \frac{3}{2\pi \ell\,s_{a}}
  \,\exp \left\{-\frac{3}{2\,s_{a}\ell}\left[\,(X-x_{1})^{2}+(Y-y_{1})^{2}\, \right] \right\}
  \nonumber \\ & & {} \times
  \frac{2}{h_{z}}\,\sum_{p=1}^{\infty}\,\sin\frac{\pi p
    Z}{h_{z}}\,\sin\frac{\pi p
    (h_{z}-\epsilon)}{h_{z}}\,\,e^{-\frac{\ell\pi^{2}p^{2}}{6 h_{z}^{2}}\,s_{a} }\,.
\end{eqnarray}
In order to obtain the partition function of all possible
conformations of the system with a \emph{particular} set of quenched
variables, we have to average over all possible spatial positions of
the junction point $\bR$:
\begin{equation}
  \label{eq:Zdisorder2}
\mathcal{Z}\,(\mathbf{\mathcal{S}}) = \int\,d \bR\, G\,\big(\br_{1},\bR;s_{a}\big)\,G\,\big(\bR,
  \br_{2};L-s_{a}\big)\,G\,\big(\br_{3},\bR;s_{b}\big)\,
  G\,\big(\bR,\br_{4};L-s_{b}\big)\,,
\end{equation}
It might be possible to compute the chain
configuration, for a specific choice of the quenched variables $s_{a}$ and $s_{b}$,  that minimizes
the free energy of a specific sample. However, in a general disordered
system such a
calculation would be deemed intractable, since the number of quenched
variables per sample is usually much larger than two\footnote{For
  example, in the Edwards-Anderson model \cite{EdwardsAnderson} for a spin glass the
  interaction $J_{ij}$ between every spin pair $\sigma_i$ and
  $\sigma_j$ is random. For $N$ spins there will be $\half N(N-1)$
  quenched variables per sample. Even numerical averaging over
  different realizations is too time-consuming \cite{Parisi}.}. 
Luckily, methods of statistical mechanics
often prove fruitful in the limit of large systems. 


\subsection{Averaging}
\label{sec:averagingdis}
In the statistical mechanics of a random system, there is a 
need to perform two distinct averages \cite{Fischer}. 
Firstly, there is the \emph{thermal}
average that is carried out for every possible conformation, in a
specific realization of $\{\mathbf{\mathcal{S}}\}$. This kind of average was
performed in \eqref{eq:Zdisorder2}, by means of a Green's function
approach. Secondly, there is the disorder or
\emph{configuration} average, which is an average over all possible
distributions of the quenched variables.      

In order to make the scenario of the random linking more
general, the \emph{thermal} average may be rewritten as follows. Let
$\mathcal{H}_{\mathbf{\mathcal{S}}}$ denote the generalized Edwards
Hamiltonian of a network subsystem composed
of two randomly linked stitches,
\begin{equation}
  \label{eq:hamilton}
  \mathcal{H}_{\mathbf{\mathcal{S}}}\,[\,\{\bR_{i}\}\,]/\bolz T \equiv \sum_{i=1}^{4}
    \frac{3}{2\ell} \int_{0}^{L}\dot{\bR}_{i}^{2}(s_{i})\,ds_{i} - 
    \int_{0}^{L}\,A\,[\bR_{i}(s_{i})]\,ds\,,
\end{equation}
where $\bR_{i}(s)$ is the position vector of the $i$th chain segment at the
arc length $s$. The Hamiltonian describes the connectivity and $\hat{z}$-confinement
of the chain system \cite{FreedEdwards}. Although the four chain pieces seem to emanate from a common
junction point at $\bR$ (Figure \ref{fig:stitchnetworkb}), the
respective arc lengths are not independent of one another, since they were formed
from just two \emph{stitches}. This adds a constraint to \eqref{eq:hamilton}:
$s_{1}=s_{a} \iff s_{2}=L-s_{a}$ and $s_{3}=s_{b} \iff s_{4}=L-s_{b}$.
 For a specific sample, the partition function is
expressed as the product of separable path integrals, 
\begin{eqnarray} 
 \label{eq:partisiealles}
 \mathcal{Z}_{\mathbf{\mathcal{S}}} &=& \int d\bR
        \int_{{\scriptscriptstyle\bR_{1}(s_{a})=\bR}}^{{\scriptscriptstyle\bR_{1}(0)=\br_{1}
        }}\int_{{\scriptscriptstyle\bR_{2}(0)=\br_{2}}}^{{\scriptscriptstyle\bR_{2}(L-s_{a})=\bR} }
        \int_{{\scriptscriptstyle\bR_{3}(s_{b})=\bR}}^{{\scriptscriptstyle\bR_{3}(0)=\br_{2}}}\int_{{\scriptscriptstyle\bR_{4}(0)=\br_{4} }}^{{\scriptscriptstyle\bR_{4}(L-s_{b})=\bR}} [\,\prod_{i=1}^{4}
        \mathcal{D}\bR_{i}(s_{i})]\,e^{-\,\frac{\mathcal{H}_{\mathbf{\mathcal{S}}}\,[R_{i}]}{\bolz T}}\,,
\end{eqnarray}
which is equivalent to the previous Green's function expression
\eqref{eq:Zdisorder2}.
The free energy of a system in some particular crosslinkage state
$\mathbf{\mathcal{S}}=\{s_a,s_b\}$ is the logarithm of \eqref{eq:partisiealles}:
 \begin{equation}
  \label{eq:extensiveFreeE}
  F\,(\mathbf{\mathcal{S}}) = -\bolz T\,\ln\,\mathcal{Z\,(\mathbf{\mathcal{S}})}\,.
\end{equation}
Let the probability of finding this particular state be
$P_{\mathbf{\mathcal{S}}}$. Then for a canonical ensemble the effective free
energy is given by 
  \begin{equation}
  \label{eq:effFreeE}
  \mathcal{F} =
  \sum_{\{\mathbf{\mathcal{S}}\}}\,P_{\mathbf{\mathcal{S}}}\,F\,(\mathbf{\mathcal{S}}),
  \qquad \text{where}\qquad \sum_{\{\mathbf{\mathcal{S}}\}}\,P_{\mathbf{\mathcal{S}}}=1\,.
\end{equation}
The effective free energy is the observable or experimental free
energy. A correct theoretical calculation of the average value of
$\mathcal{F}$ over the entire ensemble is expected to correspond to its
experimental value.  The sample free energy $F(\mathbf{\mathcal{S}})$ is an extensive
quantity and known to be \emph{self-averaging} \cite{Fischer,
  Parisi} . We
can thus expect that variations in $F(\mathbf{\mathcal{S}})$ from one
sample to the next and deviations from the average value $\mathcal{F}$, will go
to zero in the thermodynamic limit $V\to\infty$.  
In order to obtain an experimentally relevant quantity, only variables
with the self-averaging property should be disorder-averaged. 

Taking a sample
average of the partition function \eqref{eq:partisiealles} instead of
the free energy, is known as the so-called \emph{annealed} case:
 \begin{equation}
  \label{eq:annealFreeE}
  \mathcal{F}_{\rm{annealed}} =
  -\bolz T \ln\,\left[\,{\scriptstyle\sum_{\{\mathbf{\mathcal{S}}\}}}\,P_{\mathbf{\mathcal{S}}}\,\mathcal{Z}_{\mathbf{\mathcal{S}}}\right]\,,
\end{equation}
and would give the wrong physics, since the crosslinkage in
\eqref{eq:annealFreeE} is allowed to change in response to the chain
conformations. In reality, 
$\{\mathbf{\mathcal{S}}\}$ is fixed for each sample. Even if the substance
between the planar surface is deformed, the crosslinkage for a
particular sample will stay fixed in response to the
deformation. Beyond what has been said, the partition function
$\mathcal{Z}_{\mathbf{\mathcal{S}}}$ is also not suitable for
averaging since it is not 
self-averaging and not physically observable.

The correct strategy is thus to average the self-averaging free energy
$F\,(\mathbf{\mathcal{S}})$ over the distribution of the permanent, but
randomly chosen arc lengths:
\begin{eqnarray}
  \label{eq:realFreea}
  \mathcal{F}\equiv \left[F(\mathbf{\mathcal{S}}) \right]_{S} &\equiv&
  {\scriptstyle\sum_{\{\mathbf{\mathcal{S}}\}}}\,P_{\mathbf{\mathcal{S}}}\,F(\mathbf{\mathcal{S}})
  \\ \label{eq:realFreeb} &=& -\bolz T\,\int_{0}^{L}\,ds_{a}\int_{0}^{L}\,ds_{b}\,P_{\mathbf{\mathcal{S}}}\,\ln\,\mathcal{Z}(\mathbf{\mathcal{S}})\,.
\end{eqnarray}
The configuration average is denoted by
$\left[\ldots\right]_{\mathbf{\mathcal{S}}}$. Since the free energy for a specific
$\mathbf{\mathcal{S}}$ agrees with its configuration average
$\left[F(\mathbf{\mathcal{S}}) \right]_{S}$ in the thermodynamic
limit, it seems that  either quantities
may be used in further calculations. However, the average
$\left[F(\mathbf{\mathcal{S}}) \right]_{S}$ is more suitable to work
with, since it is a physically \emph{measurable}
quantity\footnote{The free energy is not the only 
  self-averaging quantity. For example, the neutron scattering form
  factor is also a useful measurable quantity to average
  \cite{Warner}.}. 
\subsection{Replicas to the rescue}
\label{sec:replicas}
Performing the average \eqref{eq:realFreeb} is
\emph{usually}\footnote{An example of a model where the disorder average has been solved
  analytically, without implementing replicas, is the random-energy
  model \cite{Derrida1, Derrida2}. In this model the independent energy levels
  $\{E_{i}\}$ are the quenched variables, and the sample average of
  the free energy is calculated by a microcanonical argument. The
  infinite range Ising model with $p$-spin quenched random
  interactions is also exactly solvable, and the
  results confirm the replica method \cite{GrossMezard}.} not feasible,
because the dependence of $\ln\,\mathcal{Z}$ on $\mathbf{\mathcal{S}}$ is often
very complicated. In order to address the averaging the following
identity,
\begin{eqnarray}
  \label{eq:replicaidentitya}
  \lim_{n \to 0}\,\frac{\mathcal{Z}^{n} -1}{n} = \ln\mathcal{Z} \qquad
  \text{or} \qquad \frac{\partial}{\partial n}\mathcal{Z}^{n}|_{n=0}=\ln\,\mathcal{Z}
\end{eqnarray}
was first employed by \cite{EdwardsAnderson,SK} in theoretical models
to predict
the spin glass phase in disordered magnetic systems.
This technical trick \eqref{eq:replicaidentitya}, called the replica method, is useful
because it is often easier to evaluate $\left[\mathcal{Z}^{n}\right]$ than
$\left[\ln\,\mathcal{Z}\right]$. Consider the partition function \eqref{eq:partisiealles} taken to the
\emph{integer} power $n$:
\begin{eqnarray}
  \label{eq:replicapartisie}
  \mathcal{Z}^{n}\,(\mathbf{\mathcal{S}}) =
  \left[\,\prod_{\alpha=1}^{n}\int\,d\bR^{(\alpha)}\,\Big(\prod_{i}\int_{i}\mathcal{D}\bR_{i}^{(\alpha)}\Big)\right]\: \exp\left\{-\sum_{\alpha=1}^{n}
  \,\mathcal{H}_{\mathbf{\mathcal{S}}}\,[\bR_{i}^{(\alpha)}]\,/\,\bolz T\right\}\,.
\end{eqnarray}
The above quantity is the partition function of $n$ independent
identical copies or replicas of the original system. In \eqref{eq:replicapartisie} the subscript $\alpha$ labels the
replicas, and not the quenched \mbox{variables $\mathbf{\mathcal{S}}$.} The chain conformation may
fluctuate from replica to replica, but the cross link constraints are
fixed for each sample replica.  
The scheme of the replica method
can be described as follows. Firstly, the quenched average
of the replicated partition function \eqref{eq:replicapartisie} for
integer $n$ must be calculated. Next, the analytic continuation of the
the resulting function, $\left[\mathcal{Z}^{n}\,(\mathbf{\mathcal{S}})
\right]_{\mathbf{\mathcal{S}}}$, should be made for an arbitrary
non-integer $n$. Lastly, the limit $n \to 0$ should be taken. In the
end, the effective free energy \label{eq:realFreea} is uncovered:
\begin{eqnarray}
  \label{eq:realFreeabulklink}
  \mathcal{F}\equiv  -\bolz T\,\lim_{n \to 0}\, \frac{1}{n}\,\Big(\,[\,\mathcal{Z}^{n}\,]_{\mathbf{\mathcal{S}}}-1\,\Big)\,.
\end{eqnarray}
The replica method might initially seem mathematically dubious since one
evaluates  $\left[\mathcal{Z}^{n}\,(\mathbf{\mathcal{S}})
\right]_{\mathbf{\mathcal{S}}}$ for integer $n$ and then extrapolates
the result to $n \to 0$.  In the pioneering decade of spin glass
theory, the
replica method was thought to be the cause of some controversial
and inconsistent results. In particular, the free energy
of the Sherrington and Kirkpatrick (SK) model \cite{SK} was shown to be
unstable and exhibit a negative entropy in the
low temperature  ($<T_{c}$) region. However, these unphysical results were a consequence
of the replica symmetry ansatz and possibly the incorrect reversal of
limits \cite{VonHemmen}, and not due to the replica procedure
itself. Alternative mathematical efforts, including replica symmetry
breaking schemes and the mean-field TAP (Thouless, Anderson and
Palmer) approach\footnote{Comprehensive
literature and reviews exist on the subject of the theory of spin
  glasses \cite{Fischer, Parisi, BinderYoung}. Recently it has been
  shown that the TAP and replica methods are equivalent in the SK
  model \cite{Cavagna}. },  agree with the SK
solution at and above the critical temperature $T_{c}$. In all cases where the calculations
can be performed by a different method, the replica approach is confirmed
and gives sensible results \cite{Derrida1, GrossMezard}. 

In polymer network
theories the use of the replica method is simpler and
undisputed \cite{Ball, BallEdwardsDoi, KKMNMc,
  KKMNEdwards}, 
because there are no competing interactions (``frustration'') or need
for replica
symmetry breaking as in the case of spin glasses\footnote{An example of
  where various methods are used to probe the disorder in a polymer
  system, is the problem of a Gaussian chain trapped in a medium with randomly frozen obstacles \cite{MuthuEdwards}. Variational
  calculations are performed with and without the breaking of replica
  symmetry, and shown to be consistent. It is also shown that
  the annealed and quenched cases are two distinctly different situations.

Furthermore, many network theory problems have been approached
without the use of replicas \cite{SolfVilgis}, and corroborate the replica results. An
example is that of determining the neutron scattering function of
polymer networks \cite{Read, ReadMc, Warner}.}.
\pagebreak
\section{The actual calculation}
\label{sec:calc}
In this section we apply all the available \emph{tools} --- the Green's
functions of Section \ref{sec:greensub} and the replica method --- to a subsystem of the stitch
network (Figure
\ref{fig:stitchnetworkb}). 
The replicated partition function of the single bulk-link system can
be abstractly expressed as:
\begin{equation}
  \label{eq:partisiedisordera}
  \mathcal{Z}^n =
  \left(\int\,d\bR^{(1)}\,\mathcal{G}^{(1)}\right)\,\left(\int\,d\bR^{(2)}\,\mathcal{G}^{(2)}\right)\,\ldots \, 
  \left(\int\,d\bR^{(n)}\,\mathcal{G}^{(n)}\right)\,
\end{equation}
where the Green's functions in \eqref{eq:greensdisex} are also
replicated 
\begin{equation}
  \label{eq:dd}
  \mathcal{G}^{(\alpha)}\equiv  G^{(\alpha)}\,\big(\br_{1},\bR^{(\alpha)};s_{a}\big)\,G^{(\alpha)}\,\big(\bR^{(\alpha)},
  \br_{2};L-s_{a}\big)\,G^{(\alpha)}\,\big(\br_{3},\bR^{(\alpha)};s_{b}\big)\,
  G^{(\alpha)}\,\big(\bR^{(\alpha)},\br_{4};L-s_{b}\big)\,.
\end{equation}
The first step of the replica procedure involves the configurational
average of $\mathcal{Z}_{\mathcal{S}}^{n}$ \eqref{eq:replicapartisie},
which in terms of Green's functions, is given by  
\begin{equation}
   \label{eq:partisiedisorderb} [\,\mathcal{Z}^n \,]_{\mathbf{\mathcal{S}}} =
   \int_{0}^{L}ds_{a}\,\int_{0}^{L}ds_{b}\,\prod_{\alpha=1}^{n}\,\left[\int\,d\bR^{(\alpha)}\,\mathcal{G}^{(\alpha)}\, 
   \right]\,P(s_a,s_b)\,.
\end{equation}
For the sake of illustrating the replica method in the simplest
manner, we choose $P(\mathbf{\mathcal{S}})$ to be the uniform
distribution, that is,
$P(\mathbf{\mathcal{S}})=1/L^{2}$. Although the bulk links are
uniformly distributed, we do not allow them
to form exactly at the ends, $s=0$ and $s=L$, where the wall links are
situated. Since the
location of the
wall links of a stitch network is by definition chosen to be
non-random, the $x$ and $y$ coordinates of the stitch ends may be
taken to coincide at the top and bottom plates respectively\footnote{This was also discussed in Section
  \ref{sec:onebulklink} and illustrated in Figure
  \ref{fig:dzero}.}. 
In the limit $\sqrt{N}\ell \gg h_{z}$, the first term in the
eigenfunction expansion \eqref{eq:greensdisex}, corresponding to
$p_{1}^{(\alpha)}=p_{2}^{(\alpha)}=p_{3}^{(\alpha)}=p_{4}^{(\alpha)}=1$
in \eqref{eq:dd},
dominates the partition function. In the case of the uniform
distribution of the quenched variables, the configurational average of
the $n$th power of the partition function  is 
\begin{eqnarray}
  \label{eq:partisienaRint}
  [\,\mathcal{Z}^{n}\,]_{\mathbf{\mathcal{S}}}&=&\frac{1}{L^2}\,(\rm{const})^{n}\, \int_{\epsilon}^{L-\epsilon}ds_{a}\,\int_{\epsilon}
  ^{L-\epsilon}ds_{b}\,
    \big[s_{a}(L-s_{a})+s_{b}(L-s_{b})
    \big]^{-n} \nonumber \\  & & { }\qquad\qquad\exp\left\{-\frac{3n\,L}{2\ell}\left[\frac{(x_{1}-x_{2})^{2}+(y_{1}-y_{2})^{2}}
      {s_{a}(L-s_{a})+s_{b}(L-s_{b})}\right] \right\}\,
\end{eqnarray}
where ``$\rm{const}$'' refers to a constant factor, dependent on the wall
spacing $h_z$, but not on the disorder. The $\epsilon$-factor in the
integration, mathematically prohibits the bulk-links from being formed
at the walls, which would result in \emph{double} crosslinking.
\subsection{The method of steepest descents}
\label{sec:steepestd}
It is convenient to transform to a new set of integration
variables,
$\sige=|s_{a}-\frac{L}{2}|$ and
\mbox{$\sigt=|s_{b}-\frac{L}{2}|$}, where
$\sigma_{i}\in(-\frac{L}{2},\frac{L}{2})$. Let the measure of the
distance between points on the top and bottom plates be denoted by
$d^2$, analogous to Section \ref{sec:onebulklink}:
$d^{2}=(x_{1}-x_{2})^{2}+(y_{1}-y_{2})^{2}$. 
After the transformation, the
exponential in 
\eqref{eq:partisienaRint} only has terms quadratic in $\sigma_{i}$:
\begin{eqnarray}
  \label{eq:Zaftertransfeng}
   [\,\mathcal{Z}^{n}\,]_{\mathbf{\mathcal{S}}}&=&\frac{1}{L^2}\,(\rm{const})^{n}\,
   \int_{\epsilon-\frac{L}{2}}^{\frac{L}{2}-\epsilon}\,d\sige\,
   \int_{\epsilon-\frac{L}{2}}^{\frac{L}{2}-\epsilon}\,d\sigt\,\,
      e^{-g\,(\sige,\sigt)}\,,
\end{eqnarray} with the exponent $g$ defined as follows  
\begin{equation}
  \label{eq:g}
 g\,(\sige,\sigt) \equiv n \ln\left[\frac{L^2}{2}-\sige^{2}-\sigt^{2}\right]+\frac{3n\,Ld^{2}}{2\ell}\,\left(\frac{L^2}{2}-\sige^{2}-\sigt^{2}\right)^{-1}\,.
\end{equation}
If the dominating factor $\frac{3n\,Ld^{2}}{2\ell}$ in $g$ is large, it is
possible to evaluate the integral by the method of steepest descents or
saddle point approximation \cite{Hassani}. The point $(\sige^*,\sigt^*)$
where $g$ is
minimized is found by making $g$ stationary with respect to $\sige$
and $\sigt$; ${\scriptstyle\frac{\partial
    g\,(\sige^{*},\sigt^{*})}{\partial \sige}}=0$ and ${\scriptstyle\frac{\partial
    g\,(\sige^{*},\sigt^{*})}{\partial \sigt^{*}}}=0$. A global minimum is found at
$(\sige^{*},\sigt^{*})=(0,0)$, only if $L <
3d^{2}/\ell$. In terms of the original arc length coordinates, the system
will thus favour a bulk link to be formed at  $s_a=L/2$ and
$s_b=L/2$.

Next, the function $g\,(\sige,\sigt)$ may be expanded in a
Taylor series around $(\sige^{*},\sigt^{*})$:
\begin{eqnarray}
  \label{eq:taylorgfa}
  g\,(\sige,\sigt) &=&  g(\sige^{*},\sigt^{*})
  +\frac{1}{2}\Big\{(\sige-\sige^{*})^{2}\:{\scriptstyle\frac{\partial^{2} g\,(\sige^{*},\sigt^{*})}{\partial\sige^{2}}} 
        +2\,(\sige-\sige^{*})\,(\sigt-\sigt^{*})\:
      {\scriptstyle\frac{\partial^{2}
          g\,(\sige^{*},\sigt^{*})}{\partial\sige\partial\sigt}}\nonumber \\ & & \qquad+
      (\sigt-\sigt^{*})^{2}\:{\scriptstyle\frac{\partial^{2}g\,(\sige^{*},\sigt^{*})}{\partial \sigt^{2}}} 
\Big\}+\ldots \\ \label{eq:taylorgfb}&\simeq& \frac{3nd^{2}}{\ell L} + n
\ln \half L^2 + \frac{2n}{L^{3}}\left(\frac{3d^{2}}{\ell}-L\right)\,(\sige^{2}+\sigt^{2})\,.
\end{eqnarray}
We choose to work with very long macromolecules, such that the contourlength
$L$ is very large (but finite) throughout the above Taylor expansion.
In this limit, the higher order
terms in \eqref{eq:taylorgfa} are insignificant, so that
\eqref{eq:Zaftertransfeng} becomes 
{\setlength\arraycolsep{1pt}
\begin{eqnarray}
  \label{eq:Zaftertransfsteepest}
   [\,\mathcal{Z}^{n}\,]_{\mathbf{\mathcal{S}}}&=&\frac{1}{L^2}\,(\rm{const} \frac{2}{L^2})^{n}\,e^{- \frac{3n\,d^{2}}{\ell L} }
   \int d\sige
   \int d\sigt\,
   e^{-
     \frac{2n}{L^{3}}(\frac{3d^{2}}{\ell}-L)\,(\sige^{2}+\sigt^{2})}\,\\
   \label{eq:nogeen}
     &=&\frac{1}{L^2}\,(\rm{const} \frac{2}{L^2})^{n}\,e^{- \frac{3n\,d^{2}}{\ell L} }
   \int d\sige
   \int d\sigt\sum_{k=0}^{\infty}{\small \frac{(-1)^{k}}{k!}}
   \left[
     \frac{2n}{L^{3}}\left(\frac{3d^{2}}{\ell}-L\right)\,(\sige^{2}+\sigt^{2})\,\right]^{k}
\end{eqnarray}}
However, since the integration limits are finite, we are forced to
revert to a series representation of the exponential in the integrand
\eqref{eq:Zaftertransfsteepest}.  The average of the replicated partition function
\eqref{eq:nogeen}, is then given by
\begin{eqnarray}
  \label{eq:finaleZ}
   [\,\mathcal{Z}^{n}\,]_{\mathbf{\mathcal{S}}}&\approx& 1 - n
   \left\{\,\frac{\ell\,\pi^{2}\,L}{3 h_{z}^{2}}+
     \frac{4\,d^{2}}{\ell\,L}-\ln\Big[{\footnotesize \frac{6\epsilon^{4}}{h_{z}^{7}}\,\left(\frac{3}{2\ell}\right)^{4}
     \frac{2}{L^{2}}}\Big]-\frac{1}{3}\right\}+\mathcal{O}\left(\ge
 n^{2}\right)\,\text{terms}
\end{eqnarray}
Lastly, by taking the
replica limit $n \to 0$ as in \eqref{eq:realFreeabulklink}, the effective free energy $\mathcal{F}$ is obtained 
\begin{eqnarray}
  \label{eq:finalef}
  \mathcal{F}/\bolz T \simeq \frac{\ell\,\pi^{2}\,L}{3 h_{z}^{2}}+
     \frac{4\,d^{2}}{\ell\,L}-\ln\Big[\frac{6\epsilon^{4}}{h_{z}^{7}}\,
     \left(\frac{3}{2\ell}\Big)^{4} \frac{2}{L^{2}}\right]-\frac{1}{3}\,,
\end{eqnarray}
which is only valid for large, but finite chain length $L$, such that
$\sqrt{\ell L} \gg h_{z}$,  and when the relation
$L < 3\,d^{2}/\ell$ holds. 
\pagebreak

\section{Deforming the \emph{stitch network}}
\label{sec:deformstitchnetw}
In Section \ref{sec:section1} we defined the free energy of the
phantom stitch network as a whole, to be the 
sum of $M$ subsystems. Each single-bulk link system has an
effective free energy $\mathcal{F}$
\eqref{eq:finalef}. If a stitch network is fabricated from long, but
finite length chains,
and the relation $ h_{z} \ll L < 3d^{2}\,/\,\ell$ holds, the replica
procedure results in the following elastic free energy:
\begin{eqnarray}
  \label{eq:fnetwork}
  \mathcal{F}_{\rm{network}}\simeq \bolz T\,M\left\{\frac{\ell\,\pi^{2}\,L}{3 h_{z}^{2}}+
     \frac{4\,d^{2}}{\ell\,L}-\ln\left[\frac{6\epsilon^{4}}{h_{z}^{7}}\,
     \left(\frac{3}{2\ell}\right)^{4} \frac{2}{L^{2}}\right]\,\right\}\,.
\end{eqnarray}
The first contribution to the free energy, $\frac{\ell\,\pi^{2}\,L}{3 h_{z}^{2}}$ is ascribed to the
$\hat{z}$-confinement, and is typical for an ideal chain trapped in a
one dimensional cavity of diameter $h_{z}$ \cite{GrosbergKhok}. The
second term, $ \frac{4\,d^{2}}{\ell\,L}$, corresponds to the free
energy of four \emph{unconfined}, ideal chains (in two dimensions) with an end-to-end
distance $d$. The last term in \eqref{eq:fnetwork} comes from the normalization factor. In
the case of a uniform distribution, $P\,(\mathcal{S})=1/L^2$, it is
possible to find a similar free energy, without the
replica method, by minimizing
$F(\mathcal{S})$ \eqref{eq:extensiveFreeE} with respect to $s_a$ and
$s_b$.

Using the above free energy \eqref{eq:fnetwork}, it is possible to investigate what happens during
strain.
Let the stitch network between plates, be strained by an
isovolumetric, uniaxial deformation $\mathbf{\Lambda}$
\eqref{eq:isomatrix}, $\bR_{i} \to \mathbf{\Lambda}\cdot \bR_{i}$. 
The stress as a function of the deformation ratio $\lambda$, can be
calculated as in Section \ref{sec:stitchelas},
\begin{equation}
  \label{eq:tensileforcebulklinkrandom}
  f_{\rm{network}} = \frac{\bolz T\,M}{V}\left(\frac{7}{\lambda}-\frac{2 
      \pi^{2}\ell \, L}{3h_{z}^{2}}\,\frac{1}{\lambda^{3}}-\frac{4\,d^{2}}{\ell\,L}
    \,\frac{1}{\lambda^{2}}\right)\,,
\end{equation}
and is found to resemble the stress-strain relationship for a network
of stitches that has an average number of $M$ \emph{non-random bulk-links}, chosen
beforehand to be at $(s_a,s_b)=(\frac{L}{2},\frac{L}{2})$.

The stitch network model served as an exercise in the application of the
replica method, and as motivation for the search of a better network
model.

\chapter{The Real Confined Network}\label{chap:realnetwork}


In this chapter we shall extend Deam and Edwards' pioneering model of
network formation \cite{DeamEdwards}, to include a
confining surface and random polymer-polymer and polymer-wall links. 

A polymer network formed at a surface is a system governed by a variety
of constraints. 
The surface at which the gel forms, plays the role of a confining
geometry that  restricts the network chain conformations and positions.
During network formation, the macromolecules form
permanent links with one another as well as with the confining
surface. 
Since the links are unbreakable, the junction points will be localized
to a certain extent dependent on the crosslink-density.
In a \emph{real} network the exact location of the links is
unknown prior to fabrication and varies from one specimen to the
next. However, any two samples having different cross-link realizations
are expected to  exhibit similar macroscopic properties. 

In previous chapters we investigated single-chain systems subject
to similar conditions of confinement and random linking. This was done
to provide the basic groundwork for understanding and approaching the
actual network (many-chain) problem. In the present chapter we 
traverse through the ``making'' of an ideal polymer network formed at a
surface, starting from a group of chains to the finished
gel-product\footnote{Note: this is done through the eyes of a
  theoretical physicist, not a chemist. In referring to an ``ideal
  network'', we mean a 3-dimensional network of phantom chains, joined
  together and fixed to the walls by random, unbreakable chemical
  bonds or crosslinks. All forces between the chains, except at the points of
  crosslinkage, are ignored and each chain is assumed to be free
  to take on any conformation in the confined region.}.

\section{The Collection of Phantom chains}
\label{sec:phantomchains}
Prior to linking, we have a system of $M$ independent chains. They are
assumed to be long, linear and have no charge. The calculations in
this chapter are performed on phantom (pure Gaussian) chains, thus neglecting
any effects due to inter- and intra-molecular forces. The term
``phantom'' emphasizes the fact that the assumption of no steric effect
is clearly unphysical. In reality each monomer segment
of a chain will possess some volume which is excluded from other
segments \cite{Ball, DoiEdwards}, and consequently the polymer will swell. 
However, in a \emph{dense} polymer solution the excluded volume interactions
are screened, and each chain is essentially ideal and Gaussian \cite{DeGennesScaling}. 
The
collection of phantom chains may thus be visualized as a rather dense,
overlapping polymer solution, called a melt. 

For a system of $M$ phantom chains, the
partition function is written as a product of $M$ Wiener path
integrals (Section \ref{sec:freegausschain}):
\begin{equation}
 \label{eq:zphantomchainsa}
        \mathcal{Z}\,\left(\{\bR_{i},\bRacc_{i};L_{i}\}\right)=
        \mathcal{N}\left\{\,\prod_{i=1}^{M}\,\int_{\bR_{i}(0)=\bRacc_{i}}^{ \bR_{i}(L_{i})=\bR_{i} }
        [\mathcal{D}\bR_{i}(s)\, ]\right\} \exp\left\{-
        \frac{3}{2\ell}\sum_{i=1}^{M} \int_{0}^{L_{i}}\bigg(\frac{\partial
            \bR_{i}(s)}{\partial s}\bigg)^{2}\,ds\right\}\,. 
\end{equation}
For the sake of generality, each polymer $i$, defined by its path
$\bR_{i}(s)$ in~\eqref{eq:zphantomchainsa}, 
has its associated contourlength
$L_{i}$. For present purposes, this extra notation is
unneccesary. Henceforth, all chains are chosen to be identical with respect to constitution and degree of
polymerization, and will have equal length $L$. The
notation $\mathcal{L}$ will refer to the total contourlength of all
the chains, that is, $\mathcal{L}=\sum_{i}^{M}\,L_{i}$. 


\section{Confining the Melt}
\label{sec:confinementformalism}
\begin{figure}[!h]
\begin{center}
  \includegraphics[width=0.6\textwidth]{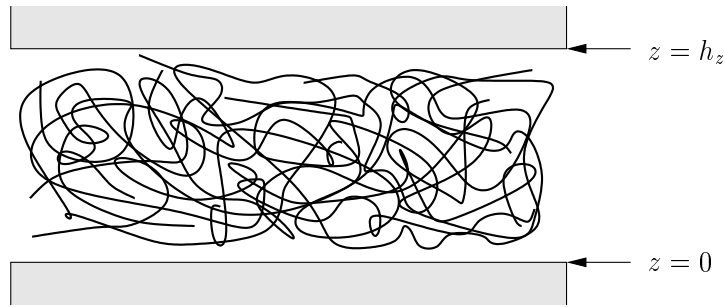}
\caption[A very dense melt of phantom chains, confined between two walls.]{\label{fig:confmelt}Simple illustration of a very dense melt of
  phantom chains, confined between two walls. There is no tendency
  towards adsorption.}   
    \end{center}
\end{figure}

The macromolecules are confined between two parallel planar surfaces
or \emph{walls}, as in 
  Figure \ref{fig:confmelt}, since symmetric problems are
  usually simpler to treat, and widen the scope of the problem. If one wall is situated at $z=0$ and the other
  at $z=h_{z}$, then the problem of a network at a single surface is retrieved by taking the
  limit $h_{z}\rightarrow\infty$.

The formalism chosen to represent the confinement, must ensure
  that all $\bR\,(s)$ vanish beyond the boundaries. The partition function
  in  ~\eqref{eq:zphantomchainsa} is modified, by noting that the path
  integral has to be carried out only in the allowable
  region~\cite{FreedEdwards}. In Section \ref{sec:singlechaininbox} a
  single chain was confined. Since the chains are independent, the
  partition function is the product of $M$ single-chain contributions \eqref{eq:greenb}:
\begin{eqnarray}
 \label{eq:zphantomchainsconfineda}
         \mathcal{Z}\,\left(\{\bR_{i},\bRacc_{i};L_{i}\}\right)&=&
        \mathcal{N}\,\left\{\prod_{i=1}^{M}\,\int_{\bR_{i}(0)=\bR_{i}' }^{ \bR_{i}(L)=\bR_{i} }
        [\mathcal{D}\bR_{i}(s)\, ]_{\rm{walls}}\,\right\}\, e^{- \frac{3}{2\ell}\,\sum_{i=1}^{M}
        \int_{0}^{L}{\big(\frac{\partial\bR_{i}(s)}{\partial s}\big)}^{2}\,ds} \\
       \label{eq:zphantomchainsconfinedb}
       &=&\mathcal{N}\int\prod_{i=1}^{M}\,[\mathcal{D}\bR_{i}(s)\,]\, e^{- \frac{3}{2\ell}\,\sum_{i=1}^{M}
        \int_{0}^{L}\dot{\bR}_{i}^{2}(s)\,ds} \nonumber\\
      & &\qquad \times \,\prod_{s}\bigg[\Theta(\bR_{i}(s)\cdot
      \hat{z})\,\Theta(h_{z}-\bR_{i}(s)\cdot\hat{z})\bigg] \,.
\end{eqnarray}
The normalization $\mathcal{N}$
refers to the number of configurations of a collection of $M$
completely free polymer chains, each starting at $\bRacc_{i}$ and ending at
$\bR_{i}$ in $N$ steps, with Kuhn step-length $\ell$. The confining walls,
which act as a constraint in the partition function, can be rewritten to
resemble an attractive potential in terms of continuous arclength
variables (Section \ref{sec:singlechaininbox}):
\begin{eqnarray}
  \label{eq:potA}
  A[\bR_{i}(s)]\equiv \ln[\Theta(\bR_{i}(s)\cdot \hat{z})]+\ln[\Theta(h_{z}-\bR_{i}(s)\cdot\hat{z})]\,, \quad\textrm{and}\quad
  \Theta\,(R\,)= \left\{ \begin{array}{ll} 1 & \textrm{if
             $R > 0$  }\\
           0 & \textrm{otherwise} \end{array} \right.
\end{eqnarray}
Since the chains are only confined in the $\hat{z}$-coordinate,
the complete partition function for the confined melt will look as follows:
\begin{eqnarray}
\label{eq:zphantomchainsconfinedc}
      \mathcal{Z}_{\rm{melt}} &=&\mathcal{N}\,\prod_{i=1}^{M}\,\,\int_{R_{xi}(0)=R_{x}'}^{R_{xi}(L)=R_{x}} [\,
        \mathcal{D} R_{xi}(s)\,]\, \exp\Big\{-\frac{3}{2\ell}
        \int_{0}^{L}\dot{R}_{xi}^{2}(s)\,ds\Big\} \nonumber\\\qquad& &
        {}\qquad \times 
 \int_{R_{yi}(0)=R_{y}'}^{R_{yi}(L)=R_{y}} [\,
        \mathcal{D} R_{yi}(s)\,]\, \exp\Big\{-\frac{3}{2\ell}
        \int_{0}^{L}\dot{R}_{yi}^{2}(s)\,ds \Big\} \nonumber\\\qquad&
        & {}\qquad \times 
\int_{R_{zi}(0)=R_{z}' }^{R_{zi}(L)=R_{z} } [\,
        \mathcal{D} R_{zi}(s)\,] \,\exp\Big\{-\frac{3}{2\ell}
        \int_{0}^{L}\dot{R}_{zi}^{2}(s)\,ds  \nonumber\\\qquad& & \qquad+
        \sum_{i=1}^{M} \,\int_{0}^{L}\,A\,[R_{zi}(s)]\,ds\Big\}\,.
\end{eqnarray}  
It might be possible to approximate the infinitely deep well represented
by $\ln\,\Theta(\bR(s)\cdot\hat{z})$ by a function that is more suitable to mathematical
manipulations.
\section{Confined Network Formation}
\label{sec:crosslinkformalism}
\begin{figure}[!h]
\begin{center}
  \includegraphics[width=0.6\textwidth]{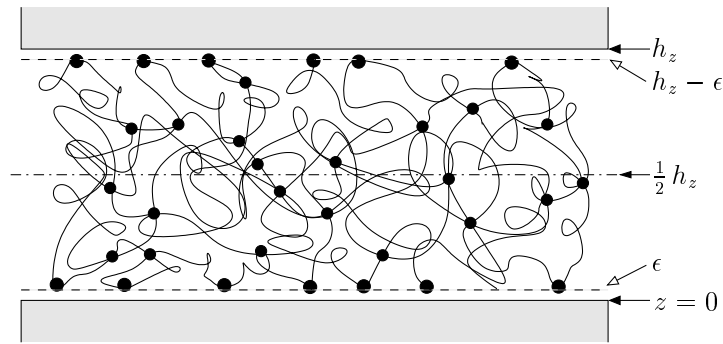}
\caption[A gel network formed at and between two, parallel planar surfaces.]{\label{fig:gelnetworkepsi}A simplified illustration of a gel
  network formed between two parallel planar surfaces of width
  $h_{z}$. The wall-links are situated an infinitesimal distance
  $\epsilon$ from the wall surface. The dash-dot line is the line of
  symmetry at $z=\half h_{z}$.}   
    \end{center}
\end{figure}

Given the collection of phantom chains, they should be crosslinked
  to each other (\emph{polymer-polymer} links) and to the confining surface
  (\emph{polymer-wall} links) in order to constitute a polymer network linked
  to the walls, Figure \ref{fig:gelnetworkepsi}. 

In practice, polymer-polymer\footnote{Henceforth, polymer-polymer links will mostly be
  denoted by the term \emph{bulk-links}, and polymer-wall links by
  \emph{wall-links}.} cross links are usually formed by joining two
segments from different chains by means of sulphur vulcanisation,
peroxide or radiation cross-linking \cite{Young}. In this problem, polymer
adsorption to a surface does not play a role whatsoever. Wall-links
are therefore synthesized in the same manner (for example irradiation) as bulk-links, but
with a lower functionality. 
In a real cross-linking process, the links
  build up in time. One link will affect the neighbouring chain density,
  therefore influencing the position of the next link, and clusters of links may
  appear. As the crosslink density is increased there is an abrupt
  change of the 
  melt from a viscous liquid to a solid, elastic gel that shows no tendency to
  flow.  At this stage --- the so-called \emph{gel-point} --- a giant cluster or network spans the whole
  sample. Theoretical explanations for how the transition
  actually takes place, still remain difficult \cite{GoldbartI,GoldbartII}. We shall only be concerned with the case of sufficiently
  high crosslink density; not a system close to the gel point. 

Here we assume
  that we are dealing with a homogeneous material where \emph{local} clustering of
  crosslinks are negligible. This is a reasonable assumption to make if
  the network is formed from a dense melt system, or the crosslinker
  density is high enough. Furthermore, it has been shown that 
  macroscopic quantities, for example the free energy, are insensitive
  to microscopic details such as inhomogeneities in
  crosslink-density~\cite{Erman}.

\subsection{The Origin of disorder}
\label{sec:disorder}
Both types of linkages are permanent, and impose certain
  topological constraints between the polymers, given that the
  crosslink density is high enough.  The
  randomness of the link formation is the origin of the quenched disorder in
  the system (Section \ref{sec:avedisorder}). \emph{What is the
    probability of this disorder?} We can imagine that the chains --- constituting
  a given, random link --- were just touching prior to the permanent
  linking.   The probability of the disorder is then given by the same
  weight as that of thermal equilibrium of the melt prior to network
  formation.  This idea of an instantaneous linking
  mechanism, was first proposed by Deam and Edwards \cite{DeamEdwards}.
\subsection{Linking Formalism}
\label{sec:instantcross}
\begin{figure}[!h]
\begin{center}
  \includegraphics[width=0.9\textwidth]{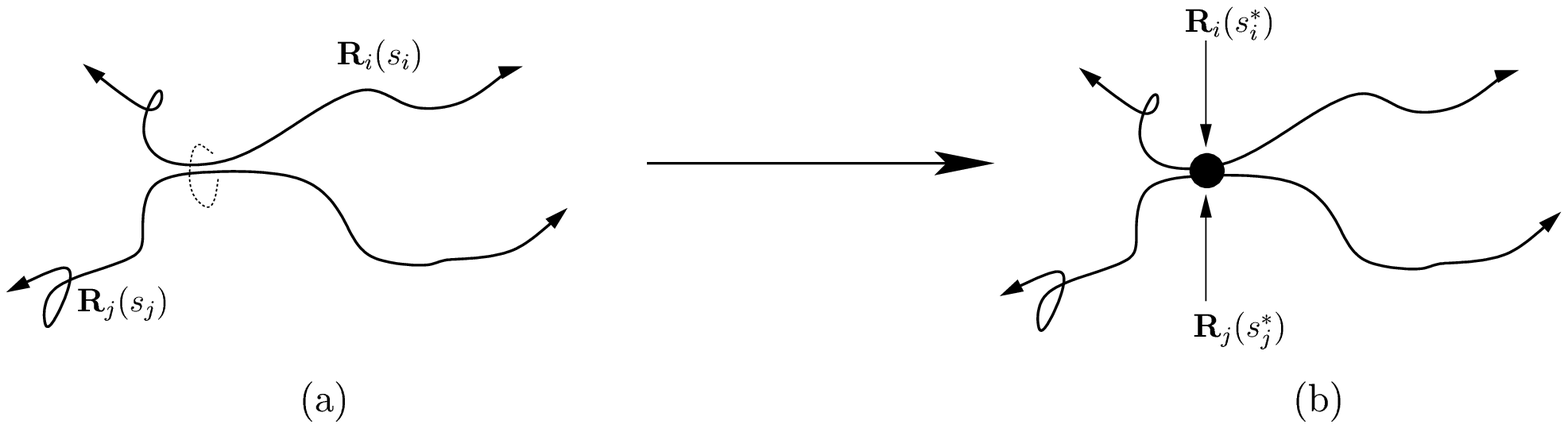}
\caption[Linking mechanism: (a) A slip link (b) Permanant link]{\label{fig:sliplink}Two polymer chains, $\bR_{i}(s_{i})$ and $\bR_{j}(s_{j})$, in close proximity (a)
  modelled by a slip-link, and (b) a permanent crosslinkage constraint \eqref{eq:constraint},  after a fast
  crosslinking procedure.}
    \end{center}
\end{figure}

If a crosslink joins arclength point $s_{i}^{*}$ on chain $i$ with point
  $s_{j}^{*}$ on chain $j$, then we have the constraint, illustrated
  in Figure \ref{fig:sliplink}(b),
  \begin{equation}
    \label{eq:constraint}
    \bR_{i}(s_{i}^{*}) = \bR_{j}(s_{j}^{*})
  \end{equation}
which is unbreakable and stays fixed at the same place on the chain.  Mathematically, we use a Dirac-delta formalism
to pick out the set of linkages, with the arc-locations of the
crosslinks specifying the crosslink topology of the network. The crosslinks are
constraints in the partition function of the system, which have to be
approximated by a more tractable potential. 

We proceed, by considering a melt with chains touching at their future
crosslink positions whilst incorporating the confinement formalism of
Section \ref{sec:confinementformalism}.
 A simplified way to think about the manner
of introducing crosslinkages, is by first adding $\Nc$ sliding
links \cite{Ball} in the bulk and $\Nw$ touching links at both walls, at random. This
is of course completely hypothetical, but will facilitate
statistical calculations of fast 
crosslinking and accommodate the chosen probability of disorder in
Section \ref{sec:disorder}.

After the \emph{touching} and \emph{sliding} links reach equilibrium with the system, they have a free
energy $F'$ given by
\begin{eqnarray}
        e^{-F'/\bolz T}=\mathcal{Z}\,(\,\{\bR_{i},\bRacc_{i},L\}\,)&=&
        \mathcal{N}\int [\,\prod_{i=1}^{M}\,
        \mathcal{D}\bR_{i}(s)\, ]_{\rm{walls}}\quad e^{- \frac{3}{2\ell}\,\sum_{i=1}^{M}
        \int_{0}^{L}{\big(\frac{\partial\bR_{i}}{\partial
            s}\big)}^{2}\,ds} \nonumber \\ 
      & & {}\times
      \Big[\,\sum_{i,j=1}^{M}\,\int_{0}^{L}ds\int_{0}^{L}ds'\,\,
        \delta(\bR_{i}(s)-\bR_{j}(s'))\Big]^{\Nc} \nonumber \\& & {
       }\times \Big[\,\sum_{i=1}^{M}\,\int_{0}^{L}ds\int dx\,dy\,\,
          \delta(\bR_{i}(s)-\mathbf{\eta}(x,y,0))\Big]^{\Nw} \nonumber\\ & & {
       }\times \Big[\,\sum_{i=1}^{M}\,\int_{0}^{L}ds\int dx\,dy\,\,
          \delta(\bR_{i}(s)-\mathbf{\eta}(x,y,h_{z}))\Big]^{\Nw}\,, \label{eq:zprenetworka}
\end{eqnarray}
with $\bfeta$ a certain vector determining the positions of the
crosslinks at the walls.
The notation $[\,\mathcal{D}\bR_{i}(s)\,]_{w}$ indicates that the path integration
for the confined ${z}$-coordinates should only be carried out in the
allowable region. Due to the $\hat{z}$ confinement the partition function \eqref{eq:zprenetworka} is
not symmetric in space. Nevertheless, in each partition function, the
crosslink arc--locations in the Dirac-delta functions are the same, 
since they are quenched variables. The crosslinks therefore act as a common combining factor between the confined
and unconfined contributions to the partition function.
\subsection{Implementing Replicas}
\label{sec:confinedphantomnetw}
\label{sec:confinedphantomnetw}
The formation of the network on the walls is done by ``freezing'' the
\emph{sliding} and \emph{touching} links (Figure \ref{fig:sliplink})
of Section~\ref{sec:instantcross}, 
such that the system reaches
its final free energy $F_{\mathcal{S}}$, given by the logarithm of:
\begin{equation}
  \label{eq:samplefree}
  e^{-F_{\mathcal{S}}\,(\{X_{i}\})/\,\bolz T}=\sum_{\{\bR_{i}\}}\,\exp\,\left(
    -\mathcal{H}_{s}\,[\,\{\bR_{i}\} \,]/\,\bolz T\right)\,.
\end{equation}
In the above equation, 
the free energy $F_{\mathcal{S}}$ is that of a specific \emph{sample}
associated with a
specific set of quenched variables $\mathcal{S}=\{X_{i}\}$. The notation
$\sum_{\{\bR_{i}\}}$ denotes the integration over all possible
conformations of the confined system with degrees of freedom frozen at
$\mathcal{S}$. In Section \ref{sec:averagingdis}
 it was shown that for a system with permanent
constraints, the free energy of the system with a specific
crosslink-state $\mathcal{S}$
should be evaluated first, followed by a disorder-average over all
possible constraints. This procedure would give the correct
experimental free energy $\mathcal{F}$ of the system:
  \begin{eqnarray}
    \label{eq:quenchedave}
    \mathcal{F}=\int\mathcal{P}_{\mathcal{S}}\,(\{X_{i}\})\,F_{\mathcal{S}}\,[\,\{X_{i}\}\,]\,dX_{i}=
    \int\mathcal{P}_{\mathcal{S}}\,(\{X_{i}\})\,\ln\,\mathcal{Z}\,[\,\{X_{i}\}\,]\,dX_{i}\,,
  \end{eqnarray}
where $\mathcal{P}_{\mathcal{S}}(\{X_{i}\})$ is the probability of arclength coordinates
being frozen at a specific set $\{X_{i}\}$. Each sample's fabrication
probability  $\mathcal{P}_{\mathcal{S}}$ --- the initial probability
of the chains when touching --- 
can be be expressed as a formation Hamiltonian\footnote{The formation
  Hamiltonian is given by the $\alpha=0$ part of the replica
  Hamiltonian $\mathcal{H}$ in \eqref{eq:hamiltonianwithchempot}.}. 

In Section \ref{sec:averagingdis} we
averaged over the randomness of formation of a \emph{single} bulk
link by means of the replica method, replicating the system $n$
times. This was done because the quenched average of a logarithm as in
\eqref{eq:quenchedave} is generally difficult to evaluate. We shall
now extend this idea to a macroscopic number of permanent
crosslinks. The probability  $\mathcal{P}_{\mathcal{S}}(\{X_{i}\})$ is
now given by the ideas of Section \ref{sec:disorder}.
In the Deam and Edwards formulation the \emph{fabrication} probability is
manifested in the \emph{zeroth} replica, coupled to the other $n$
replicas \cite{DeamEdwards}. The free energy of the network system is
then identified as the
coefficient of $n$ in the generalised partition function
$\mathcal{Z}(n)$ of $n+1$ replicas. 
Here, the generalised partition functions for the $x$ and $y$ coordinates
coincide with the existing Deam and Edwards model for
an unconfined phantom chain network \cite{DeamEdwards}. In the 
$z$ dimension, the walls act as a secondary localization of the chain
coordinates. The generalised partition
function for the $n+1$ replicas of a \emph{confined} gel is given by
\begin{eqnarray}
        e^{-\mathcal{F}(n)/\bolz T}=\mathcal{Z}(n)&=&
        \mathcal{N}\int [\,\prod_{i=1}^{M}\,
        \mathcal{D} \bR_{i}^{(0)}(s)\, ]_{\rm{w}}\,\widetilde{\int}[\,\prod_{\alpha=1}^{n}\,\prod_{i}^{M}\,
        \mathcal{D}\bR_{i}^{(\alpha)}(s)\,]_{\rm{w}}\, e^{- \frac{3}{2\ell}\,\sum_{i,\,\alpha}\,
        \int_{0}^{L}{\big(\dot{\bR}_{i}^{(\alpha)}\big)}^{2}\,ds} \nonumber \\ 
      & & { }\times
      \Big[\,\sum_{i,j=1}^{M}\,\int_{0}^{L}ds\int_{0}^{L}ds'\,\prod_{\alpha=0}^{n}\,
        \delta\,(\bR_{i}^{(\alpha)}(s)-\bR_{j}^{(\alpha)}(s'))\Big]^{\Nc} \nonumber \\& & {
       }\times \Big[\,\sum_{i=1}^{M}\,\int_{0}^{L}ds\int dx\,dy\,\prod_{\alpha=0}^{n}\,
          \delta\,(\bR_{i}^{(\alpha)}(s)-\mathbf{\bfeta}^{(\alpha)}(x,y,0))\Big]^{\Nw} \nonumber\\ & & {
       }\times \Big[\,\sum_{i=1}^{M}\,\int_{0}^{L}ds\int dx\,dy\,\prod_{\alpha=0}^{n}\,
          \delta\,(\bR_{i}^{(\alpha)}(s)-\mathbf{\bfeta}^{(\alpha)}(x,y,h_{z}))\Big]^{\Nw}\,,
        \label{eq:genzunstrained}
\end{eqnarray}
where $\bfeta^{(\alpha)}$ determines the positions of the
crosslinks at the walls. Each three-dimensional $\bfeta^{(\alpha)}$ is an element in a
$(n+1)$-dimensional vector, given by
\begin{equation}
  \label{eq:etavector}
  \text{{\boldmath $\eta$}}  = \left(\, \eta^{(0)} \,\hdots\, 
      \eta^{(\alpha)} \, \hdots \, \eta^{(n)} \,\right)^{\intercal}\, .
\end{equation}
For the sake of easing the notation, we proceed by modeling the network by one
large polymer of length $\mathcal{L}$. 

In any network synthesized from
many chains, there will be network ``defects''
\cite{Flory,Treloar}. 
\begin{figure}[!h]
\begin{center}
  \includegraphics[width=0.7\textwidth]{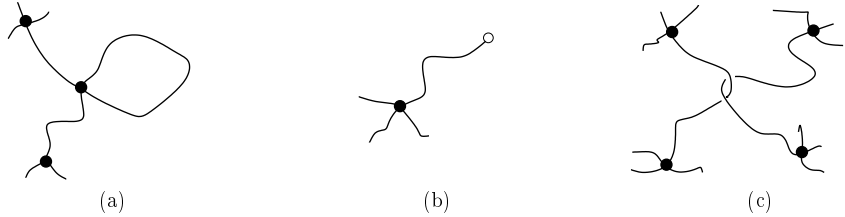}
\caption[Examples of three network imperfections.]{\label{fig:defects}Illustrating three types of network
  \emph{imperfections}: (a) A closed loop resulting from intramolecular
crosslinking; (b) A dangling end (represented by the open circle) and (c)
A trapped entanglement.}
    \end{center}
\end{figure}
The first defect, in Figure \ref{fig:defects}(a), occurs due to the linkage of
two points on the same chain, forming a closed loop that is not linked
with any other chain. The second imperfection [Figure
\ref{fig:defects}(b)] is when only one end
of a chain is attached to the network, creating a so-called
\emph{dangling} end. Dangling ends and closed loops do not
contribute to the elasticity or strength of a substance, and should be excluded when
probing the elastic properties of the gel. In contrast to the
previous defects, entanglements, Figure \ref{fig:defects}(c), should \emph{not} be omitted from 
calculations, since they can (in sufficient number) play the same role
as crosslinkages \cite{BallEdwardsDoi, VilgisEdwards}. 
However, we shall assume the net contribution of
above mentioned three defects to be negligible.  

By rewriting the Dirac-delta
constraints in \eqref{eq:genzunstrained} as pole
integrations\footnote{The constraint terms are exponentiated by implementing
  the following complex integral:
  \begin{equation}
    \label{eq:complexint}
    \textsf{B}^{N}=\frac{N!}{2 \pi i}\,\oint_{\complexnumbers}\,e^{\mu
      \textsf{B} -(N+1)\ln\mu}\,d\mu\,,
  \end{equation} where contour $\complexnumbers$ encloses the origin.}, we
arrive at the most compact formulation:
\begin{eqnarray}
  \label{eq:compactform}
  e^{-\mathcal{F}(n)/\,\bolz T}&=&
  \mathcal{N}\oint\,\frac{\Nc!\,d\mu_{\rm{c}}}{2\pi
          i}\,\oint\,\frac{\Nw!\,d\mu_{\rm{w}}}{2\pi i}
        \,\int\tilde{\int}\dots\tilde{\int}\,[\,\prod_{\alpha=0}^{n}\,\mathcal{D} \bR^{(\alpha)}(s)\,]\nonumber \\ & & {} 
        \times \,\exp\bigg\{-
        \mathcal{H}/\,\bolz T-(\Nc+1)\log\mu_{\rm{c}}-2(\Nw+1)\,\log\mu_{\rm{w}}\bigg\}\,,
\end{eqnarray}
with $\mathcal{H}$ giving the \emph{pseudo}-Hamiltonian for network
connectivity and confinement
\begin{eqnarray}
\label{eq:hamiltonianwithchempot}
\mathcal{H}/\,\bolz T &\equiv& 
     - \frac{3}{2\ell}\,\sum_{\alpha=0}^{n}\,\int_{0}^{\mathcal{L}}\,ds\left[\left(\frac{\partial\bR^{(\alpha)}(s)}{\partial
            s}\right)^{2}+\,A\,[\,\bR^{(0)}(s)\,]
          +\sum_{\alpha=1}^{n}\,\,\tilde{A}\,[\,\bR^{(\alpha)}(s)\,]\,\right]\nonumber \\ & & { } +
        \mu_{\rm{c}}
     \,\int_{0}^{\mathcal{L}}ds\int_{0}^{\mathcal{L}}ds'\,\,
       \prod_{\alpha=0}^{n}\, \delta\,(\,\bR^{(\alpha)}(s)-\bR^{(\alpha)}(s')) \nonumber \\& & {
       }+\mu_{\rm{w}}\,\int_{0}^{\mathcal{L}}ds\int dx\,dy\,\,
          \prod_{\alpha=0}^{n}\,\delta\,(\,\bR^{(\alpha)}(s)-\mathbf{\bfeta}^{(\alpha)}(x,y,0)) \nonumber\\ & & {
       }+\mu_{\rm{w}}\,\int_{0}^{\mathcal{L}}ds\int dx\,dy\,\,
          \prod_{\alpha=0}^{n}\,\delta\,(\,\bR^{(\alpha)}(s)-\mathbf{\bfeta}^{(\alpha)}(x,y,h_{z}))
\end{eqnarray}
where $\mu_{\rm{c}}$ and $\mu_{\rm{w}}$ are the chemical potentials of the bulk-
and wall-links respectively. We have thus, in a convenient manner, represented the linking
constraints as in a grand canonical ensemble. The crosslinks naturally  confine the chain(s)
to a well-defined region which must adhere to the defined region
between the plates. 

The replicas $1$ to $n$ can describe  strained versions
of the $0$'th replica, and can therefore  have different volumes and
temperature. The integrations $\widetilde{\int}$ in
\eqref{eq:compactform} imply integration over the
$n$ deformed systems, each with volume $\widetilde{V}=\mathbf{\Lambda}\, V$, with
$V$ the volume of the undeformed replica system and 
$\mathbf{\Lambda}$ the deformation tensor. After extension or compression of the system by $\mathbf{\Lambda}$, the system will conserve
its crosslinking topology that it had at fabrication, and therefore
each replica must have the same set of crosslink arclength coordinates. 
The infinite deep square well potential representing the wall confinement \eqref{eq:potA}, is
given by $A$ in the unstrained system and by $\tilde{A}$ for the strained replicas:
\begin{equation}
  \label{eq:potAstrained}
  \tilde{A}\,[\bR(s)]\equiv \ln\,[\,\Theta(\bR(s)\cdot
  \hat{z})\,]+\ln\,[\,\Theta(\lambda_{z} h_{z}-\bR(s)\cdot\hat{z})]\,, \qquad\textrm{with}\qquad
  \Theta\,(R\,)= \left\{ \begin{array}{ll} 1 & \textrm{if
             $R > 0$  }\\
           0 & \textrm{otherwise} \end{array} \right.\,.
\end{equation} 
\pagebreak

\section{Planning the variational calculation}
\label{sec:varcar}
Only a handful of statistical mechanics models have been solved
exactly. For the rest it is necessary to resort to some approximation
method. The current replica model falls in the latter class due to the
intractable path integrals in \eqref{eq:compactform}.
In this section we shall employ the Feynman variational principle,
which follows from the inequality, 
\begin{equation}
  \label{eq:feynm}
  \langle \,e^{\textsf{X}}\,\rangle \geq e^{\langle \,\textsf{X} \,\rangle}\,
\end{equation}
which is valid for any real stochastic
variable $\textsf{X}$, due to the concave nature of the $e^{\textsf{X}}$ graph \cite{Feynman}.
\subsection{Introducing transformation coordinates}
\label{sec:transformation}
It is expected that the crosslinks will physically pin down the giant
network chain in space, so that for any $s$, the coordinates
of the different replicas, $\bR^{(0)}(s)$,  $\bR^{(1)}(s)$,
$\ldots \bR^{(n)}(s)$, will be correlated in some way. The coordinates of
the undeformed (zeroth) replica, are free to be anywhere between the
constraining walls. Before we show how to mimic a crosslink
constraint, we introduce a new set of coordinates
$\{\mathbf{X}^{(0)},\mathbf{X}^{(1)},\mathbf{Y}^{(m)}\}\mid_{\{m=1\dots(n-1)\}}$,
with $\mathbf{X}^{(1)}$ a relative coordinate, and  $\mathbf{X}^{(0)}$ being the centre-of-mass coordinate of all the
replicas. The $n-1$ remaining coordinates are simply rotations in
replica space and give the deviation of the chains from the affine
position. We define the new set of coordinates in the standard way \cite{Chompff}: 
\begin{eqnarray}
  \label{eq:defcoorX0}
  X_{j}^{(0)} &=& \frac{R_{j}^{(0)} + \sum_{\alpha =
      1}^{n}\,\lambda_{j}\,R_{j}^{(\alpha)}}{(1+n\lambda_{j}^{2})^{\frac{1}{2}}} \\
  \label{eq:defcoorX1}
  X_{j}^{(1)} &=& \frac{\sqrt{n}\lambda_{j}\,R_{j}^{(0)} -\frac{1}{\sqrt{n}}\, \sum_{\alpha = 1}^{n}\,R_{j}^{(\alpha)}}{(1+n\lambda_{j}^{2})^{\frac{1}{2}}}\\
  \label{eq:defcoorYm}
  Y_{j}^{(m)} &=& \frac{1}{\sqrt{n}}\,\sum_{\alpha =
    1}^{n}\,e^{(\,2\pi i m\alpha\,)/n}\, R_{j}^{(\alpha)}\,,\qquad m = 1,2,\dots(n-1)
\end{eqnarray}
where the $j$'s are Cartesian indices. These coordinates define an
orthonormal transformation $\mathbf{\textsf{T}}$ with Jacobian equal to
one. The $\lambda_{j}$'s are the elements on the diagonal of the
deformation tensor $\mathbf{\Lambda}$. As a rule, a gel substance is
only weakly susceptible to volume change during strain, as already
mentioned in Section \ref{sec:stitchelas}. Consequently, the
$\lambda$'s are suitably defined by the isovolumetric deformation tensor
$\mathbf{\Lambda}$ given by \eqref{eq:isomatrix}.
The matrix form and other details of the transformation $\mathbf{\textsf{T}}$ are relegated to
Appendix \ref{chap:appenda}.
\subsection{Introducing a trial localization potential}
\label{sec:trialpotentialQ}
To make the problem more solvable in terms of the variational
procedure, we introduce a trial potential to simulate the Dirac-delta
crosslink constraints. Under strain the mean positions of the
crosslinks may deform affinely, but it would be wrong to enforce the
condition that all the crosslinks \textit{themselves} should deform
affinely. We therefore allow them freedom to oscillate around these affine positions. It seems reasonable to simulate the crosslink constraints by a
trial harmonic potential of the form\footnote{Note that the replica
  indices in the transformed coordinates,
  $\{\mathbf{X}^{(0)},\mathbf{X}^{(1)},\mathbf{Y}^{(m)}\}\mid_{\{m=1\dots(n-1)\}}$,
  are now denoted by $\beta$, to avoid any misconceptions. Before the transformaton
  $\mathbf{\mathsf{T}}$, the formation (unstrained) replica was
  $\alpha=0$, and the remaining ($\alpha>0$) replicas were strained versions. Now, the
  replica $\beta=0$ represents the centre-of-gravity of all the
  replicas, with the remaining ($\beta>0$) replicas being relative to
  it. Henceforth, any references to $\alpha$ indices should be
  understood as pertaining to the physical  $\mathbf{R}^{\alpha\ge 0}$
  coordinate
  model.}, first proposed in \cite{Chompff}:
\begin{equation}
  \label{eq:qpot}
  Q = \frac{1}{6\ell}\,\int_{0}^{\mathcal{L}}\,ds\,\sum_{i=x,y,z}\,q_{i}^{2}\,\sum_{\beta=1}^{n}X_{i}^{(\beta)2}
\end{equation}
with $q_{i}$ being the localization parameter and  a measure of the
limits within which each crosslink is allowed to fluctuate. If $q_{i}$ is
small, the crosslinking is weak. The
dimension of $q_{i}^{-1}$ is length$^{2}$. The inverse of $q_{i}$ defines
the mean distance in which the crosslinks are localised in the
$\hat{i}$-direction. In other words, $(q_{x}q_{y}q_{z})^{-\frac{1}{2}}$ defines the allowable volume which a
crosslink may explore. In this chapter, we assume that the type of
localization at the walls does not differ considerably from the bulk
localization. Due to the symmetry of the problem in the $x$ and $y$ coordinates, we set $q_{x}=q_{y}$, and
  henceforth work with two scalar parameters $q_{x}$ and $q_{z}$.
\pagebreak
\subsection{Notation}
\label{sec:notation}
In this section we list all the constituents\footnote{The vector
  $\nu^{(0)}$ in \eqref{eq:walllinksdef}, defined in Appendix \ref{chap:appenda}, describes the
  wall-link locations, which were defined by the vector
  $\eta$ prior to the transformation $\mathbf{\mathsf{T}}$.} of the model  that we
shall refer to frequently during the actual variational calculation. 
\begin{flushleft}
\framebox[\textwidth]{\parbox[][][c]{15.5cm}{
{\small
\begin{itemize}
\item The connectivity of the chains (approximated by one long polymer
  of length $\mathcal{L}$) is denoted by the Wiener term $\mathbb{W}$.
  \begin{equation}
    \label{eq:wienerdef}
    \mathbb{W} \equiv -\frac{3}{2 \ell} \int_{0}^{\mathcal{L}}\, \Big[
    {\mathbf{\dot{X}}}^{(0)\,2}(s) + {\mathbf{\dot{X}}}^{(1)\,2}(s)+ \sum_{m=1}^{n-1}\,|{\mathbf{\dot{Y}}}^{(m)}(s)|^{2}\Big]\,ds
  \end{equation}
\item The bulk crosslinks are denoted by $\mu_{\rm{c}}\mathbb{X}_{\rm{c}}$
 \begin{eqnarray}
    \label{eq:bulkdef}
    \mu_{\rm{c}}\mathbb{X}_{\rm{c}} & \equiv & \mu_{\rm{c}}\,\int_{0}^{\mathcal{L}}ds\int_{0}^{\mathcal{L}}ds'\,\,
       \delta\big(\,\mathbf{X}^{(0)}(s)-\mathbf{X}^{(0)}(s')\big) \,
       \delta\big(\,\mathbf{X}^{(1)}(s)-\mathbf{X}^{(1)}(s')\big)\,
      \nonumber \\ { }& & { }\times\,\prod_{m=1}^{n-1}\,\delta\big(\,\mathbf{Y}^{(m)}(s)-\mathbf{Y}^{(m}(s')\big) 
  \end{eqnarray}
\item The wall crosslinks are denoted by $\mu_{\rm{w}}\mathbb{X}_{\rm{w}}$.  [We assume 
the same chemical potential $\mu_{\rm{w}}$ for both walls.]
{\setlength\arraycolsep{2pt} 
\begin{eqnarray}
    \label{eq:walllinksdef}
    \mu_{\rm{w}}\mathbb{X}_{\rm{w}} &\equiv& \mu_{\rm{w}}\,\int_{0}^{\mathcal{L}}ds\int_{-\infty}^{+\infty}dx\,dy\,
       \bigg[\,
       \delta\big(\,\mathbf{X}^{(0)}(s)-\mathbf{\nu}^{(0)}(x,y,\epsilon)\big) \,
       \delta\big(\,\mathbf{X}^{(1)}(s)\big)\prod_{m=1}^{n-1}\,\delta\big(\,\mathbf{Y}^{(m)}(s)\big) \nonumber \\ 
       & & +\qquad
       \delta\big(\,\mathbf{X}^{(0)}(s)-\nu^{(0)}(x,y,h_{z}-\epsilon) )\big)
       \,\delta\big(\,\mathbf{X}^{(1)}(s)\big)\,
       \prod_{m=1}^{n-1}\,\delta\big(\,\mathbf{Y}^{(m)}(s)\big)\, \bigg]
  \end{eqnarray}
}
\item The trial potential is $\mathbb{Q}$.
  \begin{equation}
    \label{eq:qdef}
   \mathbb{Q} \equiv
   \sum_{i={x,y,z}}\,\frac{q_{i}^{2}\ell}{6}\int_{0}^{\mathcal{L}}\,ds\,\bigg[\,X_{i}^{(1)\,2}+\sum_{m=1}^{n-1}|Y_{i}^{(m)}|^{2}\,
   \bigg] 
  \end{equation}
\item The wall-constraints are denoted by $\mathbb{A}$.
  \begin{eqnarray}
    \label{eq:walldef}
    \mathbb{A} &\equiv & \int_{0}^{\mathcal{L}}\,ds\, \bigg\{
    A\big(T_{z}^{00}X_{z}^{(0)}+T_{z}^{10}X_{z}^{(1)}\big)  \nonumber
    \\ & & +
    \sum_{\beta=1}^{n}\,\tilde{A}\big(T_{z}^{0\beta}X_{z}^{(0)}+T_{z}^{1\beta}X_{z}^{(1)}+
    \sum_{m=1}^{n-1}\,T_{z}^{(m+1)\beta}Y_{z}^{(m)}\big)\,\bigg\}
  \end{eqnarray}
\item The constant chemical potential terms associated with the
  contour integrals, we group together as $\mathfrak{c}$, that is,
  \begin{equation}
    \label{eq:chempotdef}
    \mathfrak{c}  \equiv -(\Nc+1)\,\log\mu_{\rm{c}}-2\,(\Nw+1)\,\log\mu_{\rm{w}}.
  \end{equation} 
\end{itemize}
}}}
\end{flushleft}
In terms of the new variables $\{\mathbf{X}^{(0)},\mathbf{X}^{(1)},\mathbf{Y}^{(m)}\}$ and above
notation\footnote{The shortcut notation $[\rm{int}]$ refers to $\oint\,\frac{\Nc!d\mu_{\rm{c}}}{2\pi
          i}\,\oint\,\frac{\Nw!d\mu_{\rm{w}}}{2\pi i}\,\int\,[\mathcal{D}
        {\mathbf{X}}^{(0)}]\,[\mathcal{D} {\mathbf{X}}^{(1)}] 
        [\,\prod_{m=1}^{n-1}
        \mathcal{D} {\mathbf{Y}}^{(m)}]$}, the generalised partition function of \eqref{eq:compactform}
can be summarized as follows:
\begin{equation}
\label{eq:shortcutZa}
e^{-F(\mu_{\rm{w}}, \mu_{\rm{c}},n)/\bolz T} = \int
\,[int]\,e^{-\mathbb{W}+\mu_{\rm{c}}\mathbb{X}_{\rm{c}}+\mu_{\rm{w}}\mathbb{X}_{\rm{w}}+\mathbb{A}+\mathfrak{c}}
\equiv  \int\,[\rm{int}]\,e^{-\mathcal{H}/\bolz T}.
\end{equation}
\subsection{The variational Hamiltonian}
\label{sec:varham}
The motivation behind the variational (mean-field type) approximation
is the relative ease with which one can obtain some understanding of
the constrained system.

The crosslinks play the role of a localization potential that
restrains each chain to some mean path in the gel. If the confined gel
is strained by $\mathbf{\Lambda}$ \eqref{eq:isomatrix} the
\emph{average} trajectory of a
Gaussian chain 
will also transform affinely. However, in the strained state (corresponding
to replica indices $\alpha>0$) the chains will probably tend to deviate from
their mean paths \cite{BallEdwardsDoi}, depending on the degree of crosslinking. The fluctuations should not be
ignored, but rather simulated by a suitable trial potential. In Section \ref{sec:trialpotentialQ} an harmonic trial
potential \eqref{eq:qpot} was elected, because it is simple enough so
that the relevant path integrals in \eqref{eq:compactform} can be
calculated. 

We model the constrained system variationally by means of two variational
parameters, namely the localization parameters  $q_{x}$ and $q_{z}$.
Using the trial potential
$\mathbb{Q}$, the parameters can be found by minimizing the
variational free energy $\mathcal{F}_{\rm{var}}$.
Using the Feynman variational principle the expression \eqref{eq:shortcutZa} changes to
\begin{eqnarray}
\label{eq:shortcutZvara}
e^{-F(\mu_{\rm{w}}, \mu_{\rm{c}},n)/\bolz T} &\geq& \int \,[\rm{int}]\,e^{\langle\,
  \mathbb{Q}+\mu_{\rm{c}}\mathbb{X}_{\rm{c}}+\mu_{\rm{w}}\mathbb{X}_{\rm{w}}\,\rangle-\mathbb{W}-\mathbb{Q}+\mathbb{A}}\nonumber \\ &=& \int \,[\rm{int}]\,e^{\langle
  \mathbb{Q}+\mu_{\rm{c}}\mathbb{X}_{\rm{c}}+\mu_{\rm{w}}\mathbb{X}_{\rm{w}}\rangle}\Big(\int\,
\mathcal{G}'\Big)\nonumber \\ &\equiv&
e^{-F_{\rm{var}}'(q_{x},q_{z})/\bolz T}
\end{eqnarray}
where $\langle\ldots\rangle$ means averaging with respect to the
Green's function given by 
\begin{equation}
\label{eq:shortcutgreena}
\int\,\mathcal{G}'= \int \,[\mathcal{D} \mathbf{X}^{(0)}][\mathcal{D}
\mathbf{X}^{(1)}][\prod_{m=1}^{n-1}\,\mathcal{D}
\mathbf{Y}^{(m)}]\,\underbrace{\,e^{-\mathbb{W}-\mathbb{Q}
+\mathbb{A}}}_{e^{-\mathcal{H}_{\rm{var}}'/\bolz T}\,}\,.
\end{equation}
Calculating the Green's function in (\ref{eq:shortcutgreena}) seems
intractable at this stage, especially due to the $\mathbb{A}$ term for
the $\mathbf{X}^{(1)}$ and $\mathbf{Y}^{(m)}$ variables. For now, we circumvent this
problem by implementing another trial potential that
only contains the centre-of-mass coordinate $\mathbf{X}^{(0)}$, and
which is defined as
follows:
\begin{equation}
  \label{eq:defA0}
  \mathbb{A}_{0} \equiv \int_{0}^{\mathcal{L}}\,ds\,\ln\,\left[\,\Theta\big(T^{00} X_{z}^{(0)}(s)\big)\,\Theta\big(h_{z}-T^{00}X_{z}^{(0)}(s)\big)\,\right]
\end{equation}
such that we now choose to have to work with an exactly solvable Green's
function given by
\begin{equation}
\label{eq:shortcutgreen0}
\mathcal{G}= \int \,[\mathcal{D} \mathbf{X}^{(0)}][\mathcal{D}
\mathbf{X}^{(1)}][\prod_{m=1}^{n-1}\,\mathcal{D} \mathbf{Y}^{(m)}]\,
\underbrace{e^{\,-\mathbb{W}-\mathbb{Q}+\mathbb{A}_{0}}}_{\equiv\,
  e^{-\mathcal{H}_{\rm{var}}/\bolz T}\,}.
\end{equation}
In terms of the partition function, it is required to calculate
\begin{eqnarray}
  \label{eq:shortcutpartitionG0}
  \mathcal{Z}(n,\mu_{\rm{w}},\mu_{\rm{c}})&=& {\langle
  \,e^{\left(\mu_{\rm{c}}\mathbb{X}_{\rm{c}}+\mu_{\rm{w}}\mathbb{X}_{\rm{w}}+\mathbb{Q}+\mathbb{A}-\mathbb{A}_{0}\right) }\,\rangle}_{\mathcal{G}}
\,\left(\int\,\mathcal{G}\right)\\
\label{eq:shortcutpartitionG0geqg} &\geq & e^{{\langle
  \mu_{\rm{c}}\mathbb{X}_{\rm{c}}+\mu_{\rm{w}}\mathbb{X}_{\rm{w}}+\mathbb{Q}+\mathbb{A}-\mathbb{A}_{0}
  \rangle}_{\mathcal{G}}}\,\left(\int\,\mathcal{G}\right)\\ \label{eq:shortcutpartitionG0geqF}&\equiv&
e^{-F_{\rm{var}}(q_{x},q_{z})/\bolz T}
\end{eqnarray}
The free energy of the network can now be performed by using the
simple trial form \eqref{eq:shortcutgreen0} for the distribution of the chains. The
resultant variational free energy is then minimised with respect to the
trial function, to give the best possible approximation to the real
free energy $\mathcal{F}$ (for the specific choice of
$\mathcal{H}_{\rm{var}}$):
  \begin{equation}
    \label{eq:varproc}
    \mathcal{F} \approx \min_{\mathcal{H}_{\rm{var}}}\,\left\{\,F_{\rm{var}} \,\right\}\,.
  \end{equation}
The average in the exponent in \eqref{eq:shortcutpartitionG0geqg}
contains a \emph{reduced} wall-constraint, $\mathbb{A}-\mathbb{A}_{0}$,
which we show in Appendix
\ref{chap:appendaa} to be negligible,  
\begin{equation}
  \label{eq:ansatzAA0}
  \langle\,\mathbb{A}-\mathbb{A}_{0}\,\rangle \simeq 0\,,
\end{equation}
under the conditions of our variational
calculation and a \emph{softened} wall potential $\mathbb{A}$.
\pagebreak

\section{The variational calculation (for $\langle\mathbb{A}-\mathbb{A}_{0}\rangle=0$)}
\label{sec:varcalc}
The actual calculation of the variational free energy
$\mathcal{F}_{\rm{var}}(q_x,q_z)$ involves two steps. The first is calculating
the Green's functions needed in order to average the exponent in \eqref{eq:shortcutpartitionG0geqg}. The second is
evaluating the weighted averages. 

\subsection{The Green's functions}
\label{sec:greensfunca}
The task of computing the Green's function $\mathcal{G}$ in \eqref{eq:shortcutgreen0}
is equivalent to evaluating  three different analytically solvable
propagators\footnote{Henceforth, the bottom indices shall refer to
  the integration variable. For example: $\mathcal{G}_{m}$ indicates
  to a path integration with respect to $Y^{(m)}_{i}$, where $m \in {1,2,\dots,n-1}$. The index $i$ refers to the three Cartesian
  coordinates, except when otherwise stated. The original sample
  volume is $V$, and $V_{i}$ the box length in the $i$ coordinate.}.

For the $\hat{x}$ and $\hat{y}$ coordinates,
the Green's function corresponding to the ``centre-of-gravity'' replica
($\beta=0$), $\mathcal{G}= \int \,[\rm{int}]\,e^{-\mathbb{W}}$, satisfies the partial differential equation 
\begin{equation}
  \label{eq:diffeqx0}
  \left( \frac{\partial}{\partial
    s}-\frac{\ell}{6}\,\nabla_{X_{i}^{(0)}}^{2}\right) \,
  \mathcal{G}_{0}=\delta\,\big( X_{i}^{(0)}(s)-X_{i}^{(0)}(s')\big)
    \,\delta\,( s-s')\,,
\end{equation}
for a system in the transformed volume, namely a $(n+1)$-tuple \emph{cuboid},
given by
\begin{equation}
\label{eq:volX0}
  \widetilde{V} = V \prod_{i={x,y,z}}\,\left( 1+n \lambda_{i}^{2} \right)^{\frac{1}{2}}\,.
\end{equation}

The path integral for the remaining replicas ($\beta>0$) includes the trial
potential $\mathbb{Q}$ which mimics the fluctuations of the strained ($\alpha>0$)
from the unstrained ($\alpha=0$)
replicas. In this case, \eqref{eq:shortcutgreen0} simplifies to
evaluating  $\mathcal{G}= \int
\,[\rm{int}]\,e^{-\mathbb{W}-\mathbb{Q}}$. The solution satisfies the
differential equation
\begin{equation}
  \label{eq:diffeqx1}
  \left( \frac{\partial}{\partial
    s}-\frac{\ell}{6}\,\nabla_{X_{i}^{(\beta)}}^{2}+\frac{\ell q_{i}^{2}}{6}\,X_{i}^{(\beta)\,2}(s)\right) \,
  \mathcal{G}_{0}(X_{i\,s}^{(\beta)},X_{i\,s'}^{(\beta)};|s-s'|)=\delta\left( X_{i\,s}^{(\beta)}-X_{i\,s'}^{(\beta)}\right)
    \,\delta\big( s-s' \big)\,,
\end{equation}
and is related to the propagator of a quantum particle\footnote{The
  kernel of a quantum particle of mass $m$ in an harmonic oscillator of
  frequency $\omega$ can easily be retrieved from
  the polymer result \eqref{eq:greens1x} by the following substitution: $s\to it$,
  $\frac{3}{2\ell}\to \frac{m}{2 \hbar}$ and $q \to \frac{3\omega}{\ell}$.} in an harmonic
potential \cite{Feynman}.

For the confined $\hat{z}$ coordinate, there is an  added wall-constraint, which
exclusively influences\footnote{This follows from our choice of a Green's
  function in \eqref{eq:shortcutgreen0}. } the centre-of-mass coordinate
$X_{z}^{(0)}$. The Green's function,  $\mathcal{G}= \int
\,[\rm{int}]\,e^{-\mathbb{W}+\mathbb{A}_{0}}$, for the
``centre-of-gravity'' ($\beta=0$) replica
satisfies a
partial differential equation 
\begin{equation}
  \label{eq:diffeqreplica}
  \Big(\, \frac{\partial}{\partial
  s}-\frac{\ell}{6}\,\nabla_{X_{z}^{(0)}}^{2}-A\,[X_{z}^{(0)}(s)]\,\Big)\,\mathcal{G}_{0}(X_{z\,s}^{(0)},X_{z\,s'}^{(0)};|s-s'|)=\delta
\left( X_{z\,s}^{(0)}-X_{z\,s'}^{(0)}\right)\,\delta(s-s'),
\end{equation}
analogous to \eqref{eq:diffeqa}.

The different solutions to the various path integrals are well-known
\cite{Wiegel, Feynman}, and are summarised as follows:
\begin{flushleft}
\framebox[\textwidth]{\parbox[][345pt][c]{15.5cm}{
{\small
\begin{eqnarray}
  \label{eq:greens0x}
  \bullet \quad\lefteqn{\mathcal{G}_{0}(X_{i}^{(0)}(s),X_{i}^{(0)}(s'),|s-s'|)
     =  \mathcal{N}\,\int_{-\infty}^{\infty}
    \,
    [\,\delta X_{i}^{0}\,]\,e^{-\frac{3}{2\ell}\,\int_{0}^
    {\mathcal{L}} \dot{X}_{i}^{(0)\,2}\,ds}}\nonumber \\  
&=&
\frac{1}{V_{i}\,(1+n\lambda^{2})^{\frac{1}{2}}}+\left(\frac{3}{2\pi
    \ell |s-s'|}\right)^{1/2}\,e^{-\frac{3}{2
    \ell}\frac{\left(X_{i}^{(0)}(s)-X_{i}^{(0)}(s')\right)^{2}}{|s-s'|}}\quad \text{for}\quad i=\{x,y \}.
\end{eqnarray}
\begin{eqnarray}
  \label{eq:greens0z}
  \bullet \quad \mathcal{G}_{0}(X_{z}^{(0)}(s),X_{z}^{(0)}(s'),|s-s'|)=\mathcal{N}\,\int\,[\,\mathcal{D} X_{z}^{0}\,]\,\exp\Big\{ -\frac{3}{2\ell}\int_{0}^
    {\mathcal{L}} \,\dot{X}_{z}^{(0)\,2}\,ds\Big\}\qquad\nonumber \\
      {       }\times\, \exp\Big\{\int_{0}^
    {\mathcal{L}}\ln\,[\Theta\big(T^{00} X_{z}^{(0)}\big)\,\Theta\big(h_{z}-T^{00}X_{z}^{(0)}\big)] \,ds\Big\}\nonumber \\
=
\frac{2}{\sqrt{1+n\lambda^{2}}h_{z}}\,\sum_{p=1}^{\infty}\,e^{-\frac{\ell\pi^{2}p^{2}}{6\,h_{z}^{2}}|s-s'|}\,\sin\left[\frac{\pi
  p X_{z}^{(0)}(s)}{\sqrt{1+n\lambda^{2}}
    h_{z}}\right]\sin\left[\frac{\pi p X_{z}^{(0)}(s')}{\sqrt{1+n\lambda^{2}} h_{z}}\right]
\end{eqnarray}
\begin{eqnarray}
 \label{eq:greens1x}
  \bullet \quad \lefteqn{\mathcal{G}_{1}(X_{i}^{(1)}(s),X_{i}^{(1)}(s'),|s-s'|)=\mathcal{N}\,\int[\mathcal{D} X_{i}^{(1)}]\,e^{-\frac{3}{2\ell}\,
      \int_{0}^{\mathcal{L}}
    \dot{X}_{i}^{(1)\,2}\,ds\,-\frac{q_{i}^{2}\ell}{6}\int_{0}^{\mathcal{L}}\,X_{i}^{(1)\,2}\,ds}} \nonumber \\ 
 &=& \bigg[\frac{q_{i}}{2\pi \sinh \frac{1}{3} \ell
    q_{i}|s-s'|}\bigg]^{\frac{1}{2}}\,e^{-\frac{ q_{i} }{2}\frac{[X_{i}^{(1)\,2}(s)+X_{i}^{(1)\,2}(s')]\cosh \frac{1}{3} \ell
        q_{i}|s-s'|-2X_{i}^{(1)}(s) X_{i}^{(1)}(s')}{\sinh
        \frac{1}{3}\ell q_{i}|s-s'| } } 
\end{eqnarray}
\begin{eqnarray}
\label{eq:greensmx}
  \bullet \quad \lefteqn{\mathcal{G}_{m}(Y_{i}^{(m)}(s),Y_{i}^{(m)}(s'),|s-s'|)=\mathcal{N}\,\int[\mathcal{D} Y_{i}^{(m)}]\,e^{-\frac{3}{2\ell}\,\int_{0}^{\mathcal{L}}
    |\dot{Y}_{i}^{(m)}|^{2}\,ds-\frac{q_{i}^{2}l}{6}\int_{0}^{\mathcal{L}}\,|Y_{i}^{(m)}|^{2}\,ds}}\nonumber  \\ 
&=& \bigg[\frac{q_{i}}{2\pi
    \sinh \frac{1}{3}\ell q_{i}|s-s'|}\bigg]^{\frac{1}{2}}\,e^{-\frac{q_{i}}{2}\frac{[Y_{i}^{(m)\,2}(s)+Y_{i}^{(m)\,2}
        (s')]\cosh \frac{1}{3}\ell q_{i}|s-s'|-2Y_{i}^{(m)}(s)Y_{i}^{(m)}(s')}{\sinh \frac{1}{3}\ell q_{i}|s-s'|}} 
\end{eqnarray} }}}
\end{flushleft}
\subsection{The Averages}
\label{sec:averages}
The average
$\langle\,\mathbb{Q}+\mu_{\rm{c}}\mathbb{X}_{\rm{c}}+\mu_{\rm{w}}\mathbb{X}_{\rm{w}}\,\rangle$
in \eqref{eq:shortcutpartitionG0} in terms of the trial Hamiltonian $\mathcal{H}_{\rm{var}}$,
is evaluated by means of the Green's functions in Section
\ref{sec:greensfunca}. The full calculation is presented in \mbox{Appendix
\ref{chap:appaverages}}, and here we shall simply list the results. 

\begin{flushleft}
\underline{\textbf{The Bulk Crosslinks}   $\langle \mu_{\rm{c}}\mathbb{X}_{\rm{c}}\rangle$}
\end{flushleft}
The weighted average of the polymer-polymer link contribution, defined
in \eqref{eq:bulkdef} and calculated in Appendix \ref{sec:bulkappend}, is
given by
\begin{eqnarray}
\langle\, \mu_{\rm{c}}\mathbb{X}_{\rm{c}}\,\rangle &=&
  \mu_{\rm{c}}\,\langle\mathbb{X}_{\rm{c} x} \mathbb{X}_{\rm{c} y}\mathbb{X}_{\rm{c}
    z}\rangle \nonumber\\  \label{eq:completeXcb}
&=& \left \langle \mu_{\rm{c}}\, \int_{0}^{\mathcal{L}}ds\int_{0}^{\mathcal{L}}ds'\,\,
       \prod_{i=x,y,z}\delta\big(\,X_{i}^{(0)}(s)-X_{i}^{(0)}(s')\big) \,
       \delta\big(\,X_{i}^{(1)}(s)-X_{i}^{(1)}(s')\big)\,\right.
      \nonumber \\ & & \left.\qquad\times\,\prod_{m=1}^{n-1}\,\delta\big(\,Y_{i}^{(m)}(s)-Y_{i}^{(m}(s')\big) \right\rangle \\ 
          &=& \frac{3 \mu_{\rm{c}}}{2
  \sqrt{1+n\lambda^{2}}\,h_{z}} \,
\left(\frac{q_{z}\,q_{i}^{2}}{8 \pi^{3}}\right)^{n/2}\,
    \left[
    \frac{\mathcal{L}^{2}}{A\left(1+n/\lambda\right)}\,+
    \frac{3\mathcal{L}}{2 \pi \ell}\,\ln\left(\frac{\mathcal{L}}{\ell_{\rm{c}}}\right)\right]\,, \label{eq:completeXce}
\end{eqnarray}
where $A$ is the cross-sectional area of the undeformed volume
containing the sample. The chain \emph{cutoff}-length $\ell_{\rm{c}}$, over
which the chain is not flexible but stiff, is of the order of magnitude
of the Kuhn length $\ell$. In order to compute the average
$\langle\mathbb{X}_{\rm{c} z}\rangle_{\mathcal{G}_{0}}$ of the
\emph{zero}th replica, we limited the calculations to the limit of
total chain length larger that the spacing between the plates, that is
$h_{z}\ll \sqrt{\mathcal{L}\,\ell}$ of \eqref{eq:limitf1}.

\begin{flushleft}
\underline{\textbf{The Wall Crosslinks}   $\langle \mu_{\rm{w}} \mathbb{X}_{\rm{w}}\rangle$}
\end{flushleft}
The wall crosslinks are completely specified by the vectors $\nu^{(0)}(x, y,\epsilon)$
and $\nu^{(0)}(x,y,h_{z}-\epsilon)$, for crosslinks situated an
infinitesimal distance $\epsilon$ from the wall ``surface'':
\begin{eqnarray}
  \label{eq:nudefeps}
  \nu^{(0)}(x,y,\epsilon) &\equiv&\left(x \,\sqrt{1+n/\lambda},y \,\sqrt{1+n/\lambda},\epsilon\,\sqrt{1+n\lambda^{2}}\right) \\
  \label{eq:nudefheps}
  \nu^{(0)}(x,y,h_{z}-\epsilon) &\equiv&\left(x \,\sqrt{1+n/\lambda},y \,\sqrt{1+n/\lambda},(h_{z}-\epsilon)\sqrt{1+n\lambda^{2}}\right)\,.
\end{eqnarray}
The weighted average of the polymer-wall link contribution  calculated
in Appendix \ref{sec:appwall} is
\begin{eqnarray}
\langle \,\mu_{\rm{w}}\mathbb{X}_{\rm{w}}\,\rangle &=&
  \mu_{\rm{w}}\,\langle\mathbb{X}_{\rm{w} x} \mathbb{X}_{\rm{w} y}\mathbb{X}_{\rm{w}
    z}\rangle \nonumber \\  \label{eq:completeXwb}
&=&\left\langle 2\mu_{\rm{w}}\,\int_{0}^{\mathcal{L}}ds\int_{-\infty}^{+\infty}dx\,dy\,
       \, \delta\big(\,X_{x}^{(0)}(s)-\nu_{x}^{(0)}\big)
       \,\delta\big(\,X_{y}^{(0)}(s)-\nu_{y}^{(0)}\big) \,
       \delta\big(\,X_{z}^{(0)}(s)-\nu_{z}^{(0)}\big) \right.\nonumber
     \\ & & { } \left. \times 
       \delta\big(\,X_{x}^{(1)}\big)\,
       \delta\big(\,X_{y}^{(1)}\big)\,
       \delta\big(\,X_{z}^{(1)}\big)
           \prod_{m=1}^{n-1}\,\delta\big(\,Y_{x}^{(m)}\big)\,\delta\big(\,Y_{y}^{(m)}\big)\,\delta\big(\,Y_{z}^{(m)}\big) 
       \right\rangle \\
   \label{eq:completeXwd} & \simeq &
   \frac{16 \mu_{\rm{w}}\pi^{2}\epsilon^{2}\mathcal{L}}{h_{z}^{3}\,\left(1+n\lambda^{2}\right)^{1/2}}\, 
 \left(\frac{q_{z}}{\pi}\right)^{n/2}\,\left(\frac{q}{\pi}\right)^{n}\,,
\end{eqnarray}
which is valid only if $h_{z} \ll \mathcal{L}$ and
$\frac{\ell\,q_{i}}{3} \ge 1$.
\begin{flushleft}
\underline{\textbf{The harmonic trial potential} $\langle \,\mathbb{Q}\,\rangle$}  
\end{flushleft}
The average over the harmonic trial potential consists of $n$
identical terms (Appendix \ref{sec:qappend}) that differ only in terms of  the variational
parameters $q_{x}$ and $q_{z}$.
\begin{eqnarray}
  \label{eq:completeQ}
   \langle \,\mathbb{Q} \, \rangle &=& \left\langle
   \sum_{i={x,y,z}}\,\frac{q_{i}^{2}\ell}{6}\int_{0}^{\mathcal{L}}\,ds\,\bigg[\,X_{i}^{(1)\,2}+\sum_{m=1}^{n-1}|Y_{i}^{(m)}|^{2}\,
   \bigg] \right\rangle\\ \label{eq:completeQb}
&=&  \frac{n}{4}\,\left( 3+ \frac{\ell}{3}
  q_{z}\mathcal{L}\coth\frac{\ell}{3} q_{z}\mathcal{L}
  +\frac{2\ell}{3} q_{x}\mathcal{L}\coth\frac{\ell}{3}
  q_{x}\mathcal{L} \right)\\\label{eq:completeQfin}
&=& \frac{n}{4}\left(3+\frac{\ell \mathcal{L}}{3}(q_{z}+2q_{x})
\right)\quad \textrm{for}\quad \frac{q_{i}\,\ell\mathcal{L}}{3} \geq 1\,.
\end{eqnarray}

\subsection{The variational Free energy}
\label{sec:varfreenergyfirst}
The variational free energy \eqref{eq:shortcutpartitionG0geqF} of $n+1$
replica systems is given by
\begin{eqnarray}
  \label{eq:partitiona}
  \lefteqn{\mathcal{Z}_{\rm{var}}(n,q_{x},q_{z}) = e^{-\mathcal{F}_{\rm{var}}(n,
      q_x, q_z)/\bolz T}} \\
 &=&  e^{{\langle
  \mu_{\rm{c}}\,\mathbb{X}_{\rm{c}}+\mu_{\rm{w}}\mathbb{X}_{\rm{w}}+\mathbb{Q}+\mathbb{A}-\mathbb{A}_{0}
  \rangle}_{\mathcal{G}}}\,\left(\int\,\mathcal{G}\right) \\
\label{eq:partitionb} &=&
\oint\,\frac{\Nc!d\mu_{\rm{c}}}{2\pi i\,\mu_{\rm{c}}^{\Nc+1}}
\,e^{ \mu_{\rm{c}}\,
  \langle\,\mathbb{X}_{\rm{c}}\,\rangle}\,\oint\,\frac{\Nw!d\mu_{\rm{w}}}{2
  \pi i \,\mu_{\rm{w}}^{\Nw+1}}
\,e^{ \mu_{\rm{w}}\, \langle\,\mathbb{X}_{\rm{w}}\,\rangle}\,e^{\frac{n}{4}\left(3+\frac{\ell \mathcal{L}}{3}(q_{z}+2q_{x})
\right)} \nonumber \\ & & \times 
\left(\frac{2\pi}{q_{z}}\right)^{\frac{n}{2}}\,\left(\frac{2\pi}{q_{x}}\right)^{n}\,e^{-\frac{\ell \pi^{2}
    L}{6(1+n\lambda_{z}^{2})\,h_{z}^{2}}}\,
  \frac{8V\prod
    \sqrt{1+n\lambda^{2}}}{\pi^{2}}\,e^{-\frac{n\,\ell}{2}\left(2\,q_{x}+q_{z}\right)}
\end{eqnarray}
In the above expression the integrals over the chemical potentials
$\mu_{\rm{c}}$ and $\mu_{\rm{w}}$ can be evaluated using \eqref{eq:chempotdef}
and the identity \eqref{eq:complexint}
\cite{DeamEdwards}, such that the variational generalised partition function
can be written as follows: 
{\setlength\arraycolsep{1pt}
\begin{eqnarray}
\label{eq:partitionc} \mathcal{Z}_{\rm{var}} &=& 
 \langle\,\mathbb{X}_{\rm{c}}\,\rangle^{\Nc}
\langle\,\mathbb{X}_{\rm{w}}\,\rangle^{\Nw}\,e^{\frac{n}{4}\left[3-\frac{\ell \mathcal{L}}{3}(q_{z}+2q_{x})
\right]}
\Big(\frac{2\pi}{q_{z}}\Big)^{\frac{n}{2}}\Big(\frac{2\pi}{q_{x}}\Big)^{n}\,e^{-\frac{\ell \pi^{2}
    L}{6(1+n\lambda_{z}^{2})\,h_{z}^{2}}}\,
  \frac{8V\prod \sqrt{1+n\lambda^{2}}}{\pi^{2}}\\
  \label{eq:partitiond} &=&
  \left\{\frac{3 \mu_{\rm{c}}}{2}\,
    \left[
    \frac{\mathcal{L}^{2}}{V\prod_{i}\sqrt{1+n\lambda_{i}^{2}}}\,+
    \frac{3\mathcal{L}}{2 \pi
      \ell h_{z}\sqrt{1+n\lambda_{z}^{2}}}\,\ln\Big(\frac{\mathcal{L}}{\ell_{\rm{c}}}\Big)\right]\right\}^{\Nc}\left(\frac{q_{z}\,q_{x}^{2}}{8 \pi^{3}}\right)^{\frac{n(\Nc-1)}{2}}\nonumber \\
 & &{}\left\{\frac{16
     \mu_{\rm{w}}\pi^{2}\epsilon^{2}\mathcal{L}}{h_{z}^{3}\,\left(1+n\lambda^{2}\right)^{1/2}}
 \left(\frac{q_{z}\,q_{x}^{2}}{\pi^{3}}\right)^{n/2}\right\}^{\Nw}e^{\frac{n}{4}\left[3-\frac{\ell \mathcal{L}}{3}(q_{z}+2q_{x})\right]}
 e^{-\frac{\ell \pi^{2} L}{6(1+n\lambda_{z}^{2})\,h_{z}^{2}}} \frac{8V\prod \sqrt{1+n\lambda^{2}}}{\pi^{2}}
\end{eqnarray}}
In Deam and Edwards \cite{DeamEdwards} the replica limit $n \to 0$ is
taken prior to minimizing the free energy with respect to $q$. The
variational free energy $\mathcal{F}_{\rm{var}}(q_x,q_z)$ of the original
gel system can be identified
as the coefficient of $n$ as follows:
\begin{eqnarray}
  \label{eq:partitionf}
  -\mathcal{F}_{\rm{var}}/\bolz T &=& \frac{\left.{\partial\,\mathcal{Z}_{\rm{var}}/{\partial
   n}}\right|_{n=0}}{\mathcal{Z}_{\rm{var}}(n=0)}
\\ \label{eq:partitionfb} &=&
\frac{1}{2}\frac{\Nc}{(1+c/\rho)}\left[\,\sum_{i=x,y,z}\lambda_{i}^{2}+\frac{c\,\lambda_{z}^{2}}{\rho}\,\right]+\frac{(\Nw-1)\,
  \lambda_{z}^{2}}{2}
 - 
  \frac{\ell\,{\pi }^2\,\lambda_{z}^{2}\,\mathcal{L}}{6\,h_{z}^{2}}
  \nonumber \\ & & { }  + \frac{\ell \mathcal{L}}{12}\,\left(q_{z}+2q_{x}\right) -
  \frac{(\Nc+\Nw-1)}{2}\sum_{i}\,\ln\left(\frac{q_{i}}{2\pi} \right)
\end{eqnarray}
where
\begin{equation}
  \label{eq:rhoetc}
  c = \frac{3}{2\pi \ell\,h_{z}} \ln \frac{\mathcal{L}}{\ell_{\rm{c}}}
  \quad \textrm{and}\quad \rho=\mathcal{L}/V.
\end{equation}
Next the free energy should be minimized with respect to $q_{x}$ and
$q_{z}$ to find the best approximation (for the given trial
Hamiltonian function in \eqref{eq:shortcutgreen0}) to the real free
energy of the system.  The resultant stationary points are isotropic,
and deformation-independent:
\begin{equation}
  \label{eq:qwaardes}
  q_{x}\, = \, \frac{6\,(\Nw+\Nc)}{\ell\,\mathcal{L}}\, =
  \, q_{z}\,.
\end{equation}
Substituting the $q$-values in \eqref{eq:partitionfb} we find that the free energy on deformation, $\mathcal{F}$, has the
following upper bound:
\begin{eqnarray}
  \label{eq:finalevryenergie}
  \mathcal{F} &\leq& \bolz T\,\left[
\frac{1}{2}\frac{\Nc}{(1+c/\rho)}\Bigg \{\,\sum_{i=x,y,z}\lambda_{i}^{2}+\frac{c\,\lambda_{z}^{2}}{\rho}\,\Bigg\}+\frac{(\Nw-1)\,
  \lambda_{z}^{2}}{2}
 - 
  \frac{\ell\,{\pi }^2\,\lambda_{z}^{2}\,\mathcal{L}}{6\,h_{z}^{2}}\right.
  \nonumber \\ & & \left. { }  + \frac{3 \Nw}{2} -
  \frac{(\Nc+\Nw-1)}{2}\sum_{i}\,\ln\left(\frac{3(\Nw+\Nc)}{\pi\ell\,\mathcal{L}} \right)\right]\,.
\end{eqnarray}
\pagebreak

\section{Results}
\label{sec:resultsrealconfined}
In this chapter we extended the Deam-Edwards theory \cite{DeamEdwards}
to a polymer
network that has formed at and between two confining, parallel plane surfaces. 

The free energy on deformation, including only the
$\lambda_{i}$-dependent terms in \eqref{eq:finalevryenergie}, is
\begin{eqnarray}
  \label{eq:finalevryenergielambda}
  \mathcal{F} &\leq& \bolz T\,\left[
\frac{1}{2}\frac{\Nc}{(1+c/\rho)}\Bigg
\{\,\sum_{i=x,y,z}\lambda_{i}^{2}+\frac{c\,\lambda_{z}^{2}}{\rho}\,\Bigg\}+
\frac{\Nw\,\lambda_{z}^{2}}{2}
 - \frac{\ell\,\pi^2\,\lambda_{z}^{2}\,\mathcal{L}}{6\,h_{z}^{2}}\,\right]\,
\end{eqnarray}
where $\Nc$ and $\Nw$ are the total
number of bulk-links and wall-links respectively. The above result is
only valid when the crosslink density is sufficiently high, such that
we are definitely dealing with a \emph{solid} gel. Secondly, the
chain-density of the melt, and the effective contour length $\mathcal{L}$ of
the polymer
network contributing to the elasticity should be greater than the
spacing between the wall $h_{z}$.

In this model the fluctuations of the crosslinks were simulated with
an harmonic potential and variational parameters $q_{x}$ and
$q_{z}$. The physical significance of these parameters is that they
play the role of localizing the fluctuations a distance proportional to
$q_{x}^{-\half}$ in the $x$ and $y$ dimensions, and  $q_{z}^{-\half}$ in
the $z$ dimension. 
The fluctuation volume that each crosslink may explore
can thus be estimated as  
\begin{equation}
  \label{eq:flucvol}
  q^{-3/2}=\left[\frac{\ell\,\mathcal{L}}{6(\Nc+\Nw)}
  \right]^{\frac{3}{2}}\,, 
\end{equation}
since the localisation parameters \eqref{eq:qwaardes} are isotropic and
$\lambda$-independent.

The above result is similar to the Deam and Edwards result
\cite{DeamEdwards} for an
unconfined phantom network, with no wall-links. In the case of a free,
phantom network the localization
parameter is given by the above result \eqref{eq:qwaardes} with no wall-links ($\Nw
\equiv 0$), that is $q_{i}\, = \,
\frac{6\,\Nc}{\ell\,\mathcal{L}}$. The Deam and Edwards result for
the free energy on deformation 
\begin{eqnarray}
  \label{eq:DeamEdwardsfreeEnergy}
  \widetilde{\mathcal{F}}_{\rm{DE}} &\leq& \bolz T \left[
\frac{1}{2}\frac{\Nc}{(1+c/\rho)}\,\sum_{i=x,y,z}\lambda_{i}^{2}\,\right]\\
  \label{eq:rhoetcDE}
  & & \text{with}\qquad c \equiv \left(\frac{3}{2\pi \ell}\right)^{\frac{3}{2}}\,\frac{1}{2\sqrt{\ell_{\rm{c}}}} 
  \qquad \textrm{and}\qquad \rho\equiv\mathcal{L}/V\,,
\end{eqnarray}
is consistent with the confined-network result
\eqref{eq:finalevryenergielambda} in the limit of a gel only confined
by a single wall \mbox{($h_{z} \to \infty$)}, and having no
wall-links ($\Nw=0$).

From \eqref{eq:finalevryenergielambda} the stress\footnote{The stress
  is the elastic force per unit area of the undeformed
cross-section of the sample. The stress was also computed for simpler systems
(Sections \ref{sec:stitchelas} and \ref{sec:onebulklink}).} $f$, in
the specific case of a uniaxial, isovolumetric deformation
\eqref{eq:isomatrix}, is given by
\begin{eqnarray}
  \label{eq:stressrealconfinednetw}
  f = \frac{1}{V}\,\frac{\partial\,\mathcal{F}}{\partial\lambda} = \frac{\bolz T}{V}\,\left[
\frac{\Nc}{(1+c/\rho)}\left\{\,\underline{\,\lambda-\frac{1}{\lambda^2}}+\frac{c}{\rho}\,\lambda\right\}+\left(\,\Nw
 - \frac{\ell\,\pi^2\,\mathcal{L}}{3\,h_{z}^{2}}\,\right)\,\lambda\,\right]\,.
\end{eqnarray}
The underlined part in \eqref{eq:stressrealconfinednetw} coincides
with the well-known result of the classical theory of high elasticity
of polymer networks \cite{JamesGuth}, apart from the different front factor, which may
be attributed to wasted loops \cite{DeamEdwards}. 

The wall-confinement
introduces new terms in the free energy of the network, dependent on the spacing $h_{z}$ between the
walls. For example, the typical confinement term $\frac{\ell \pi^{2}
  \mathcal{L}}{6\,h_{z}^{2}}$, is consistent with earlier results for a
stitch network
(Section \ref{sec:deformstitchnetw}) and a
single chain (Section \ref{sec:stitchmedium}) with an effective
contourlength $\mathcal{L}$. 

To summarize: 
So far, we have applied the Deam and Edwards idea of homogeneous
crosslinking and consequently obtained a similar deformation free energy
(apart from confinement terms). It was found that the harmonic potential $\mathbb{Q}$,
controlling the fluctuations away from the mean (affine), depends on
the inverse distance between $(\Nc+\Nw)$ crosslinks and \emph{not}
on the strain $\lambda$. This means that the affine macroscopic
deformation of the gel-plate system, results in a microscopic
deformation of the chains, which consists of an affine contribution and
a \emph{non-affine} fluctuation contribution. 

Unfortunately the model employed in this chapter is too crude to differentiate
between the types of crosslinking. This resulted in the wall-links and
bulk-links being treated in the same manner. Intuitively the degree of
localization, manifested in the values $q_{x}$ and $q_{z}$, is
expected to depend on the spatial coordinate.

In the next chapter a network, formed from polymer brushes, is
investigated. Within this framework, it is
possible to distinguish, in a simple albeit concise manner, between the localization that each type of
crosslink imposes.

\chapter{The Brush Network }\label{chap:brushnetw}
\begin{figure}[h]
   \centering
   \includegraphics[width=0.3\textwidth]{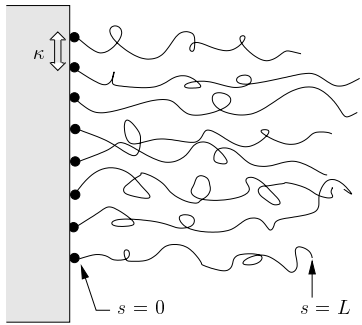}
   \caption[Simplified illustration of a polymer brush attached to a solid surface.]{\label{fig:brush}Simplified illustration of a brush,
     consisting of polymer chains attached to a flat, solid surface
     permanently, with average anchor distance $\kappa$ between chains.}
 \end{figure}

In this chapter we consider a network formed from two grafted polymer
brushes. The term \emph{polymer brush} was invented by de Gennes
\cite{DeGennes80} to
describe an architecture in which polymer chains are terminally
tethered to a surface at a high density. This is the case when the
separation $\kappa$ between anchors, is much smaller than the coil dimension, as shown in Figure \ref{fig:brush}.

The anticipated behaviour (within the replica scheme) of a surface-attached,
crosslinked polymer includes some kind of localization
of the polymer. In Chapter \ref{chap:realnetwork} we  employed the
idea of
Deam and Edwards 
\cite{DeamEdwards}, which treats this localization to be homogeneous and
translation invariant, that is $q_{x}=q_{z}=\mathit{constant}$ in the trial
potential $\mathbb{Q}$ of Section~\ref{sec:trialpotentialQ}. The
\emph{average} polymer
network between planar surfaces is symmetric about $z=\half h_{z}$
(Figure \ref{fig:gelnetworkepsi}), but
certainly not translationally invariant in the $z$-direction. Moreover, we expect the polymer to be
localized differently near the surface than in the bulk region away
from the surfaces\footnote{This is especially so if the distance between the plates
becomes sufficiently large. There exists also the freedom of choosing 
the bulk and surface crosslink densities independently.}. This implies that the variational
parameters $q_{i}$ should depend on the spatial height $z(s)$, of a polymer
segment $s$, between the plates. The only coordinate in the transformed replica
system (Section \ref{sec:transformation}), which represents physical position of the chain segments, is the
``centre-of-mass'' coordinate $\mathbf{X}^{(0)}$. The remaining coordinates
  $\{\mathbf{X}^{(1)},\mathbf{Y}^{(m)}\}\mid_{\{m=1\dots(n-1)\}}$ are simply relative coordinates. 
  Mathematically, it is thus possible to
distinguish between different localizations, by using a trial 
potential of the form
\begin{equation}
  \label{eq:trialintro}
  \mathbb{Q} \thicksim \int
  ds\,q^{2} \left(z(s)\right)\,\sum_{\beta=1}^{n}\mathbf{X}^{(\beta)2}\,,\quad \alpha > 0\,,
\end{equation}
with the localization parameter $q$ a function \mbox{of $z(s)\equiv
  \hat{z}\cdot\mathbf{X}^{(0)}\,(s)$}. 
Unfortunately, the path integration of
an harmonic potential, which is dependent on the arclength coordinate $s$ and 
a mixture of configurational coordinates of \emph{different} replicas,
seems intractable.

However, in a brush network each stretched-out chain is only attached
by one \emph{endpoint} ($s=0$) to the grafting surface.
Consequently, we choose the average
localization to be solely dependent on the arc
distance $s$, and not the spatial position of the chain. It is with
this simplification in mind, that we set out to investigate
\emph{inhomogeneous} localization, in terms of a brush architecture. 


\section{Formation of the Brush Network}
\label{sec:formbrushnetw}
In order to obtain a surface-attached polymer network,
the chains of two polymer brushes should be sufficiently crosslinked. Each \emph{chain} has the
same molecular weight $N$ (they are monodisperse) and contourlength $L$, such that $L=N\ell$.
Each \emph{brush} consists of $\frac{1}{2}\Nw$ phantom
chains that are permanently attached via endpoints to a solid surface at very high
density, that is $\kappa \ll \sqrt{\ell L}$, Figure \ref{fig:brush}. Attachment occurs in the
\emph{absence} of adsorption: the chains are not attracted to the
wall, as in Figure \ref{fig:wells}(a). In
practice, real chains will tend to stretch perpendicular to the surface
due to overcrowding, until the loss in entropy balances the enthalpic
gain of segment dilution. We again choose to work with phantom
(Gaussian) chains, which are not subject to excluded volume effects and therefore
overcrowding should not affect them.

\begin{figure}[h]
   \centering
   \includegraphics[width=0.4\textwidth]{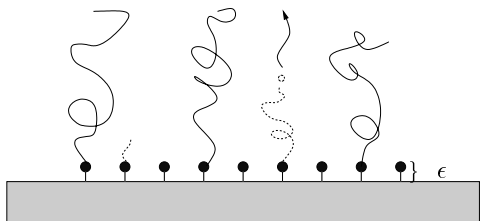}
   \caption[Schematic illustration of 
     polymers grafting from a solid surface.]{\label{fig:grafting}Schematic (cross-section) illustration of 
     polymers grafting from a solid surface. The black circles
     represent the initiator molecules.}
     \end{figure}

The best way to realize a brush network, is to grow chains directly at the
surface of the wall. This approach is called \emph{grafting} and is
shown in Figure \ref{fig:grafting}. 

Firstly the surface is covered
with a monolayer of \emph{initiator} molecules, from which long
molecules grow like a lawn\footnote{We assume, as in previous
  chapters, that the polymer chains are linked an infinitesimal
  distance $\epsilon$ away from the wall surface. In this case, the $\epsilon$
  can be interpreted as the \emph{size} of the initiator molecules, from which the chains
  are grafted.}. Secondly, the network is formed, by
crosslinking the chains. There are a multitude of chemical fabrication
techniques that can be used to effect crosslinking \cite{Young}. For example, if the molecules have photoreceptive groups,
the chains are crosslinked by irradiation with UV light \cite{Prucker,
imtek}.  The formation of a
brush network thus differs from the one
considered in Chapter \ref{chap:realnetwork}, because surface
attachment and crosslinking are not performed simultaneously, but in
subsequent steps. 

\subsection{Confinement formalism}
\label{sec:brushconf}
\begin{figure}[h!]
\centering 
\mbox{\subfigure[]{\epsfig{figure=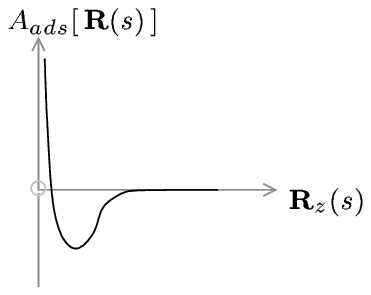,width=0.3\textwidth}}\qquad\qquad\qquad
  \subfigure[]{\epsfig{figure=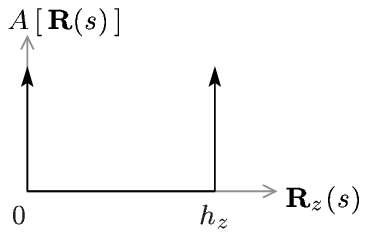,width=0.3\textwidth}}}
\caption[Spatial profiles of the potential associated with adsorption, and
confinement between two plates.]{Characteristic plots of the spatial profiles of two \underline{different} situations: (a) A potential well corresponding to a hard,
  adsorbing wall; (b) An infinite square well potential, for
  confinement in the $\hat{z}$ direction between two
  confining plates of width $h_{z}$.}
\label{fig:wells}
\end{figure}

Prior to network formation, we have
two brushes, that is, a
dense system of $\Nw$ independent chains, already attached to the
wall surfaces. The planar surfaces are brought within a distance
$h_{z}$ of each other, so as to confine the grafted polymers in the $\hat{z}$
direction perpendicular to the surface. Mathematically, the
confinement is represented by an infinite square well potential $A$, defined
in \eqref{eq:potA} and depicted in Figure \ref{fig:wells}(b). 

The partition function for a collection of $\Nw$
confined, grafted chains is given by:
\begin{eqnarray}
 \label{eq:zphantomchainsconfinedabrush}
        \mathcal{Z}\,\left(\{\bR_{i},\bR_{i}';L\}\right)&=&
        \mathcal{N}\,\left\{\,\prod_{i=1}^{\Nw}\,\int_{\bR_{i}(0)=\bR_{i}' }^{ \bR_{i}(L)=\bR_{i} } [
        \mathcal{D}\bR_{i}(s)\, ]_{\text{walls}}\right\}\, e^{- \frac{3}{2\ell}\,\sum_{i=1}^{\Nw}
        \int_{0}^{L}{\big(\frac{\partial\bR_{i}(s)}{\partial s}\big)}^{2}\,ds} \\
       \label{eq:zphantomchainsconfinedbbrush}
       &=&\mathcal{N}\,\prod_{i=1}^{\Nw}\,
       \int\,[\mathcal{D}\bR_{i}(s)\,]\, e^{-
         \int_{0}^{L}
         \left[\,\frac{3}{2\ell}\,\dot{\bR}_{i}^{2}(s)-A\,(\bR_{i}(s))\,\right]\,ds} \,,
    \end{eqnarray}
where $\mathcal{N}^{-1}$ refers to the statistical weight of $\Nw$ free
polymer chains, each starting at $\bR_{i}'$ (on the wall) and ending at
$\bR_{i}$ in $L/ \ell$ steps. 
\subsection{Crosslink formalism}
\label{sec:brushlink}
\begin{figure}[h]
   \centering
   \includegraphics[width=0.6\textwidth]{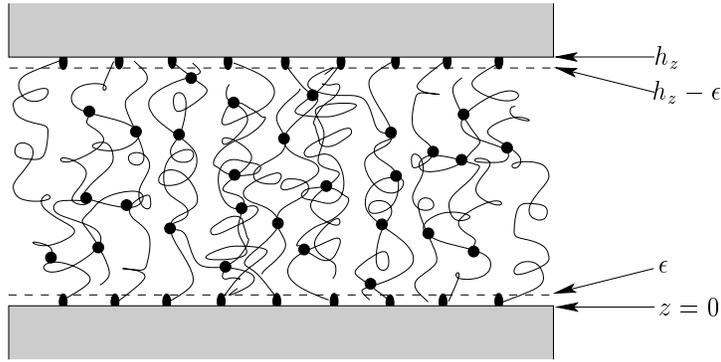}
   \caption[A brush network confined between two parallel planar surfaces.]{An illustrated, simplified view of a brush network. The
     walls are a distance, $h_{z}$, apart and the long polymer chains
     are fixed at $\epsilon$ from the surface. }\label{fig:brushnetworkwithtext}
 \end{figure}

After confining the polymer brushes, the chains should be
crosslinked to each other, in order to constitute a surface-attached
network, Figure \ref{fig:brushnetworkwithtext}. Let the set of crosslink
locations on the chains be denoted by  $\{\mathcal{S}\}$. These links
are formed randomly, such that the
crosslink positions  $\{\mathcal{S}_{a}\}$ of gel sample $a$ will differ
from $\{\mathcal{S}_{b}\}$  of sample $b$.
However, once the links are formed, they are assumed
to be permanent within a specific sample. 

A cross linkage that joins points $s_{i}^{*}$ and $s_{j}^{*}$ on
chains labeled $i$ and $j$ respectively, is
described mathematically by a Dirac delta constraint,
$\delta\,\big[\bR_{i}(s_{i}^{*})-\bR_{j}(s_{j}^{*})\big]$. A
\emph{product} of $\Nc$ $\delta$--constraints stipulates the
topology of a network having $\Nc$ crosslinks\footnote{
A more suitable notation for $s_{i}^{*}$ mentioned above, would be
$s_{ij}^{k}$, where $i$ and $j$ are the chains involved in the linkage, and $k$ the
degeneracy of the pair.  For example: A small network, consisting of three
chains, could have the following incidence set
\mbox{$\{\mathcal{S}\}=\{(s_{11}^{1},s_{11}^{2}),
(s_{12}^{1},s_{21}^{1}),(s_{12}^{2},s_{21}^{2}),(s_{23}^{1},s_{32}^{1}),(s_{13}^{1},s_{31}^{1})\,\}$.}
\begin{flushleft}
   \includegraphics[width=0.8\textwidth]{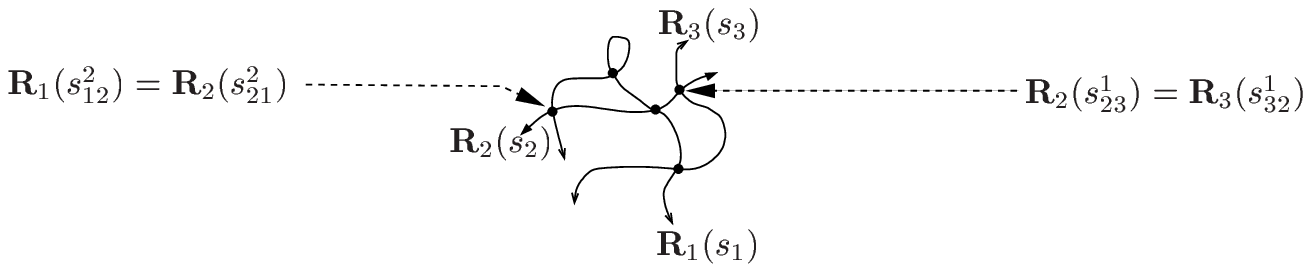}
\end{flushleft}
}. Since the set  $\{\mathcal{S}\}$ of crosslinkages is \emph{unknown} prior to
network formation, we assume that the linkage positions are
distributed according to a probability distribution
$\mathcal{P}(\mathcal{S})$. Let the distribution of the crosslinkages in the
resultant network, be given by the probability of the crosslinkages an
\emph{instant} before linking, that is, 
\begin{eqnarray}
        \mathcal{P}\,(\{\mathcal{S}\}) &\propto&
        \int\, [\,\prod_{i=1}^{\Nw}\,
        \mathcal{D}\bR_{i}(s)\, ]\quad e^{-\sum_{i=1}^{\Nw}\,\int_{0}^{L}
         \left[\,\frac{3}{2\ell}\,\dot{\bR}_{i}^{2}(s)-A\,(\bR_{i})\,\right]\,ds} 
        \nonumber \\ 
      & & {}\times
      \underbrace{\prod_{\{\mathcal{S}\}}\,
        \delta[\,\bR_{i}(s_{i}^{*})-\bR_{j}(s_{j}^{*})]\Big]}_{\text{Bulk}}\nonumber \\& & {
       }\times \,\prod_{i=1}^{\Nw/2}
          \underbrace{\delta[\,\bR_{i}(s=0)-\mathbf{\eta}(x_{i},y_{i},\epsilon)]\,}_{\,\text{Wall at }z=\epsilon}
          \,\,\underbrace{
          \delta[\,\bR_{i}(s=0)-\mathbf{\eta}(x_{i}',y_{i}',h_{z}-\epsilon)]\,}_{\text{Wall at }z=h_{z}-\epsilon} \,,\label{eq:zprebrushnetworka}
\end{eqnarray}
where the $\eta$'s are wall-link position vectors [see Section \ref{sec:instantcross}].
 This simplified \emph{history-dependent} situation, first proposed by
Edwards \cite{Chompff}, is equivalent to a system where the future
crosslinks are just touching prior to an instantaneous
linking\footnote{In Chapter \ref{chap:realnetwork} the \emph{choice} of
  probability distribution was done more comprehensively, but also
  amounts to an instantaneous ``freezing'' of links.}.
\subsection{Replica formalism}
\label{repform}
Since the set $\{\mathcal{S}\}$ is fixed for a specific sample, the set
and its linking history
will remain unaltered if the specimen is deformed in some way. Let
$F(\{\mathcal{S}\})$ be the free energy of a particular sample that has
been subject to strain $\mathbf{\Lambda}$ (Section \ref{sec:stitchelas}), after network formation.
The effective free energy of elasticity, $\mathcal{F}(\mathbf{\Lambda})$, is obtained by taking the
quenched average of the sample free energy over all possible
realizations of the arclength locations $\{\mathcal{S}\}$,
\begin{eqnarray}
    \label{eq:quenchedaveb}
    \mathcal{F}(\mathbf{\Lambda})=\int\mathcal{P}(\{\mathcal{S}\})\,F(\{\mathcal{S}\})\,d\mathcal{S}_{i}=
  \int\mathcal{P}(\{\mathcal{S}\})\,\ln\,\mathcal{Z}(\{\mathcal{S}\})\,d\mathcal{S}_{i}\,.
  \end{eqnarray}
The averaging of the logarithm in \eqref{eq:quenchedaveb} is facilitated
by using the mathematical identity \eqref{eq:replicaidentitya}, $Z^{n}
= 1 + n \ln\,Z + \ldots$, and replicating the system $n$ times.
At this point, it is possible to incorporate the probability distribution
$P(\{\mathcal{S}\})$ of $\{\mathcal{S}\}$, when the chains are touching
\emph{before} strain, as a \emph{zeroth} replica. The effective free
energy can then be identified as the coefficient of $n$ in the
following generalized partition function \cite{DeamEdwards}:
\begin{eqnarray}
    \label{eq:coeff}
    e^{-F\,(n)/\bolz T}
  =\int\mathcal{Z}^{(0)}(\{\mathcal{S}\})\,\left[\,\mathcal{Z}(\{\mathcal{S}\})\,\right]^{n}\,\exp\left(F_{0}/\bolz T\right)
\equiv \mathcal{Z}\,(n)\,.
  \end{eqnarray}
The factor\footnote{The free energy $F_{0}$ coincides with $F'$ of a
  slipping link system in
  \eqref{eq:zprenetworka}, apart from
  different naming of constants and wall-link integrations.}, $e^{F_{0}/\bolz T}$ in \eqref{eq:coeff}, is the normalization factor of the
probability $P(\mathcal{S})$, and the statistical weight of a system
where the arclength variables are not constrained to be a fixed \mbox{set
$\{\mathcal{S}\}$}.  Note that the  crosslink  points
$\{\mathcal{S}\}$ are common to \emph{all} the replicas, and so we can easily
average over them. 
 Since we are
interested in a resulting network with a sufficiently high crosslink
density, we assume that each long macromolecule is involved in 
numerous crosslinkages. The constraint --- averaged over all possible
positions of the links --- that picks out the $\Nc$
crosslinks, is given by
\begin{equation}
  \label{eq:diracconst}
  \mathit{Crosslink \,constraint} \equiv
  \left[\,\sum_{i=1}^{\Nw}\,\sum_{j=1}^{\Nw}\,\int_{0}^{L}ds\,\int_{0}^{L}ds'\,\,\prod_{\alpha=0}^{n}\,\delta
\left[\,\bR_{i}^{(\alpha)}(s)-\bR_{j}^{(\alpha)}(s') \,\right]\, \right]^{\Nc}\,.
\end{equation}
The wall-links are ``replicated'' in a similar manner
[see \eqref{eq:genzwithchempotbrush}]. 

We now proceed as in Chapter \ref{chap:realnetwork}, by calculating the generalized Gibbs formula of
$n+1$ replicas, 
\begin{eqnarray}
\label{eq:genzwithchempotbrush}
        \mathcal{Z}(n)&=&
        \mathcal{N}\int [\,\prod_{i=1}^{\Nw}
        \mathcal{D} \bR_{i}^{(0)}(s)\, ]\,\widetilde{\int}[\,\prod_{i=1}^{\Nw}\,\prod_{\alpha=1}^{n}\,
        \mathcal{D} \bR_{i}^{(\alpha)}(s)\,]\, e^{-\sum_{\alpha}\,\sum_{i=1}^{\Nw}\,\int_{0}^{L}
         \left[\,\frac{3}{2\ell}\,\dot{\bR}_{i}^{2}(s)-A\,(\bR_{i})\,\right]\,ds}
        \nonumber \\ 
      & & { }\times
      \Big[\,\sum_{i=1}^{\Nw}\,\sum_{j=1}^{\Nw}\,\int_{0}^{L}\,ds\int_{0}^{L}\,ds'\,\prod_{\alpha=0}^{n}\,\,
        \delta[\,\bR_{i}^{(\alpha)}(s)-\bR_{j}^{(\alpha)}(s')]\,\Big]^{\Nc}\nonumber \\& & {
       }\times \Big[\,\sum_{i=1}^{\Nw}\,\int dx\,dy\,\prod_{\alpha=0}^{n}\,\,
          \delta[\,\bR_{i}^{(\alpha)}(0)-\mathbf{\eta}^{(\alpha)}(x,y,0)]\Big]^{\half \Nw} \nonumber\\ & & {
       }\times \Big[\,\sum_{i=1}^{\Nw}\,\int dx\,dy\,\prod_{\alpha=0}^{n}\,\,
          \delta[\,\bR_{i}^{(\alpha)}(0)-\mathbf{\eta}^{(\alpha)}(x,y,h_{z})]\Big]^{\half \Nw}\,,
\end{eqnarray}
where the notation $\int$ refers to integration in an unstrained
\emph{zeroth} replica system, and $\widetilde{\int}$ to integration in
the remaining $n$ deformed replicas.  The bulk crosslink points have
the freedom to form anywhere along the length of the chains.
However, the \emph{wall} linking always takes place at $s=0$ for
each chain, but the precise location on the $xy$-plane surfaces is
unknown. Therefore, the wall-link constraint
in
\eqref{eq:genzwithchempotbrush} is written
\emph{without} an $s$-integration, in contrast with
the situation previously formulated in \eqref{eq:zprenetworka}. 

The crosslink constraints are exponentiated by introducing chemical
potentials \eqref{eq:complexint} $\mu_{\rm{c}}$ and $\mu_{\rm{w}}$ for the bulk and wall linkages
respectively. This leads to the most compact notation for the
generalized partition function,
\begin{eqnarray}\label{eq:compactbrush} \mathcal{Z}(n) &\equiv&
           \mathcal{N}\oint\,\frac{\Nc!\,d\mu_{\rm{c}}}{2\pi
          i}\,\oint\,\frac{\Nw!\,d\mu_{\rm{w}}}{2\pi i}
        \,\int\tilde{\int}\dots\tilde{\int}\,[\,\prod_{\alpha=0}^{n}\,\mathcal{D}\,\bR^{(\alpha)}(s)\,]\nonumber \\ & & {} 
        \times \,\exp\bigg\{- \mathcal{H}(n)/\bolz T-\Nc\log\mu_{\rm{c}}-\Nw\log\mu_{\rm{w}}\bigg\}\,,
\end{eqnarray} 
where $\mathcal{H}$ is a \emph{pseudo}-Hamiltonian for network
connectivity and confinement similar to the previous Hamiltonian
encountered in \eqref{eq:hamiltonianwithchempot}.
The next task at hand is to evaluate the path
integrals in \eqref{eq:compactbrush}. Thereafter, 
the free energy of deformation $\mathcal{F}(\mathbf{\Lambda})$ is recovered by
taking the 
the replica limit \eqref{eq:replicaidentitya}.
\pagebreak

\section{The Variational Approach}
\label{sec:varapproach}
Evaluation of the difficult path integrals in \eqref{eq:compactbrush}
will be tackled by the usual
Feynman variational procedure \cite{Feynman}, together with the
replica coordinate transformation \eqref{eq:DeamEdwardsTmatrix} of
Chapter \ref{chap:realnetwork}.  
\subsection{Trial potential for the brush network}
\label{sec:trialbrush}
The reason for investigating a \emph{brush} network in the first
place, is the prospect of an inhomogeneous, but integrable (trial)
potential, that can simulate the Dirac-delta constraints.  With the \emph{brush} network we capitalize on the fact that it is the
endpoints, at arc location $s=0$, that are permanently fixed to the
walls. We do this by introducing \emph{two} constant localization parameters. The
first, $q_0$, describes the localization within a certain region near
the wall attachment, and $q_1$ the bulk region, as
illustrated by Figure~\ref{fig:qline}:
\begin{figure}[h!]
   \centering
   \includegraphics[width=0.6\textwidth]{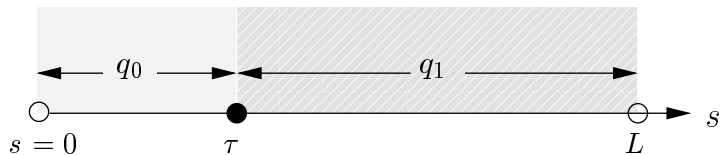}
   \caption[Depicting the domains of the variational parameters in \eqref{eq:twoqpot}.]{\label{fig:qline}The $s$-line depicts the domains of the
     variational parameters, $q_{0}$, $q_{1}$, and $\tau$ in \eqref{eq:twoqpot}.}
 \end{figure}

It is now possible to simulate the cross-link constraints by the following
trial harmonic potential [compare with \eqref{eq:qpot}],
 \begin{equation}
  \label{eq:twoqpot}
  Q = \frac{1}{6\ell}\,\sum_{\beta=1}^{n}\left(\,
    \int_{0}^{\tau}\,ds\,q_{0}^{2}\,\mathbf{X}^{(\beta)2}
    +
    \int_{\tau}^{L}\,ds\,q_{1}^{2}\,\mathbf{X}^{(\beta)2}\,\right),\quad \beta \in [1, 2,\ldots,n]
\end{equation} 
and thus model the system variationally by means of \emph{three} variational
parameters, $q_0$, $q_1$ \mbox{and $\tau$}. In \eqref{eq:twoqpot} the chain
conformations were written in terms of more convenient, transformed coordinates
$\{\mathbf{X}^{(0)},\mathbf{X}^{(1)},\mathbf{Y}^{(m)}\}\mid_{\{m=1\dots(n-1)\}}$,
as previously defined by the transformation matrix
$\mathbf{\mathsf{T}}$ in
\eqref{eq:DeamEdwardsTmatrix}. 

\pagebreak

In terms of the new set of coordinates, the generalized partition function can be summarized\footnote{The shortcut notation $[int]$ refers to $\oint\,\frac{\Nc!d\mu_{\rm{c}}}{2\pi
          i}\,\oint\,\frac{\Nw!d\mu_{\rm{w}}}{2\pi i}\,\int\,[\delta
        {\mathbf{X}}^{(0)}]\,[\delta {\mathbf{X}}^{(1)}] 
        [\,\prod_{m=1}^{n-1}
        \delta {\mathbf{Y}}^{(m)}]$.\\ The notation $\mathfrak{c}$
        represents the constant chemical potential terms associated with the
  contour integrals, that is,
     $\mathfrak{c}  \equiv -(\Nc+1)\,\log\mu_{\rm{c}}-(\Nw+1)\,\log\mu_{\rm{w}}$.} as follows:
\begin{eqnarray}
\label{eq:shortcutZabrush}
e^{-F(\mu_{\rm{w}}, \mu_{{\rm{c}}},n)/\bolz T} &=&\prod_{i=1}^{\Nw}\, \int
\,[{\rm{int}}]\,e^{-\mathbb{W}_{i}+\mu_{\text{c}}\,\mathbb{X}_{{\rm{c}} i}+\mu_{{\rm{w}}}\mathbb{X}_{{\rm{w}}
  i}+\mathbb{A}_{i}+\mathfrak{c}}
\\
&\equiv& \prod_{i=1}^{\Nw}\,
\int\,[{\rm{int}}]\,e^{-\mathcal{H}_{i}/\bolz T}
\end{eqnarray}

In the above and future succinct expressions, we employ the
following notation:
\begin{flushleft}
\framebox[\textwidth]{\parbox[][][c]{15.5cm}{
{\small
\begin{itemize}
\item The connectivity of the chains, each of length $L$,
  is denoted by the Wiener term $\mathbb{W}$.
  \begin{equation}
    \label{eq:wienerdefb}
    \mathbb{W}_{i} \equiv -\frac{3}{2 \ell} \int_{0}^{\mathcal{L}}\, \Big[
    {\mathbf{\dot{X}}}_{i}^{(0)\,2}(s) + {\mathbf{\dot{X}}}_{i}^{(1)\,2}(s)+ \sum_{m=1}^{n-1}\,|{\mathbf{\dot{Y}}}_{i}^{(m)}(s)|^{2}\Big]\,ds
  \end{equation}
\item The bulk crosslinks are denoted by $\mu_{\rm{c}}\mathbb{X}_{\rm{c}}$
 \begin{eqnarray}
    \label{eq:bulkdefb}
    \mu_{\rm{c}}\,\mathbb{X}_{{\rm{c}}i} & \equiv & \mu_{\rm{c}}\,\sum_{j=1}^{\Nw}\,\int_{0}^{\mathcal{L}}ds\int_{0}^{\mathcal{L}}ds'\,\,
       \delta\big(\,\mathbf{X}_{i}^{(0)}(s)-\mathbf{X}_{j}^{(0)}(s')\big) \,
       \delta\big(\,\mathbf{X}_{i}^{(1)}(s)-\mathbf{X}_{j}^{(1)}(s')\big)\,
      \nonumber \\ { }& & {  }\times\,\prod_{m=1}^{n-1}\,\delta\big(\,\mathbf{Y}_{i}^{(m)}(s)-\mathbf{Y}_{j}^{(m}(s')\big) 
  \end{eqnarray}
\item The wall cross-links are denoted by $\mu_{\rm{w}}\mathbb{X}_{\rm{w}}$. The
  variables $\bnu$ stand for the transformed vectors describing the
  wall-linking [see also Appendix \ref{chap:appenda}].
\begin{eqnarray}
    \label{eq:walllinksdefb}
    \mu_{\rm{w}}\mathbb{X}_{{\rm{w}} i} &\equiv& \mu_{\rm{w}}\,\int_{-\infty}^{+\infty}dx\,dy\,
       \bigg[\,
       \delta\big(\,\mathbf{X}_{i}^{(0)}(0)-\mathbf{\nu}^{(0)}(x,y,\epsilon)\big) \,
       \delta\big(\,\mathbf{X}_{i}^{(1)}(0)\big)\prod_{m=1}^{n-1}\,\delta\big(\,\mathbf{Y}_{i}^{(m)}(0)\big) \nonumber \\ 
       & & +\qquad
       \delta\big(\,\mathbf{X}_{i}^{(0)}(0)-\bnu^{(0)}(x,y,h_{z}-\epsilon) )\big)
       \,\delta\big(\,\mathbf{X}_{i}^{(1)}(0)\big)\,
       \prod_{m=1}^{n-1}\,\delta\big(\,\mathbf{Y}_{i}^{(m)}(0)\big)\, \bigg]
  \end{eqnarray}
\item The trial potential is $\mathbb{Q}$.
  \begin{equation}
    \label{eq:qdefb}
   \mathbb{Q}_{i} \equiv \frac{\ell}{6}\left(\,q_{0}^{2}\,
    \int_{0}^{\tau}\,ds\left[\,{\mathbf{X}}_{i}^{(1)2}+\sum_{m=1}^{n-1}|{\mathbf{Y}}_{i}{(m)}|^{2}\right]+q_{1}^{2}
    \int_{\tau}^{L}\,\left[\,{\mathbf{X}}_{i}^{(1)2}+\sum_{m=1}^{n-1}|\mathbf{Y}_{i}^{(m)}|^{2}\,\right]\,\right) 
  \end{equation}
\item The wall-constraints are denoted by $\mathbb{A}$.
  \begin{eqnarray}
    \label{eq:walldefb}
    \mathbb{A}_{i} &\equiv & \int_{0}^{\mathcal{L}}\,ds\, \bigg\{
    A\big(T_{z}^{00}X_{iz}^{(0)}+T_{z}^{10}X_{iz}^{(1)}\big)  \nonumber
    \\ & & +
    \sum_{\alpha=1}^{n}\,\tilde{A}\big(T_{z}^{0\alpha}X_{iz}^{(0)}+T_{z}^{1\alpha}X_{iz}^{(1)}+
    \sum_{m=1}^{n-1}\,T_{z}^{(m+1)\alpha}Y_{iz}^{(m)}\big)\,\bigg\}
  \end{eqnarray}
\end{itemize}
}}}
\end{flushleft}

If a second \emph{trial} potential $\mathbb{A}_{0}$
\eqref{eq:defA0} is
introduced\footnote{Note that $A_{0}$ is not \emph{really} a trial
  potential, since it has no variational parameters like
  $\mathbb{Q}$ has. However, since it only depends on the
  ``centre-of-mass'' coordinate $\mathbf{X}^{(0)}$, it is integrable.},
analogous to  Section \ref{sec:varcalc}, we obtain the following generalized partition function:
\begin{eqnarray}
  \label{eq:partitionG0brush}
  \mathcal{Z}(n, q_0, q_1,\tau)&=& \prod_{i=1}^{\Nw} {\langle
  \,e^{\mu_{\rm{c}}\mathbb{X}_{\rm{c}}+\mu_{\rm{w}}\mathbb{X}_{\rm{w}}+\mathbb{Q}+\mathbb{A}-\mathbb{A}_{0} }\,\rangle}_{\mathcal{G}}
\,\left(\int\,\mathcal{G}\right)\\
\label{eq:partitionG0geqbrush} &\geq & \prod_{i=1}^{\Nw}\, e^{{\langle
  \mu_{\rm{c}}\mathbb{X}_{\rm{c}}+\mu_{\rm{w}}\mathbb{X}_{\rm{w}}+\mathbb{Q}+\mathbb{A}-\mathbb{A}_{0}
  \rangle}_{\mathcal{G}}}\,\left(\int\,\mathcal{G}\right) \\\label{eq:partitionG0geqbrushc} &\equiv& 
  e^{-F_{\rm{var}}(n, q_0, q_1,\tau)/\bolz T}\,.
\end{eqnarray}
where the notation $\langle \ldots \rangle_{\mathcal{G}}$ indicates averaging with
respect to the Green's function $\mathcal{G}$.  We will
again make the Ansatz \eqref{eq:ansatzAA0},
$\langle\,\mathbb{A}-\mathbb{A}_{0}\,\rangle\,\simeq 0$. The
variational parameters $q_0$, $q_1$ and $\tau$ are recovered 
by minimizing the variational free energy, $F_{\rm{var}}$.

\subsection{Averaging}
\label{sec:averagingb}
In order to obtain the variational free energy $F_{\rm{var}}$ in
\eqref{eq:partitionG0geqbrushc},  the average
$\langle\,\mathbb{Q}+\mu_{\rm{c}}\mathbb{X}_{\rm{c}}+\mu_{\rm{w}}\mathbb{X}_{\rm{w}}\,\rangle$
is evaluated in terms of the Green function,
\begin{equation}
\label{eq:shortcutgreenbrush}
\mathcal{G}=\int \,[\delta \mathbf{X}^{(0)}][\delta
\mathbf{X}^{(1)}][\prod_{m=1}^{n-1}\,\delta \mathbf{Y}^{(m)}]\,
\underbrace{\,e^{-\mathbb{W}-\mathbb{Q}+\mathbb{A}_{0}}\,}_{\equiv e^{-\mathcal{H}_{\rm{var}}/\bolz T}}\,,
\end{equation}
which can be factored, depending on the replica index and Cartesian
coordinate, into well-known Green's functions listed in Section
\ref{sec:varcalc}. Since we are working with a model
having two parameters dependent on $s$, the averages should be  performed
carefully for each region, for example:
\begin{equation}
  \label{eq:avetwoparametersa}
  \langle \,f(s)\, \rangle = \left\{ \begin{array}{lll}
      \frac{\int\, dX_{s}\,dX_{\tau}\, dX_{L}\,
\mathcal{G}(X_{s},X_{0};s,q_{0})\,\mathit{f}(s)\,\mathcal{G}(X_{\tau},X_{s};\tau-s,q_{0})\,\mathcal{G}(X_{L},X_{\tau};L-\tau,q_{1})
}{\int\,dX_{\tau}\,dX_{L}\,
\mathcal{G}(X_{\tau}\,,X_{0}\,;\tau,\,q_{0})\,\mathcal{G}(X_{L}\,,X_{\tau}\,;L-\tau,\,q_{1})}
& \qquad &\textrm{if } 0 < s < \tau \\
& \qquad & \\ 
  \frac{\int\,dX_{\tau}\, dX_{s}\, dX_{L}\,
\mathcal{G}(X_{\tau},X_{0};\tau,q_{0})\,\mathcal{G}(X_{s},X_{\tau}\,
;s-\tau,q_{0})\,\mathit{f}(s)\,
\mathcal{G}(X_{L},X_{s} ;L-s,q_{1})}{\int\,dX_{\tau}dX_{L}\,
\mathcal{G}(X_{\tau},\,X_{0};\,\tau,q_{0})\,\,\mathcal{G}(X_{L},\,X_{\tau};\,L-\tau,\,q_{1})}
& \qquad &\textrm{if } \tau < s < L\,.
\end{array} \right.
\end{equation}
This section contains the final results of the averages, but the full calculation is relegated to  Appendix \ref{chap:appendc}.
\pagebreak
\begin{flushleft}
\underline{\textbf{The Bulk Cross-links   $\langle \mu_{\rm{c}}
    \mathbb{X}_{\rm{c}}\rangle$}}
\end{flushleft}
The weighted average of the crosslink contribution, defined in
\eqref{eq:bulkdefb} and calculated in Appendix \ref{sec:bulkappendb},
is given by the following lengthy expression:
\begin{eqnarray}
\label{eq:completeXcabrush}
\lefteqn{\therefore\,\langle\, \mu_{\rm{c}}\mathbb{X}_{\rm{c}}\,\rangle =
  \mu_{\rm{c}}\,\langle\mathbb{X}_{\text{c} x} \mathbb{X}_{\text{c} y}\mathbb{X}_{\text{c}
    z}\rangle }\nonumber\\ 
      &=&  \label{eq:completeXccbrush}
      \frac{3\mu_{\rm{c}}}{2}\Bigg[\left(\frac{q_{0}}{2\pi} \right)^{\frac{3n}{2}}\,
          \left\{\frac{2 \Nw\tau^{2}}{V \prod
              \sqrt{1+n\lambda_{i}^{2}}}+\frac{3\left(\ell_{\rm{c}}-\sqrt{\ell_{\rm{c}}^{2}+\tau^{2}}
           +\frac{\tau}{2}\,\ln\big[\frac{\sqrt{\ell_{\rm{c}}^{2}+\tau^{2}}+\tau}{\sqrt{\ell_{\rm{c}}^{2}+\tau^{2}}-\tau}\big]
        \right)}{\pi \ell h_{z}\sqrt{1+n\lambda_{z}^{2}}} \right\} \nonumber \\ & & {}  + \left(\frac{q_{1}}{2 \pi}
        \right)^{\frac{3n}{2}}\Bigg\{\frac{2 \Nw(L-\tau)^{2}}{V \prod
            \sqrt{1+n\lambda_{i}^{2}}}+\frac{3}{\pi
            \ell h_{z}\sqrt{1+n\lambda_{z}^{2}}}\bigg(\ell_{\rm{c}}-\sqrt{\ell_{\rm{c}}^{2}+(L-\tau)^{2}} \nonumber \\ 
          & &  +\frac{L}{2}\ln\big[\frac{\sqrt{\ell_{\rm{c}}^2+(L-\tau)^{2}}+L-\tau }
          {\sqrt{\ell_{\rm{c}}^2+(L-\tau)^{2}}-L+\tau}\big]+\frac{\tau}{2}\ln\big [\frac{\sqrt{\ell_{\rm{c}}^2+(L-\tau)^{2}}-L+\tau }
          {\sqrt{\ell_{\rm{c}}^2+(L-\tau)^{2}}+L-\tau}\big] \bigg) \Bigg\}
           \\& & { }  + \left(\frac{q_{0}\,q_{1}}{\pi(q_{0}+q_{1})}
        \right)^{\frac{3n}{2}}\Bigg\{\frac{4 \Nw \tau(L-\tau)}{V \prod
            \sqrt{1+n\lambda_{i}^{2}}}-\frac{3}{\pi
            \ell
            h_{z}\sqrt{1+n\lambda_{z}^{2}}}\bigg(\ell_{\rm{c}}+\sqrt{\ell_{\rm{c}}^{2}+L^{2}}-\sqrt{\ell_{\rm{c}}^{2}+\tau^{2}}
      \nonumber \\ 
          & & -\sqrt{\ell_{\rm{c}}^{2}+(L-\tau)^{2}} -L\ln \big[\frac{\sqrt{\ell_{\rm{c}}^2+(L-\tau)^{2}}-L+\tau }
          {\sqrt{\ell_{\rm{c}}^2+L^{2}}-L}\big]-\tau\ln \big[\frac{\sqrt{\ell_{\rm{c}}^2+\tau^{2}}-\tau }
          {\sqrt{\ell_{\rm{c}}^2+(L-\tau)^{2}}-L+\tau}\big] \bigg) \Bigg\} \Bigg]\nonumber\,,
\end{eqnarray}
where the constant $\ell_{\rm{c}}$ is the chain
\emph{cut-off} length\footnote{The continuous $s$ representation of a
macromolecule, is only acceptable for distances larger than $\ell_{\rm{c}}$. At
distances smaller than $\ell_{\rm{c}}\approx \ell$, the chain will appear
stiff due to actual monomer bonds.}.
Equation \eqref{eq:completeXccbrush} is valid when the polymer brush chains are long, and
when the relation, \mbox{$\sqrt{N\ell} \gg h_{z}$} holds. 
\begin{flushleft}
\underline{\textbf{The Wall Cross-links   $\langle \mu_{\rm{w}}
  \mathbb{X}_{\rm{w}}\rangle$}}
\end{flushleft}
The brush chains are attached an infinitesimal distance $\epsilon$
from the planar surface. It is
shown in Appendix \ref{chap:appenda} that any wall-link, after the
transformation $\mathbf{\mathsf{T}}$ \eqref{eq:DeamEdwardsTmatrix}, is
specified by the following two vectors,
\begin{equation}
  \label{eq:nudefeps}
  \bnu^{(0)}(x,y,\epsilon) \equiv \left(x \,\sqrt{1+n/\lambda},y \,\sqrt{1+n/\lambda},\epsilon\,\sqrt{1+n\lambda^{2}}\right) 
\end{equation}
\begin{equation}
  \label{eq:nudefheps}
  \bnu^{(0)}(x,y,h_{z}-\epsilon) \equiv \left(x \,\sqrt{1+n/\lambda},y \,\sqrt{1+n/\lambda},(h_{z}-\epsilon)\sqrt{1+n\lambda^{2}}\right)\,,
\end{equation}
for the original bottom and top surfaces respectively. The weighted
average of the wall-links, calculated in  Appendix \ref{sec:appwallb},
is
\begin{eqnarray}
    \label{eq:walllinksalles}
   \langle \mu_{\rm{w}}\mathbb{X}_{\rm{w}}\rangle &\equiv& \left\langle \mu_{\rm{w}}\,\int_{-\infty}^{+\infty}dx\,dy\,
       \bigg[\,
       \delta\big(\,\mathbf{X}_{i}^{(0)}(0)-\mathbf{\nu}^{(0)}(x,y,\epsilon)\big) \,
       \delta\big(\,\mathbf{X}_{i}^{(1)}(0)\big)\prod_{m=1}^{n-1}\,\delta\big(\,\mathbf{Y}_{i}^{(m)}(0)\big) \right.\nonumber \\ 
       & &\left. +{} 
       \delta\big(\,\mathbf{X}_{i}^{(0)}(0)-\bnu^{(0)}(x,y,h_{z}-\epsilon) )\big)
       \,\delta\big(\,\mathbf{X}_{i}^{(1)}(0)\big)\,
       \prod_{m=1}^{n-1}\,\delta\big(\,\mathbf{Y}_{i}^{(m)}(0)\big)\,
       \bigg]\right\rangle \\\label{eq:walllinksallesb}
     &=& \frac{4\,\mu_{\rm{w}}\,\pi^{2} \epsilon^{2}}{h_{z}^{3}}\,.
\end{eqnarray}
The fact that the wall links in a brush network only occur at $s=0$,
is portrayed in the above $q$-independent constant  average, in contrast with
the result for a ``normal'' surface-attached network \eqref{eq:completeXwd}.
\begin{flushleft}
\underline{\textbf{The harmonic trial potential $\langle
    \,\mathbb{Q}\,\rangle$}} 
\end{flushleft}
The average of the harmonic trial potential for inhomogeneous
localization is given by the following expression:
\begin{eqnarray}
  \label{eq:completeQbrush}
  \langle \,\mathbb{Q} \, \rangle &=& \left\langle
     \frac{\ell\,q_{0}^{2}}{6} \int_{0}^{\tau}ds
       \big(\,\mathbf{X}_{i}^{(1)\,2}+\sum_{m=1}^{n-1}|\mathbf{Y}_{i}^{(m)}|^{2}\,
   \big) +\frac{\ell\,q_{1}^{2}}{6} \int_{\tau}^{L}ds\big(\,\mathbf{X}_{i}^{(1)\,2}+\sum_{m=1}^{n-1}|\mathbf{Y}_{i}^{(m)}|^{2}\,
   \big) \right\rangle \\ \label{eq:completeQbrusha}
 &=& 
  \frac{n\,\ell}{8}\left(2\,[\,(L-\tau)\,q_{1}^{2}+\tau \,q_{0}^{2}\,]\sinh \frac{\ell q_{0}}{3}\tau\,
  \cosh\frac{\ell
    q_{1}}{3}(L-\tau) \right.\nonumber \\ & & {  }\left. +\quad Lq_{0}q_{1}\bigg\{\sinh\frac{\ell}{3}\left([L-\tau]q_{1}+\tau
  q_{0}\right)+\sinh\frac{\ell}{3}\left([L-\tau]q_{1}-\tau q_{0}\right) \bigg\}\right)
\nonumber \\ & & { } \times\quad \left(q_{0}\cosh \frac{\ell q_{0}}{3}\tau\,
  \cosh\frac{\ell q_{1}}{3}(L-\tau)+q_{1}\cosh \frac{\ell q_{1}}{3}(L-\tau)\,
  \cosh\frac{\ell q_{0}}{3}\tau\right)^{-1} \,,
\end{eqnarray}
When the hyperbolic functions in \eqref{eq:completeQbrusha} are written
as exponentials, the average simplifies to 
\begin{equation}
\langle\,\mathbb{Q}\,\rangle \simeq
\frac{n\,\ell}{4}\left[\left(q_{0}-q_{1}\right)\tau +q_{1}\,L\right]\,,
\end{equation} 
which is valid when the conditions, $q_{0}\ge\frac{3}{\ell\tau}$ and
$q_{1}\ge\frac{3}{\ell\,(L-\tau)}$, are satisfied. 

In the limit of homogeneous localization, that is $q_{0} \equiv q_{1}$,
the average \eqref{eq:completeQbrush} reduces to the familiar result
given in \eqref{eq:completeQfin}.
\section{The variational Free energy}
\label{sec:varfreenergybrush}
In order to obtain an approximate free energy of the original gel
system (Figure \ref{fig:brushnetworkwithtext}), we first have to find values of $q_{0}$, $q_{1}$ and $\tau$ that
minimize the variational free energy $F_{\rm{var}}$. At these stationary points, denoted by
$\{q_{0}^{*},q_{1}^{*},\tau^{*}\}$, the variational free energy
is expected be a good upper bound for the real free energy
of a \emph{replicated} system. In the last step the replica limit, which amounts to $n
\to 0$, should be
taken. However, for polymer network theories it is generally accepted
that the replica limit and the minimization procedure may be safely
interchanged \cite{Ball, DeamEdwards}. We shall approach the
problem in the manner of Deam and Edwards, by first taking the replica limit to
obtain the variational free energy $\widetilde{\mathcal{F}}$ of a single system, followed by
minimizing the resultant free energy. 
\pagebreak

The variational free energy \eqref{eq:partitionG0geqbrushc} in
$3\,(n+1)$ replica space is
obtained by assembling all the averages of Section \ref{sec:averagingb}:
\begin{eqnarray}
  \label{eq:partitionbrush}
  \mathcal{Z}_{\rm{var}}\,(n)&=& e^{-F_{\rm{var}}(n, q_0, q_1,\tau)/\bolz T} \\
 &=& \prod_{i=1}^{\Nw}\, e^{{\langle
  \mu_{\rm{c}}\mathbb{X}_{\rm{c}}+\mu_{\rm{w}}\mathbb{X}_{\rm{w}}+\mathbb{Q}+\mathbb{A}-\mathbb{A}_{0}
  \rangle}_{\mathcal{G}}}\,\left(\int\,\mathcal{G}\right) \\
\label{eq:partitionbrushb} &=&\prod_{i=1}^{\Nw}
\oint\,\frac{\Nc!\,d\mu_{\rm{c}}}{\mu_{\rm{c}}^{\Nc+1}}
\,e^{\,\mu_{\rm{c}}\, \langle\,\mathbb{X}_{\rm{c}}\,\rangle}\,\oint\,\frac{\Nw!\,d\mu_{\rm{w}}}{\mu_{\rm{w}}^{\Nw+1}}
\,e^{\,\mu_{\rm{w}}\, \langle\,\mathbb{X}_{\rm{w}}\,\rangle}\,e^{\frac{n\,\ell}{4}\left[\left(q_{0}-q_{1}\right)\tau +q_{1}\,L\right]} \nonumber \\ & & \times \left(
\frac{4\,\epsilon}{h_{z}}\,e^{-\frac{\ell \pi^{2}
    L}{6(1+n\lambda_{z}^{2})\,h_{z}^{2}}}\,
  \left(\frac{4\,q_{0}}{q_{0}+q_{1}}\right)^{\frac{3n}{2}}\,e^{-\frac{n\,\ell}{2}\left[\left(q_{0}-q_{1}\right)\tau +q_{1}\,L\right]}\right)
\\ \label{eq:partitionbrushc} &=& \prod_{i=1}^{\Nw}
 \langle\,\mathbb{X}_{\rm{c}}\,\rangle^{\Nc}
\langle\,\mathbb{X}_{\rm{w}}\,\rangle^{\Nw}
\,e^{-\frac{n\,\ell}{4}\left[\left(q_{0}-q_{1}\right)\tau +q_{1}\,L\right]}
\,e^{-\frac{\ell \pi^{2}
    L}{6(1+n\lambda_{z}^{2})\,h_{z}^{2}}}\,
  \left(\frac{4\,q_{0}}{q_{0}+q_{1}}\right)^{\frac{3\,n}{2}}\,\frac{4\,\epsilon}{h_{z}}\,.
\end{eqnarray}

In Section \ref{sec:averagingb} some of the averages were
simplified in the limit $\sqrt{N\ell} \gg h_{z}$. For a dense network
in a relatively
narrow confinement, it is expected that
the localization near the walls should not be appreciably different
from that further away in the ``bulk'' region. 
We portray this expectation\footnote{This substitution is also
  chosen to simplify the mimimization task at hand.} by letting
\begin{equation} 
  \label{eq:qnulens}
  q_{0} \,\equiv \,(1+\gamma)\,q_{1}\,,\qquad \gamma \in \realnumbers\,.
\end{equation}
With the above  substitution, the variational partition function becomes
\begin{eqnarray}
  \label{eq:partitionbrushgamma}
  \mathcal{Z}_{\rm{var}} &=& \prod_{i=1}^{\Nw}
 \left(\frac{q_{1}}{2\pi} \right)^{\frac{3n}{2}}\langle\,\mathbb{X}_{\rm{c}}\,\rangle^{\Nc}
\langle\,\mathbb{X}_{\rm{w}}\,\rangle^{\Nw}
\,e^{-\frac{n\,q_{1}\,\ell}{4}\left[\gamma\,\tau+L\right]}
\,e^{-\frac{\ell \pi^{2}
    L}{6(1+n\lambda_{z}^{2})\,h_{z}^{2}}}\,
  \left(\frac{4\,(1+\gamma)}{2+\gamma}\right)^{\frac{3\,n}{2}}\,\frac{4\,\epsilon}{h_{z}}.
\end{eqnarray}
\subsection{The Replica limit}
\label{sec:brushreplica}
The variational free energy of the \emph{original} three dimensional system,
say $\widetilde{\mathcal{F}}$, can be
identified as the coefficient of $n$ in \eqref{eq:coeff}, by 
employing the
following procedure \cite{DeamEdwards},
\begin{eqnarray}
  \label{eq:partitionf}
  -\widetilde{\mathcal{F}}/ \bolz T \quad \equiv \quad
\frac{\left.{\partial\,\mathcal{Z}_{\rm{var}}/{\partial
   n}}\right|_{n=0}}{\mathcal{Z}_{\rm{var}}(n=0)}\,,
\end{eqnarray}
which results in the following variational free energy for the brush network:
\begin{eqnarray}
\lefteqn{\widetilde{\mathcal{F}}(q_{1},\gamma,\tau)/\bolz T  =
}\nonumber \\
& &  
  \frac{\ell \Nw}{4}\,\left( L + \gamma\,\tau \right)\,q_{1} - 
  \frac{\ell\,L\,{\lambda}^2\,\Nw\,{\pi }^2}{6\,h_{z}^{2}} - 
  \frac{3}{2}\,\Nw\,\ln \left(\frac{4\,\left( 1 + \gamma \right)} {2 + \gamma}\right) - 
  \frac{3}{2}\,\Nc\,\Nw\,\ln \left(\frac{q_{1}}{2\,\pi }\right) \nonumber
  \\ & & {} - \Nc\,\Nw
  \left[ \frac{3}{2}\,\ln \left(1 + \gamma\right)\,\left\{
      \frac{3}{\pi\ell\,h_{z}}\Big(\ell_{\rm{c}} - 
            \sqrt{\ell_{\rm{c}}^{2} + \tau^2} + 
            \frac{\tau}{2}\,\ln \left[\frac{\sqrt{\ell_{\rm{c}}^{2} +
                  \tau^2}+\tau}{\sqrt{\ell_{\rm{c}}^{2} +
                  \tau^2}-\tau}\right] \Big) + \frac{2 \Nw\tau^2}{V}\right\}
         \right. \nonumber \\ & & {} \left. + 
       \frac{3}{2}\,\ln \left[\frac{2\,\left( 1 + \gamma \right) }{2 + \gamma}\right]\,
          \left\{ \frac{4 \Nw\,\left( L - \tau \right) \,\tau}{V}- \frac{3}{\pi\ell\,h_{z}}\Big(\ell_{\rm{c}} + \sqrt{\ell_{\rm{c}}^{2} + L^2} - 
            \sqrt{\ell_{\rm{c}}^{2} + ( L - \tau )^2} 
            - \sqrt{\ell_{\rm{c}}^{2} + \tau^2} \nonumber \right. \right. \\
           & & {} \left. \left.- L\,\ln \left[\frac{\sqrt{\ell_{\rm{c}}^{2} + ( L - \tau) ^2} -L+ \tau}
                 {\sqrt{\ell_{\rm{c}}^{2} + L^2}-L}\right] - 
            \tau\,\ln \left[\frac{\sqrt{\ell_{\rm{c}}^{2} + \tau^2}-\tau}
                 {\sqrt{\ell_{\rm{c}}^{2} + ( L - \tau)^2} -L+\tau}\right]\Big) \right\} \right.
             \nonumber\\
             & & \left.  -
             \frac{\Nw L^2}{V}\sum_{i=x,y,z}\lambda_{i}^{2}-\frac{3\lambda_{z}^{2}}{2\pi\ell h_{z}}\left(\ell_{\rm{c}} - 
     \sqrt{{\ell_{\rm{c}}}^2 + L^2} + 
     \frac{L}{2}\,\ln \left[\frac{\sqrt{\ell_{\rm{c}}^{2} +
           L^2}+L}{\sqrt{\ell_{\rm{c}}^{2} +
           L^2}-L}\right]\right) \nonumber \right]
              \\\label{eq:lastfree} & & \times \left(\frac{2 \Nw L^2}{V} + \frac{3}{\pi\ell h_{z} }\left(\ell_{\rm{c}} - 
     \sqrt{{\ell_{\rm{c}}}^2 + L^2} + 
     \frac{L}{2}\,\ln \left[\frac{\sqrt{\ell_{\rm{c}}^{2} +
           L^2}+L}{\sqrt{\ell_{\rm{c}}^{2} +
           L^2}-L}\right]\right)\right)^{-1}\,.
\end{eqnarray}
At this point, all that remains to be done, is to minimize the above
expression with respect to the variational parameters $\{q_{1},\gamma,\tau\}$.
\subsection{The minimization problem}
\label{sec:minprob}
The formidable task of
finding the stationary points $\{q_{1}^{*},\gamma^{*},\tau^{*}\}$, where $\widetilde{\mathcal{F}}$ has a global
minimum, is equivalent to solving the following equations simultaneously,
\begin{eqnarray}
  \label{eq:solvevgla}
  \frac{\partial \widetilde{\mathcal{F}}(q_{1}^{*},\gamma^{*},\tau^{*})}{\partial q_{1}} &=& 0 \\
  \label{eq:solvevglb}
  \frac{\partial \widetilde{\mathcal{F}}(q_{1}^{*},\gamma^{*},\tau^{*})}{\partial \gamma} &=& 0 \\
  \label{eq:solvevglc}
  \frac{\partial
    \widetilde{\mathcal{F}}(q_{1}^{*},\gamma^{*},\tau^{*})}{\partial
    \tau} &=& 0 \,\quad \text{on region} \,\,\tau \in (\ell_{\rm{c}},\,L-\ell_{\rm{c}})\,,
\end{eqnarray}
together with the condition that the Hessian $\mathsf{H}$ \cite{Gradsteyn} of $\widetilde{\mathcal{F}}$,
\begin{equation}
  \label{eq:hessian}
 \textsf{H} =  
  \left|\begin{array}{ccc}%
  \frac{\partial^{2}\widetilde{\mathcal{F}}}{\partial q_{1}^{2}}
  &\frac{\partial^{2}\widetilde{\mathcal{F}}}{\partial
    q_{1}\partial \gamma}&
  \frac{\partial^{2}\widetilde{\mathcal{F}}}{\partial
    q_{1}\partial \tau} \\
 \frac{\partial^{2}\widetilde{\mathcal{F}}}{\partial \gamma \partial q_{1}}
  &\frac{\partial^{2}\widetilde{\mathcal{F}}}{
    \partial \gamma^{2}}&
  \frac{\partial^{2}\widetilde{\mathcal{F}}}{\partial
    \gamma \partial \tau}\\
 \frac{\partial^{2}\widetilde{\mathcal{F}}}{\partial \tau \partial q_{1}}
  &\frac{\partial^{2}\widetilde{\mathcal{F}}}{\partial \tau
    \partial \gamma}&
  \frac{\partial^{2}\widetilde{\mathcal{F}}}{\partial
    \tau^{2}}                           \end{array}\right|_{(q_{1},\gamma,\tau)=(q_{1}^{*},\gamma^{*},\tau^{*})}
\end{equation}
and any second derivative, for example
$\frac{\partial^{2}\widetilde{\mathcal{F}}}{\partial q_{1}^{2}}|_{(q_{1}^{*},\gamma^{*},\tau^{*})}$, are
both  positive at the stationary point. Solving the problem
analytically seems intractable due to the transcendental equation
\eqref{eq:solvevglc}. However, it is possible to find solutions,
$q_{1}^{*}$ and $\gamma^{*}$, for \eqref{eq:solvevgla} and
\eqref{eq:solvevglb}. The localization parameters are found to be
\begin{equation}
  \label{eq:locstuff}
  q_{1}^{*}=\frac{6\,\Nc}{\ell\,(L+\gamma^{*})}\qquad \text{and}\qquad  q_{0}^{*}=\frac{6\,\Nc\,(1+\gamma^{*})}{\ell\,(L+\gamma^{*})}\,,
\end{equation}
where $\gamma^{*}$ is still a function of $\tau$. The extremum
points of $\widetilde{\mathcal{F}}$ are calculated numerically. Next, the value
$\tau^{*}$, which minimizes the free energy $\widetilde{\mathcal{F}}$, is
found by plotting the free energy as a function of $\tau$, and
identifying $\tau^{*}$ as the value, on region $(\ell_{\rm{c}},\,L-\ell_{\rm{c}})$, where $\widetilde{\mathcal{F}}(\tau)$ has a
global minimum. 

\section{Analysis and Results}
\label{sec:analysis}
In this section we probe the physics of the system, without
beforehand obtaining
closed form solutions for the variational parameters
$\{q_{1},\tau,\gamma\}$ that satisfy  the minimization equations
\eqref{eq:solvevgla}---\eqref{eq:solvevglc}. 
\subsection{The homogeneous brush limit}
\label{sec:homobrushnetw}
In a homogeneous brush network the crosslinks impose a uniform
localization on the system. This situation is obtained, by letting $\tau=0$ or
$\gamma=0$ in \eqref{eq:locstuff} and \eqref{eq:lastfree} respectively. In this
limit we obtain the usual Deam and Edwards \cite{DeamEdwards}
localization parameters
\begin{equation}
  \label{eq:locstuffDE}
  q_{1}^{*}=\frac{6\,\Nc}{\ell\,L}=q_{0}^{*}\,,
\end{equation}
and the following free energy of deformation
\begin{eqnarray}
\mathcal{F}_{\text{uniform}}/\bolz T  &\ge&
  \frac{3 \Nc\,\Nw}{2}\,\left( 1 + \ln\left[\frac{3\Nc}{\pi\ell L}\right]\right) - 
  \frac{\pi^{2}\ell\,L\,\lambda_{z}^2\,\Nw}{6\,h_{z}^{2}} - 
  \frac{3}{2}\,\Nw\,\ln 2 \nonumber
  \\ & & {} +\frac{1}{2}\,\frac{
    \Nc\,\Nw\,\left[\sum_{i=x,y,z}\lambda_{i}^{2}+\frac{c}{\rho}\lambda_{z}^{2}\right]}{\left(\,1+c/\rho\,\right)}\\
 \label{eq:locstuffDEsc}
 &=&  \frac{\half
    \Nc\,\Nw\,\left[\sum_{i=x,y,z}\lambda_{i}^{2}+\frac{c}{\rho}\lambda_{z}^{2}\right]}{\left(\,1+c/\rho\,\right)}
  -\frac{\pi^{2}\ell\,L\,\lambda_{z}^2\,\Nw}{6\,h_{z}^{2}}+ \stackrel{\lambda-\text{independent}}{\scriptstyle\text{ terms}}\,,\\
\label{eq:locstuffpar}\!\!\!\text{with  } \rho &=& \frac{\Nw L}{V}\
\text{and } c \equiv  \frac{3}{2\pi\,\ell \,h_{z}\,L}\left(\ell_{\rm{c}}
  - {\scriptstyle \sqrt{{\ell_{\rm{c}}}^2 + L^2}}+\frac{L}{2}\,\ln \Big[{\scriptstyle \frac{\sqrt{\ell_{\rm{c}}^2 + L^2}+L}
          {\sqrt{\ell_{\rm{c}}^{2} + L^2}-L}}\Big] \right)\,.
\end{eqnarray}
In the last step \eqref{eq:locstuffDEsc} we show only the important $\lambda$-dependent terms.
In the limit of uniform localization, we obtain a
deformation free
energy 
closely resembling the result \eqref{eq:finalevryenergielambda} of the
previous chapter\footnote{Here, the total number of crosslinks in the
  sample is given by $\Nw \Nc$.}.
The free energy $\mathcal{F}_{\text{uniform}}$ has the
typical Deam and Edwards deformation term
$\half\,\Nc\sum\,\lambda_{i}^{2}$ and front factor $\sim (1+c/\rho)^{-1}$ [see \eqref{eq:DeamEdwardsfreeEnergy}], apart from a
different $c$-value which now also depends on $h_{z}$. The second term
in \eqref{eq:locstuffDEsc} represents the typical confinement of a
macromolecule between planar surfaces, and is common to the gel
described in Chapter \ref{chap:realnetwork}, as well as the stitch network
of Chapter \ref{chap:disorder}.

\pagebreak
\subsection{The single chain limit}
\label{sec:singlechainbrush}
\begin{figure}[h!]
\centering
\subfigure[]{\psfrag{F}[][l]{$\widetilde{\mathcal{F}}$}\psfrag{t}[][l]{$\,\,\tau$}
{\includegraphics[width=0.3\textwidth]{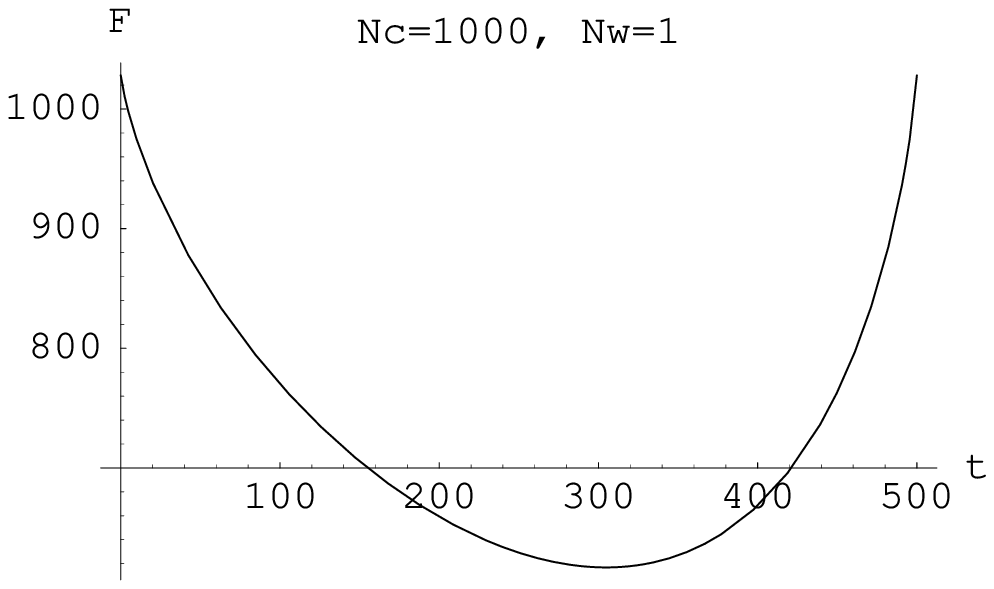}}}\qquad \quad\subfigure[]{\psfrag{q}[][l]{$q$}\psfrag{s}[][l]{$\,\,s$}
{\includegraphics[width=0.3\textwidth]{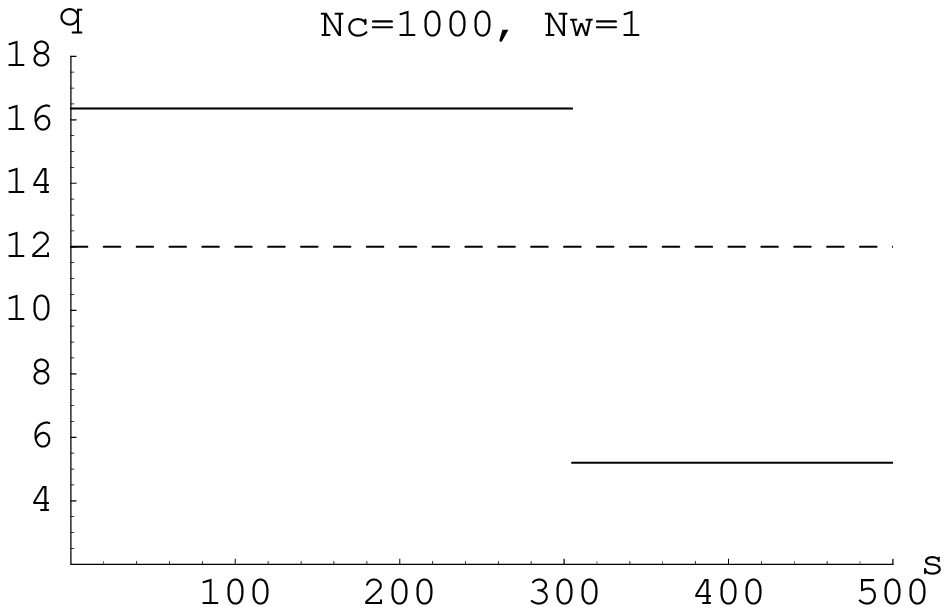}}}\caption[Illustrating
the case of only one wall-link, $\Nw\equiv 1$.]{\label{fig:Nwone}Illustrating
the case of only one wall-link, $\Nw\equiv 1$. 
(a) The plot of the variational free energy of the brush network
$\widetilde{\mathcal{F}}$ versus $\tau$ exhibits a global
minimum at $\tau^{*}\simeq 305$. (b) Plot of the localization
parameters, $q_{0}$ and $q_{1}$ \eqref{eq:locstuff}, versus $s$ for the
  brush network (solid line), in comparison with a uniformly localized bulk
  network (dashed line), \eqref{eq:locstuffDE}. Constant plot
  parameters: $L=500$, $h_{z}=20$, $\ell=\ell_{\rm{c}}=1$,
  $\rho=\frac{\Nw L}{V}=0.4$, $\lambda=1$.}
\end{figure}

In Section \ref{sec:formbrushnetw} we mentioned that a polymer brush is
(by  definition) a very dense array of polymer chains, that is
$\Nw\gg 1$. However, if the crosslink density in the bulk is
sufficiently high, we find a  global minima for the free energy
$\widetilde{\mathcal{F}}$, even if we only have a brush network
synthesized from one chain, as in Figure \ref{fig:Nwone}(a). 
When $\Nw=1$, we have a single chain  network with
one wall-link. The effect of the wall linkages relative to the bulk
linkages, is then expected to be
negligible, so that we are essentially dealing with a confined
\emph{bulk} network.  A bulk network can be thought of as
one giant chain, having no wall-links.
 
From Figure
\ref{fig:Nwone}(b) we can infer that  the localization of a  bulk
network is expected to be the \emph{average} of the inhomogeneous
brush network localization contributions, $q_{0}$ and $q_{1}$. 
\pagebreak
\subsection{Localization of the polymer brush network}
\label{locpolbrushnetw}
\begin{figure}[h!]
\centering
\subfigure[]{\psfrag{F}[][l]{$\widetilde{\mathcal{F}}$}\psfrag{t}[][l]{$\,\tau$}\psfrag{Nc=1000,Nw=2000}[][l]{$\Nc=2000$,$\,\Nw=1000$}
{\includegraphics[width=0.3\textwidth]{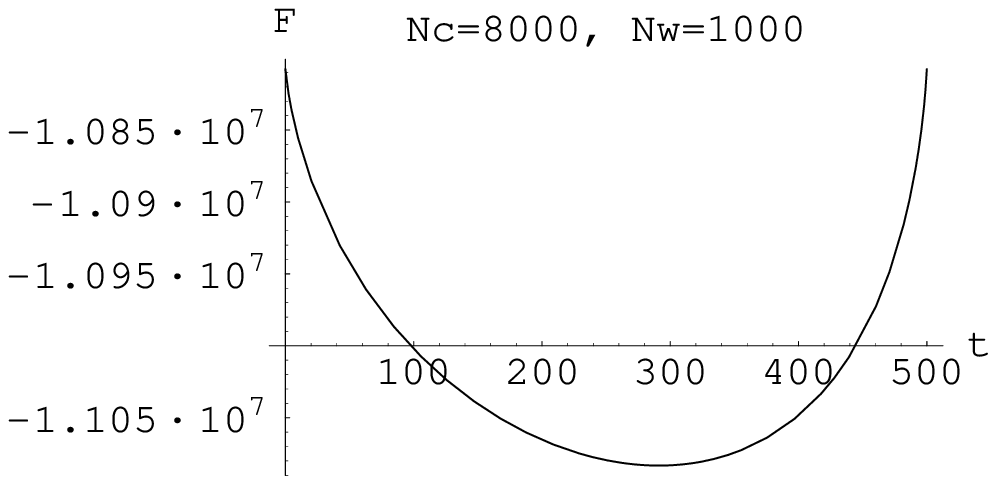}}}\quad\subfigure[]{\psfrag{gamma}[][r]{$\qquad{\scriptstyle \gamma^{*}=0.40}$}\psfrag{gam}[][l]{$\,\gamma$}\psfrag{t}[][l]{$\,\,\tau$}
{\includegraphics[width=0.3\textwidth]{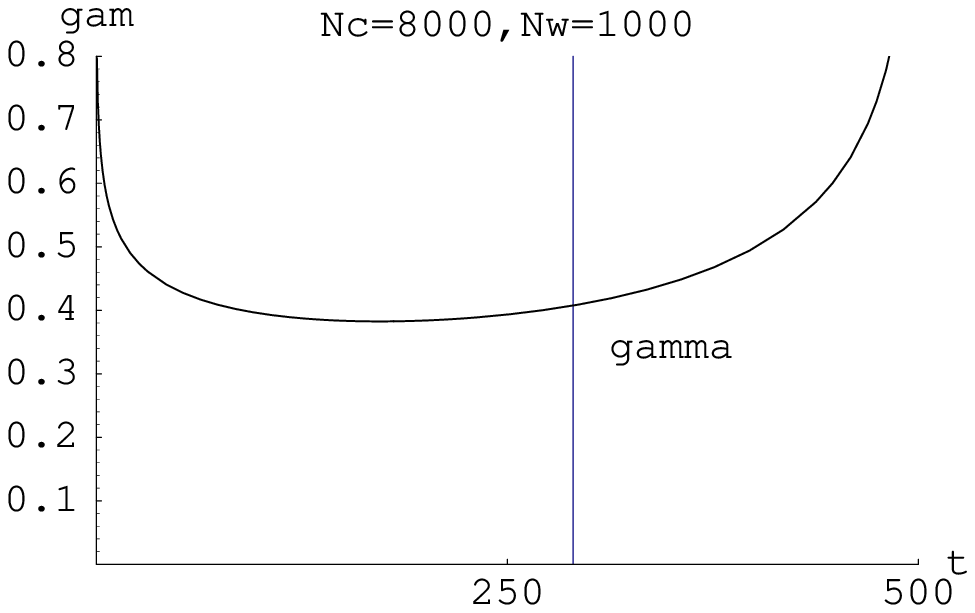}}}\quad\subfigure[]{\psfrag{q}[][l]{$\,q$}\psfrag{s}[][l]{$\,s$}
{\includegraphics[width=0.3\textwidth]{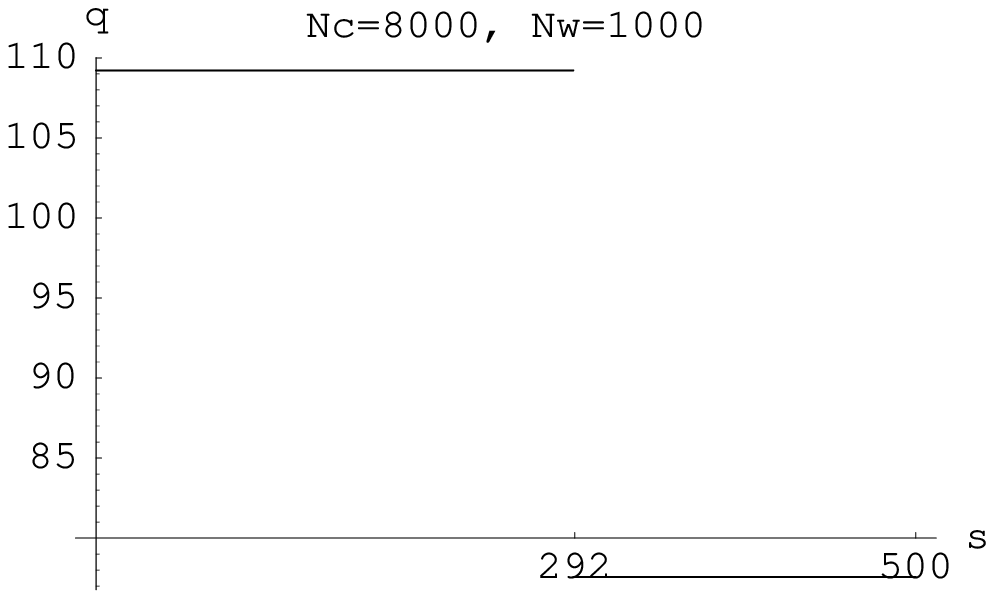}}}\\
\subfigure[]{\psfrag{F}[][l]{$\widetilde{\mathcal{F}}$}\psfrag{t}[][l]{$\,\tau$}
{\includegraphics[width=0.3\textwidth]{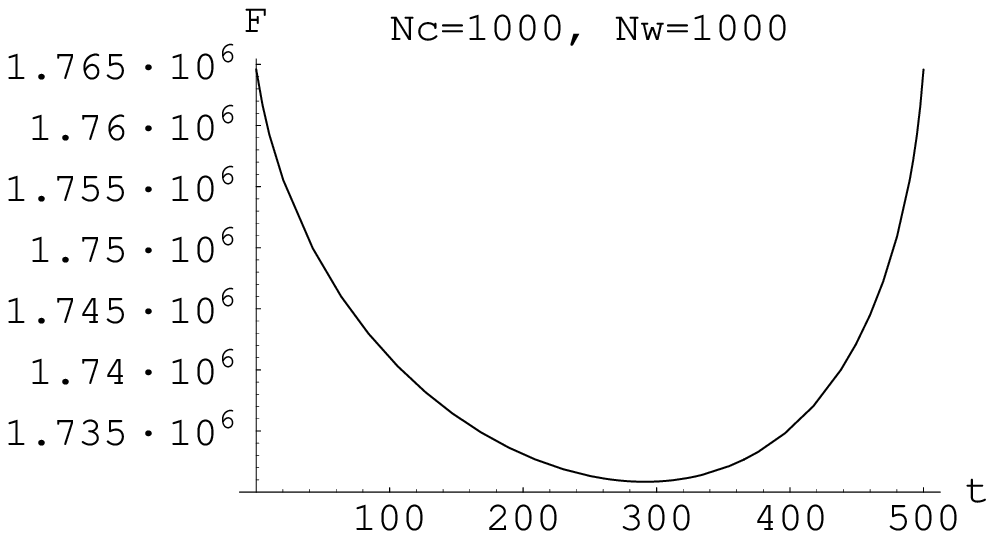}}}\quad\subfigure[]
{\psfrag{gam}[][l]{$\gamma$}\psfrag{gamma}[][r]{$\qquad{\scriptstyle \gamma^{*}=0.41}$}\psfrag{t}[][l]{$\,\,\tau$}
{\includegraphics[width=0.3\textwidth]{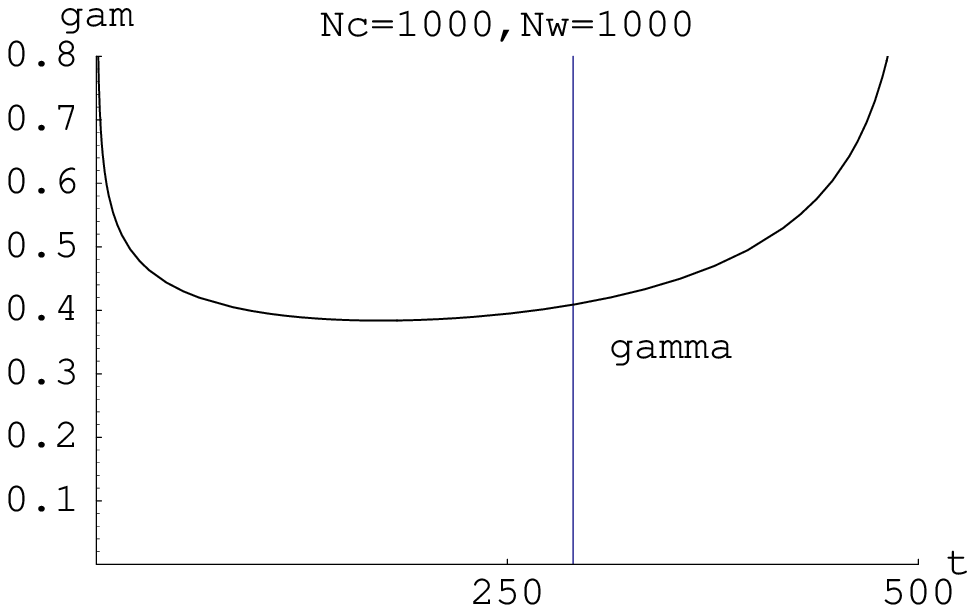}}}\quad\subfigure[]{\psfrag{q}[][l]{$\,q$}\psfrag{s}[][l]{$\,s$}
{\includegraphics[width=0.3\textwidth]{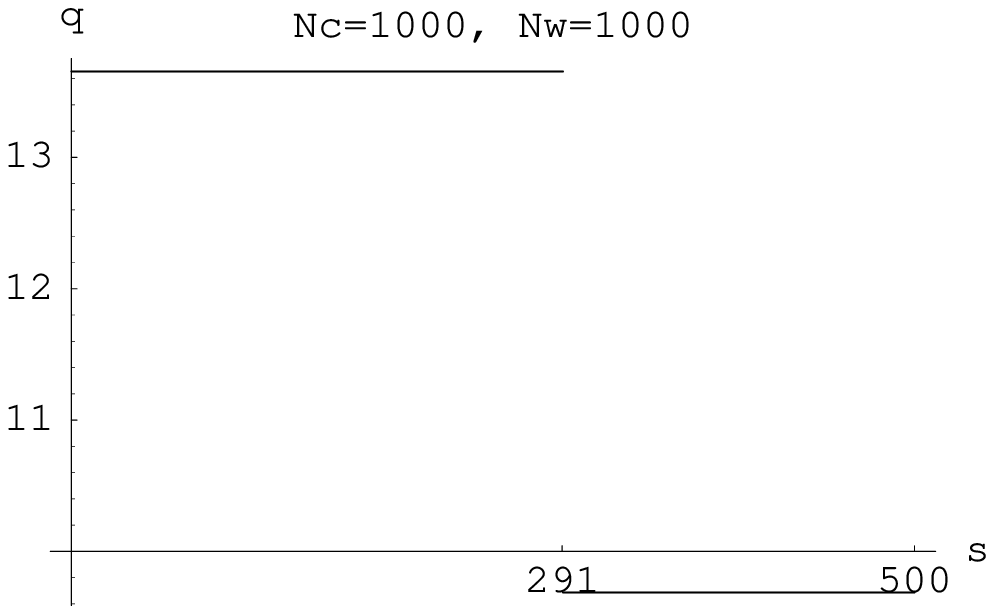}}}\\\subfigure[]{\psfrag{F}[][l]{$\widetilde{\mathcal{F}}$}\psfrag{t}[][l]{$\,\tau$}
{\includegraphics[width=0.3\textwidth]{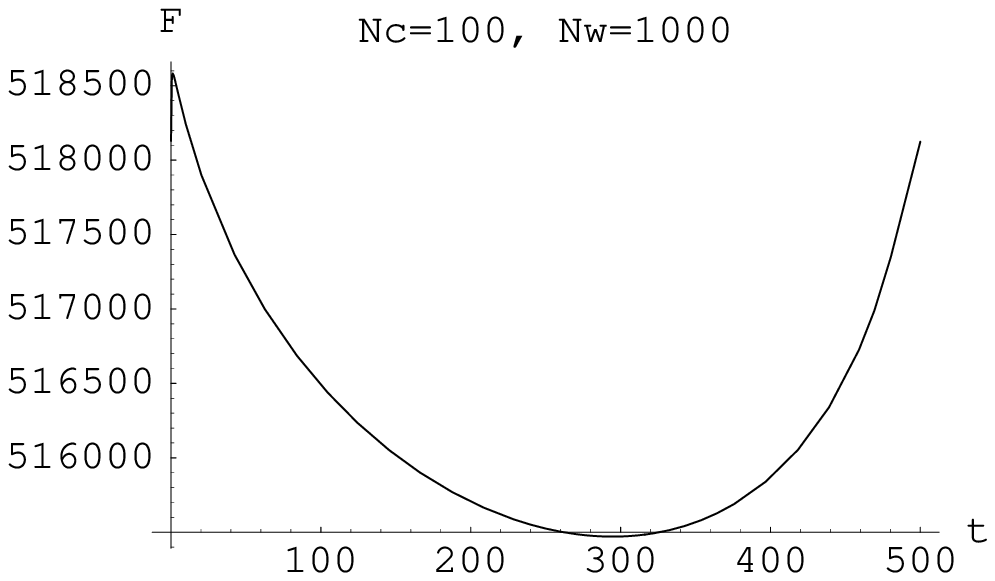}}}\quad\subfigure[]{\psfrag{gamma}[][r]{$\qquad{\scriptstyle \gamma^{*}=0.42}$}\psfrag{gam}[][l]{$\,\gamma$}\psfrag{t}[][l]{$\,\,\tau$}
{\includegraphics[width=0.3\textwidth]{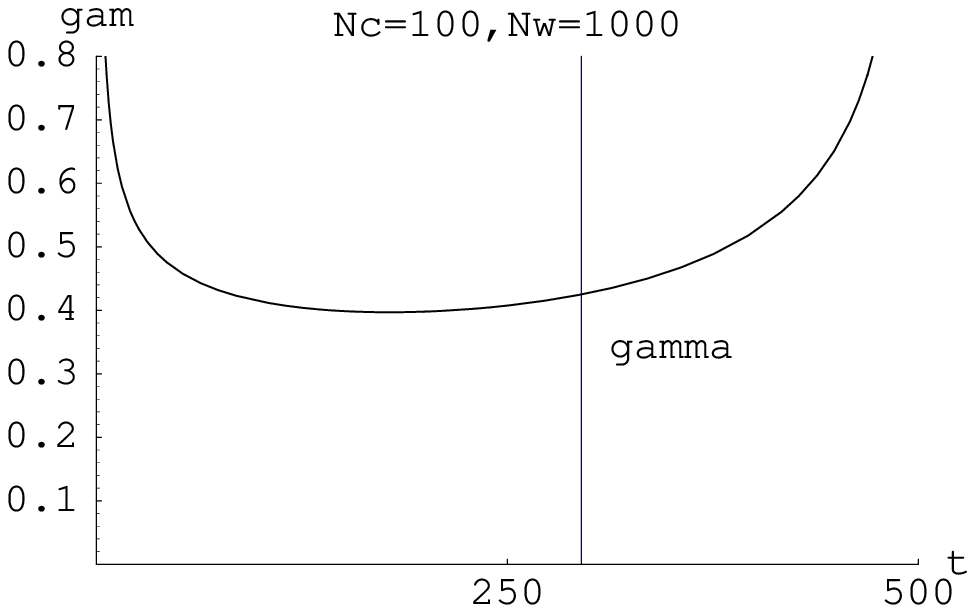}}}\quad\subfigure[]{\psfrag{q}[][l]{$\,q$}\psfrag{s}[][l]{$\,s$}
{\includegraphics[width=0.3\textwidth]{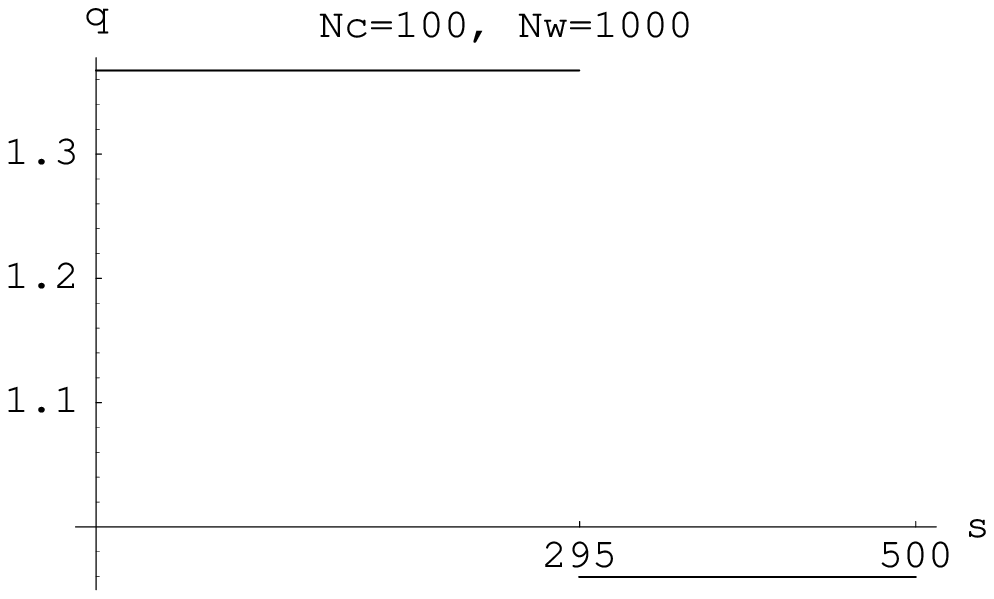}}}\caption[Investigating
the influence of the crosslink density on the localization of the polymer.]{\label{fig:localfigfgamq}Investigating
the influence of the crosslink density ${\Nc}$ on the localization of
the polymer. The leftmost column of graphs depicts the variational free
energy as a function of $\tau$. Global minima were found at
$\tau^{*}\simeq 291$ in (a), $\tau^{*}\simeq 292$ in (d) and
$\tau^{*}\simeq 295$ in
(g). These values are represented by vertical lines in the second
column. The value $\gamma^{*}$ is given by the intersection
of the vertical line with the $\gamma(\tau)$-curve. The rightmost
column illustrates that $q_0>q_1$. Also, the localization of the
polymer decreases as the number of bulk-links decreases
\eqref{eq:locstuff}. Plots were rendered with the following kept constant: $L=500$,
$h_{z}=20$, $\lambda=1$, $\ell=\ell_{\rm{c}}=1$, $\rho=0.4$.}
\end{figure}

Next, we examine the graphs in Figure \ref{fig:localfigfgamq} to see how the linking density $\Nc$ 
influences the localization of the polymer.  Each row in Figure \ref{fig:localfigfgamq}
corresponds to a different value of $\Nc$. The first observation that can be made, is that the chains in the ``surface''
region ($q_{0}$) are localized to a greater extent than the chains
away from the
walls ($q_{1}$). This follows from the fact that $\gamma^{*}$ is
positive in the middle column of Figure \ref{fig:localfigfgamq}, and so we have that $q_{1}<q_{0}$ from
\eqref{eq:locstuff}. The measure of the fluctuations of the crosslinks from their
affine positions, given by the
$q$-values, is greater for a chain that is lightly crosslinked
($\Nc< \Nw$) as shown in Figure \ref{fig:localfigfgamq}, than for a network where
$\Nc\geq \Nw$. Since $\gamma^{*}<1$, the  $q$-values
\eqref{eq:locstuff} are primarily
influenced by the number of crosslinks $\Nc$.

\pagebreak
\subsection{Confinement and the free energy}
\label{sec:confiFreeEn}
When a macromolecule is placed in a geometric confinement, the total number
of its possible chain conformations is reduced. This leads to a decrease
in the entropy, or equivalently, to an increase of the free energy. 

\begin{figure}[h!]
\centering
\mbox{\subfigure[]{\psfrag{F}[][l]{$\widetilde{\mathcal{F}}$}\psfrag{t}[][l]{$\,\,\tau$}\psfrag{h}[][l]{$h_{z}$}
{\includegraphics[width=0.4\textwidth]{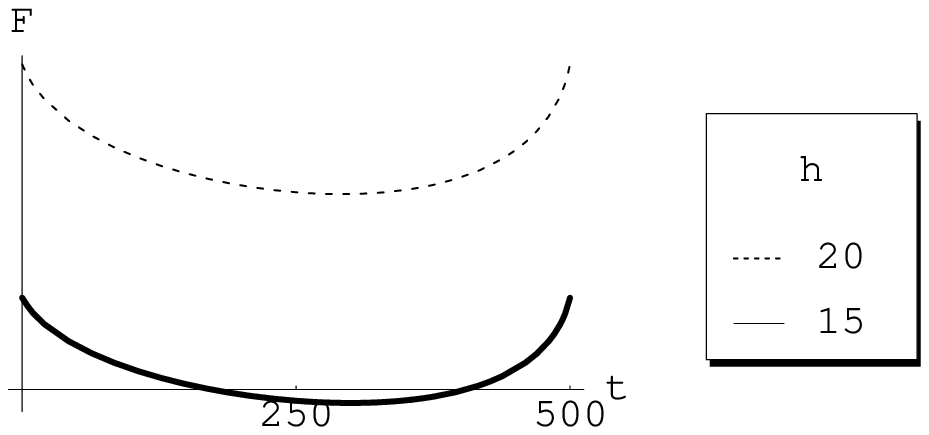}}}\qquad\subfigure[]{\psfrag{F}[][l]{$\widetilde{\mathcal{F}}$}\psfrag{t}[][l]{$\,\,\tau$}
\psfrag{r}[][l]{$\rho$}
{\includegraphics[width=0.4\textwidth]{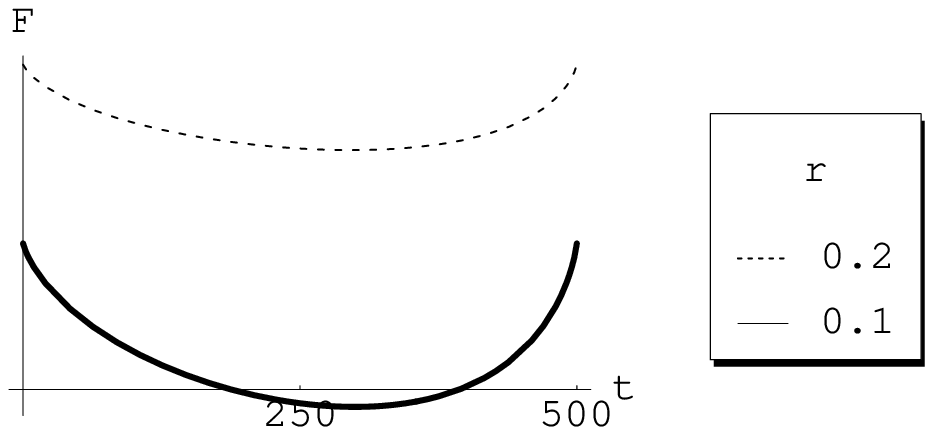}}}}\caption[Investigating
the influence of (a) the wall confinement $h_{z}$ and (b) the polymer density $\rho$.]{\label{fig:volum}Illustrations
of
the variational free energy $\widetilde{\mathcal{F}}$ as a function of
$\tau$, for  (a) different values of $h_{z}$ at constant volume $V$, and (b) different values
of polymer chain density $\rho=\frac{\Nw\,L}{V}$ for the
the gel. Constant plot parameters were: $L=500$,
$\Nc=\Nw=1000$, $\ell=\ell_{\rm{c}}=1$ and $\lambda=1$.}
\end{figure}

However, the case of a confined, surface-attached network is more
involved, since the 
crosslink constraints also influence the network chain conformations. We
find that the entropy of the brush network \emph{increases} as the
distance between the planar surfaces gets narrower, with the volume of
the sample kept \emph{constant}. This phenomenon is
illustrated in
Figure \ref{fig:volum}(a), where we plot the variational free energy
$\widetilde{\mathcal{F}}(\tau)$ for two values of the wall separation
$h_{z}$.

 At a sufficiently high cross link density and wall
separation, the chains in a brush network may be thought of as highly
stretched strings. When $h_{z}$ decreases, the tension in the chains will
decrease, thereby allowing more possible conformations to the network
chains, and an associated increase in the entropy. 

In this model the network is not confined in the $xy$-plane. This
means that the average size of the chain is smaller than the $x$ and $y$
dimensions of the volume containing the gel sample. An increase in the
box volume is equivalent to a decrease in the polymer chain
density. As a result the free energy (entropy) of the network
decreases (increases), when the polymer density decreases, as shown in Figure \ref{fig:volum}(b). The difference in
the two graphs, lies in the fact that the chains are only confined and
fixed in the $z$-dimension.

\pagebreak
\subsection{Elasticity}
\label{sec:elasticity}
It is possible to calculate the stress-strain relationship for the
polymer brush network, even though a closed-form of the free energy cannot
be obtained. This is so, because the localization parameters $q_0$ and
$q_1$ and the parameter $\tau$ do not depend on the affine
deformation of the system, that is, they are
$\lambda$-\emph{independent}. By differentiating the variational free
energy in \eqref{eq:lastfree} with respect to $\lambda$, one obtains the following stress-strain
relationship:
\begin{eqnarray}
  \label{eq:stressbrushnetw}
  f\,(\lambda) & = &
  \frac{\bolz T}{V}\,
  \left[\frac{\Nc\,\Nw}{\left(\,1+c/\rho\,\right)}\,\left\{\lambda-1/\lambda^{2}+\frac{c}{\rho}\lambda\right\}
  -\frac{\pi^{2}\ell\,L\,\Nw}{3\,h_{z}^{2}}\,\lambda\right]\,,
\end{eqnarray}
where the polymer density $\rho$ and the factor $c$ are defined in \eqref{eq:locstuffpar}.
 
The factor $\lambda - 1/\lambda^{2}$ gives the form of the 
characteristic stress-strain plot so typical of the
classical theory of rubber elasticity \cite{Treloar}.  However, the
deformation  factor is altered by the constant front
factor $g\equiv(1+ c/\rho)^{-1}$, analogous to the result
of the previous chapter \eqref{eq:stressrealconfinednetw}, and the Deam and Edwards
\cite{DeamEdwards} result for
an unconfined network. The above result is consistent with the
stress-strain relation obtained in  \eqref{eq:stressrealconfinednetw}, for a surface-attached network
where the wall-links and bulk-links are formed
simultaneously. In the latter network, the wall-links have the freedom
to form anywhere along the length of the polymer chains, in contrast
with a brush network. This difference is portrayed in the
$\Nw\lambda$ term in  \eqref{eq:stressrealconfinednetw}, which is
absent from the brush network stress-strain equation. 

As an example, we
list  numerical estimates of the front factor constants for the
Deam and Edwards model (unconfined network), and the confined
network models of the previous and present chapter:
\begin{table}[h]
  \begin{center}
      \begin{tabular}[]{|c|c|c|c|}
        \cline{2-4}
        \multicolumn{1}{c|}{ } & \multicolumn{1}{c|}{\textbf{Unconfined}}& 
        \multicolumn{2}{c|}{\textbf{Confined}}\\
        \cline{2-4}
        \multicolumn{1}{c|}{ } & Deam Edwards  & Real Confined Network & Brush Network \\
        \hline\hline
       $c$ & 0.17 & 0.31 & 0.14\\
        $\rho$  & 0.4 & 0.4  & 0.4  \\
        $\textsf{front factor}\,g$  & $0.70$ & 0.56 & $0.74$
        \\
        \hline
      \end{tabular}      
       \caption{\label{tab:table1}The $g$-factors give a \underline{rough}
         estimate of the percentage fraction of crosslinks
         that are elastically-active. The estimates in the example were
  computed for a network fabricated from $\Nw=1000$ chains of
  chain length $L=500$, chain
  cut-off length $\ell_{\rm{c}}=1=\ell$, and
  confinement $h_{z}=20$, where lengths are in units of $\ell$.}
       \end{center}
\end{table}

The front factor alters the classic \cite{JamesGuth} elastic constant
of $\Nw\Nc \bolz T$, since the number of elastically-able
crosslinks are reduced from $\Nw\Nc$ to $g\,\Nw\Nc$. This phenomena can
be attributed to the fact that there was no concise way of
excluding network defects, like closed loops, from the start. Note that
the $\lambda$ term containing the  front factor should not be translated into the elastic modulus of
the system. In the new theories 
there are clearly other $\lambda_{z}$-dependent
terms due to the confinement and wall-links, which will \emph{also} play a
role in the elasticity of the system. 

The second,
$h_{z}$-dependent term in \eqref{eq:stressbrushnetw} coincides with
the stress of a single chain, of length $\Nw L$, confined between
walls of width $h_{z}$. Furthermore, the term $\frac{c}{\rho}\lambda$,
seems to be a distinctive feature of a surface attached network, with
the constant $c$ depending only on the method of network formation.

In this chapter we  introduced \emph{inhomogeneous} crosslinking in the
simplest manner, by means of
a confined surface-attached network fabricated from two polymer
brushes. We saw that the polymer is localized to a greater degree in
the ``surface'' region. The surface region is defined by
  $s<\tau^{*}$, which turned out to be larger than expected. However,
  this should be ascribed to the fact that the  calculations are only
  valid for the case when we consider relatively large macromolecules, $\sqrt{L\ell} \gg h_{z}$.
The stress-strain relationship (which contains
the macroscopic information of the network) is not affected by
the inhomogeneity that was introduced in the model. 



   \chapter{Concluding Remarks}\label{chap:concl}
In this thesis we described three different confined,
surface-attached network models: the stitch network, the real confined
network and lastly the brush network. 

The stitch network in Chapter 3 captured the essence of confinement, and laid the
groundwork (together with Chapter 2) --- in terms of the relevant Green's function approach ---
for the more sophisticated models in subsequent chapters. Above all,
the stitch network served as a preliminary exercise in treating
\emph{disorder}.  The origin of the quenched disorder
was the random, albeit permanent, crosslinking in the system. It was
shown in Chapter 3 that one can handle the random crosslinking
mathematically by applying the \emph{replica} method. The
controversial problems sometimes associated with spin glass systems,
are not applicable to our simple polymer networks, and it was sufficient
to employ only the replica-symmetric approximation. The stitch
network, can be described as a \emph{sum} of smaller star-polymers (or two stitches), with
four arms emanating from fixed positions on the parallel walls. Each
bulk-link favoured the arrangement corresponding to minimum tension of
the chains, namely at half of the chain contour lengths. 
It is
therefore not surprising that the stitch network lacked the essential
(classical) features of a typical network free energy. 

In Chapter 4, we implemented the ideas of confinement and disorder of
the first two chapters, to develop a more acceptable model of
network formation. This was done by adapting an existing model of
network formation of Deam and Edwards \cite{DeamEdwards}, to incorporate
two parallel, confining surfaces (walls) and random
wall-linking. The network was formed from  pre-existing,
confined macromolecules, that were crosslinked simultaneously to
each other (polymer-polymer links) and to the confining surface
(polymer-wall links). Calculations were performed under the following
assumptions:
\begin{enumerate}
\item The network was fabricated from a sufficiently dense melt, such
  that excluded volume effects were ignored. Consequently, in the
  \emph{phantom} network, the chains could intersect each other and
  the possibility of entanglements was neglected.
\item The crosslink density was sufficiently high, such that the gel
  was solid and not near the sol-gel transition point.
\item The distance between the confining walls, $h_{z}$ was smaller
  than $\sqrt{\mathcal{L}\ell}$, the effective size of the macroscopic
  network chain. 
\item The infinite square potential, which represents the hard wall
  potential, was \emph{softened} in the framework of a variational calculation.
\end{enumerate}
Since we employed an harmonic potential identical to that of Deam and
Edwards, with isotropic, homogeneous
localization, we found the localization parameters to be equal,
$q_{x}=q_{z}$, and  analogous to
that of an unconfined network model. They were namely strain
independent, and proportional to the mean square radius of gyration of
a
chain piece between two junction points. In terms of localization, the
wall-links were treated as bulk-links, $q = \frac{6 (\Nc+\Nw)}{\ell \mathcal{L}}$,
where $\mathcal{L}$ is the effective contourlength of the giant
network polymer, $\ell$ the Kuhn steplength, and $\Nc$ and $\Nw$ is the total number of bulk-links and
wall-links, respectively.
 The quantity $q^{-1/2}$ gives
a measure of the fluctuation of the network chains from the mean
affine deformation path. In other words, it defines the relative
diameter of a tube in which each chain is confined due to the
surrounding crosslink constraints. 
 
In Chapter 5 we investigated a surface-attached network, fabricated
via an instantaneous crosslinking of two pre-existing polymer
brushes. This architecture was specifically chosen  to
facilitate an \emph{inhomogeneous} localization scheme. This scheme
modeled the effect of the crosslink and wall constraints by two
constant localization parameters, which were arclength-dependent. It
was found that each chain is localized to a greater degree near the
surface at which it is attached, than far away from its grafting
surface. The great advantage of this scheme, is in the fact that the
wall-links were not treated as bulk-links, but played a role in the
arc length distance over which each constant localization takes effect.

Common to both the brush network and the real confined network, was
the general form (apart from constants) of 
the stress-strain relationship $f(\lambda)$. Under the influence of an uniaxial,
isovolumetric deformation, we obtain the following expression:
\begin{displaymath}
 f(\lambda) = \frac{\bolz T}{V}\,\left[
\frac{\Nc }{(1+c/\rho)}\left\{\,\underline{\,\lambda-\frac{1}{\lambda^2}}+\frac{c}{\rho}\,\lambda\right\}
 - \frac{\ell\,\pi^2\,\mathcal{L}}{3\,h_{z}^{2}}\,\lambda\,\right]\,,
\end{displaymath}
with $c$ a constant dependent on the chain length and confinement, and
$\rho$ the polymer density.
The underlined part gives the characteristic curve of classic rubber
elasticity models: a type of Hooke's law for small $\lambda$'s and a
sharp rise in slope for larger strains. However, this slope is characterized by a
\emph{different} elastic modulus than encountered in the usual unconfined
models. There are two new additions to the result of Deam and Edwards
for an unconfined phantom model. The first is the $c/\rho$ term, due
to wall-links and the second, $h_{z}$-dependent term, due to the
confinement. The denominator $(1+c/\rho)$ gives a measure of the
elastically active crosslinks in the system, and was also encountered
in the Deam and Edwards model. 

In all of the calculations, we assumed a phantom model. Thus, an obvious
improvement of the model would be to include  trapped
entanglements and excluded volume effects. This extension could for
example be based on a
non-Gaussian tube model or slip-link model \cite{VilgisEdwards} with non-affine deformation.

The variational calculation is extremely difficult when the
localization potential depends on the relevant spatial position, say $Z(s)$,
between the parallel plates. For this reason, we chose a simple
arclength-dependence. However, it might still be possible to find a
potential that is tractable, and more representative of the
problem. 

Another future consideration is the study of the effect of 
\emph{fillers} in the network. These volume-giving components can 
influence the network's response to deformation, by reducing the
internal stress. It is known that a filler, like carbon black,
leads to the reinforcement of  network structure \cite{Heinrich},
meaning increased stiffness, elastic modulus, tensile strength and
abrasion resistance. The study of these improved properties, in the
context of 
confinement and surface-attachments, may have a range of possible
applications, and would be a worthwhile research endeavour.


  \appendix

\chapter{Coordinate transformation}\label{chap:appenda}
The coordinates $X_{j}^{(\beta)}(s) = T_{j}^{\beta\alpha}\,R_{j}^{(\alpha)}\,(s)$  define an
orthogonal transformation with Jacobian equal to
one\footnote{Cartesian coordinates $x$, $y$, $z$ are represented by
  $j$. The index $m$ runs from $1$ to $n-1$; $\therefore m=1
  \Leftrightarrow \beta=2$.}. In matrix form
the transformation $\mathbf{\textsf{T}}$ may look as follows:
\begin{equation}
  \label{eq:DeamEdwardsTmatrix}
  \mathbf{\textsf{T}}_{j} = \left( \begin{array}{cccc}(1+n\lambda_{j}^{2})^{-\frac{1}{2}}  &
    \lambda_{j}\,(1+n\lambda_{j}^{2})^{-\frac{1}{2}} &
  \lambda_{j}\,(1+n\lambda_{j}^{2})^{-\frac{1}{2}} & \ldots \\
    \sqrt{n}\, \lambda_{j}\,(1+n\lambda_{j}^{2})^{-\frac{1}{2}}  &
  -\frac{1}{\sqrt{n}}\,(1+n\lambda_{j}^{2})^{-\frac{1}{2}}  &
-\frac{1}{\sqrt{n}}\,(1+n\lambda_{j}^{2})^{-\frac{1}{2}} & \ldots \\
0 & \frac{1}{\sqrt{n}}\,e^{(\,2\pi i m\alpha\,)/n} &
\frac{1}{\sqrt{n}}\,e^{(\,2\pi i m\alpha\,)/n} &
\ldots \\
\vdots & \vdots & \vdots & \ddots
 \end{array} \right)\, .
\end{equation} 
This is the most symmetric transformation method \cite{DeamEdwards}, but there also
exist other transformations, notably in \cite{BallEdwardsDoi}, where all entries
are chosen to be real. The factor $(1+n\lambda_{j}^{2})^{1/2}$ ensures
the orthonormality of the transformation.

The wall-crosslink position vectors $\mathbf{\eta}^{(\alpha)}$  are transformed in the same way as
the polymer chain coordinates $\bR^{(\alpha)}$ by the transformation
$\mathbf{\textsf{T}}$ \eqref{eq:DeamEdwardsTmatrix} as follows:
\begin{eqnarray}
  \label{eq:defcoornu0}
  \nu_{j}^{(0)} &=& \frac{\eta_{j}^{(0)} + \sum_{\alpha =
      1}^{n}\,\lambda_{j}\,\eta_{j}^{(\alpha)}}{(1+n\lambda_{j}^{2})^{\frac{1}{2}}} \\
  \label{eq:defcoornu1}
  \nu_{j}^{(1)} &=& \frac{\sqrt{n}\lambda_{j}\,\eta_{j}^{(0)}
    -\frac{1}{\sqrt{n}}\, \sum_{\alpha = 1}^{n}\,\eta_{j}^{(\alpha)}}
  {(1+n\lambda_{j}^{2})^{\frac{1}{2}}}\\
  \label{eq:defcooromegam}
  \omega_{j}^{(m)} &=& \frac{1}{\sqrt{n}}\,\sum_{\alpha =
    1}^{n}\,e^{(\,2\pi i m\alpha\,)/n}\, \eta_{j}^{(\alpha)}\,,\qquad m = 1,2,\dots(n-1)\,.
\end{eqnarray}
The above
calculation can be done explicitly for each Cartesian coordinate. For example:
\begin{eqnarray}
  \label{eq:nuz0}
  \nu_{x}^{(0)} &=& \frac{x + \sum_{\alpha =
      1}^{n}\,\lambda_{x}^{2}\,x}{(1+n\lambda_{x}^{2})^{\frac{1}{2}}} = \sqrt{1+n\lambda_{x}^{2}}\,x \\
  \label{eq:nuz1}
  \nu_{x}^{(1)} &=& \frac{\sqrt{n}\lambda_{x}\,x
    -\frac{1}{\sqrt{n}}\, \sum_{\alpha = 1}^{n}\,\lambda_{x}\,x}
  {(1+n\lambda_{x}^{2})^{\frac{1}{2}}} = 0\\
  \label{eq:omegam}
  \omega_{x}^{(m)} &=& \frac{\lambda_{x}\,x}{\sqrt{n}}\,e^{(\,2\pi
    i m\,)/n}\,\left( \frac{1-e^{2\pi i m} }{1-e^{(\,2\pi
    i m\,)/n}}\right)= 0,\qquad m = 1,2,\dots(n-1)\,.
\end{eqnarray}
All except the \emph{zeroth} replica variables vanish during the $\mathbf{\textsf{T}}$ transformation. 

After substitution of the coordinate transformation
\eqref{eq:DeamEdwardsTmatrix}, the pseudo-Hamiltonian of the
generalised partition function can be written as follows:
\begin{eqnarray}
  \label{eq:genzonlyzdimtransformed}
  -\frac{\mathcal{H}_{z}}{\bolz T}  &=&
  -\frac{3}{2\ell}\,\int_{0}^{\mathcal{L}}\dot{X}_{z}^{(0)\,2}\,ds 
  -\frac{3}{2\ell}\,\int_{0}^{\mathcal{L}}\dot{X}_{z}^{(1)\,2}\,ds
  -\frac{3}{2\ell}\,\sum_{m=1}^{n-1}\,\int_{0}^{\mathcal{L}}\dot{Y}_{z}^{(m)\,2}\,ds
\nonumber
  \\ & &
  {}+\int_{0}^{\mathcal{L}}A\,\left[T_{z}^{0\alpha}\,X_{z\,s}^{(0)}+T_{z}^{1\alpha}\,X_{z\,s}^{(1)}\right]\,ds 
 \nonumber \\& & { } 
  +\sum_{\alpha=1}^{n}\,\int_{0}^{\mathcal{L}}A'\,\left[T_{z}^{0\alpha}\,X_{z\,s}^{(0)}+T_{z}^{1\alpha}\,X_{z\,s}^{(1)}
        +\,\sum_{m=1}^{n-1}\,T_{z}^{(m+1)\alpha}\,Y_{z\,s}^{(m)} \right]  \,ds \nonumber \\
          & & { } +\mu_{\text{c}}\,\int_{0}^{\mathcal{L}}ds\int_{0}^{\mathcal{L}}ds'\,\,
       \prod_{\alpha=0}^{n}\,\,
       \delta\,\Big[\,T_{z}^{0\alpha}\left(X_{z\,s}^{(0)}-X_{z\,s'}^{(0)}\right)+T_{z}^{1\alpha}\left(X_{z\,s}^{(1)}-X_{z\,s'}^{(1)}\right)
        \nonumber \\ & & \qquad  +\,\sum_{m=1}^{n-1}\,T_{z}^{(m+1)\alpha}\left(Y_{z\,s}^{(m)}-Y_{z\,s'}^{(m)}\right) \Big] 
       \nonumber \\& & { }+\mu_{\text{w}}\ \int_{0}^{\mathcal{L}}ds\int dx\,dy\,\Bigg\{
          \prod_{\alpha=0}^{n}\,\,\delta\Big[\,T_{z}^{0\alpha}\left(X_{z}^{(0)}(s)-\nu_{z}^{(0)}(x,y,\epsilon)\right)+
            T_{z}^{1\alpha}\left(X_{z}^{(1)}(s)-\nu_{z}^{(1)}(x,y,\epsilon\right)
        \nonumber \\ & & \qquad
        +\,\sum_{m=1}^{n-1}\,T_{z}^{(m+1)\alpha}\left(Y_{z}^{(m)}(s)
           -\omega_{z}^{(m)}(x,y,\epsilon)\right) \Big] +
          \prod_{\alpha
            =0}^{n}\,\,\delta\Big[\,T_{z}^{0\alpha}\left(X_{z}^{(0)}(s)-\nu_{z}^{(0)}(x,y,h_{z}-\epsilon)\right)  \nonumber \\
          & & \qquad + \,
            T_{z}^{1\alpha}\left(X_{z}^{(1)}(s)-\nu_{z}^{(1)}(x,y,h_{z}-\epsilon)\right) 
        \nonumber \\ & & \qquad
        +\,\sum_{m=1}^{n-1}\,T_{z}^{(m+1)\alpha}\left(Y_{z}^{(m)}(s)
           -\omega_{z}^{(m)}(x,y,h_{z}-\epsilon)\right) \Big]
         \,\Bigg\} \\ &=& -\frac{3}{2\ell}\,\int_{0}^{\mathcal{L}}\dot{X}_{z}^{(0)\,2}\,ds 
  -\frac{3}{2\ell}\,\int_{0}^{\mathcal{L}}\dot{X}_{z}^{(1)\,2}\,ds
  -\frac{3}{2\ell}\,\sum_{m=1}^{n-1}\,\int_{0}^{\mathcal{L}}\dot{Y}_{z}^{(m)\,2}\,ds
\nonumber
  \\ & &
  {}+\int_{0}^{\mathcal{L}}A\,\left[T_{z}^{0\alpha}\,X_{z\,s}^{(0)}+T_{z}^{1\alpha}\,X_{z\,s}^{(1)}\right]\,ds 
 \nonumber \\& & { } 
  +\sum_{\alpha=1}^{n}\,\int_{0}^{\mathcal{L}}A'\,\left[T_{z}^{0\alpha}\,X_{z\,s}^{(0)}+T_{z}^{1\alpha}\,X_{z\,s}^{(1)}
        +\,\sum_{m=1}^{n-1}\,T_{z}^{(m+1)\alpha}\,Y_{z\,s}^{(m)} \right]  \,ds \nonumber \\
          & & { } +\mu_{\text{c}}\,\int_{0}^{\mathcal{L}}ds\int_{0}^{\mathcal{L}}ds'\,\,
       \delta\,\left[\,X_{z\,s}^{(0)}-X_{z\,s'}^{(0)}\right]\,\delta\left[\,X_{z\,s}^{(1)}-X_{z\,s'}^{(1)}\right]
        \,\prod_{m=1}^{n-1}\,\delta\left[Y_{z\,s}^{(m)}-Y_{z\,s'}^{(m)}\right] 
       \nonumber \\& & { }+\mu_{\text{w}}\ \int_{0}^{\mathcal{L}}ds\int dx\,dy\,\Bigg\{
          \delta\Big[\,X_{z}^{(0)}(s)-\nu_{z}^{(0)}(x,y,\epsilon)\Big]\,
            \delta\Big[\,X_{z}^{(1)}(s)\Big]\,
        \prod_{m=1}^{n-1}\,\delta\Big[Y_{z}^{(m)}(s)\Big] \nonumber \\
          & & \qquad  +
                   \,\delta\left[\,X_{z}^{(0)}(s)-\nu_{z}^{(0)}(x,y,h_{z}-\epsilon)\right]  \, \delta\left[\,X_{z}^{(1)}(s)\right] 
        \,\prod_{m=1}^{n-1}\,\delta\left[\,Y_{z}^{(m)}(s) \right]
         \,\Bigg\}\,. \label{eq:genzonlyzdimtransformedb}
\end{eqnarray}
Only the $\hat{z}$-dimension was shown, since this is the part of the
free energy, which embodies all the constraints at this stage. 
In \eqref{eq:genzonlyzdimtransformedb} we applied the fact that
$\nu^{(\beta)} = 0$ for $\beta \neq 0$, as was seen in
\eqref{eq:genzonlyzdimtransformed}.

  \chapter{The average $\langle \mathbb{A}-\mathbb{A}_{0}\rangle$ in the
variational principle}
\label{chap:appendaa}
We have to calculate the following:\footnote{Note that there is a
  summation convention implied with respect to the $\beta$'s, and
  $a\equiv (T_{z}^{00})^{-1}=\sqrt{1+n\lambda_{z}^{2}}$. Also, since
  the transformation $\mathsf{T}$ is orthogonal, we immediately wrote
  $(T_{z}^{-1})^{\alpha \beta}$ as its transponent $T_{z}^{\beta \alpha}$.}
{\setlength\arraycolsep{1pt}
\begin{eqnarray}
  \label{eq:apotensiaal}
  e^{\mathbb{A}}&=&\Theta(T_{z}^{\beta
    0}\,X_{z}^{(\beta)})\,\Theta(h_{z}-T_{z}^{\beta 0}\,X_{z}^{(\beta)})
  \prod_{\alpha=1}^{n}\,\Theta(T_{z}^{\beta
    \alpha}\,X_{z}^{(\alpha)})\,\Theta(\lambda_{z}h_{z}-T_{z}^{\beta
    \alpha}\,X_{z}^{(\beta)})\\ 
  \label{eq:a0potensiaal}
  \!\!\!\text{and}\quad  e^{
    \mathbb{A}_{0}}&=&\Theta\,(T_{z}^{00}\,X_{z}^{(0)})\,\Theta\,(h_{z}-T_{z}^{0 0}\,X_{z}^{(0)}) = 
    \Theta(X_{z}^{(0)})\,\Theta\,(\,a
      h_{z}-X_{z}^{(0)})\,.
\end{eqnarray}}
Now, $\mathbb{A}$ is a very hard potential that we choose to
\emph{soften}, by writing
\begin{equation}
  \label{eq:scalarphi}
  \Theta\,(Z)\,\Theta(h-Z) \simeq
  e^{-\phi^{2}\,\left(1-\Theta(Z)\,\Theta(h-Z)\right)}\,,
\end{equation}
where $\phi$ is a big enough scalar, and $Z$ stands for the
transformed variables such that
{\setlength\arraycolsep{1pt}
\begin{eqnarray}
  \label{eq:apotensiaalsoften}
  e^{\langle\,\mathbb{A}\rangle} &\longrightarrow&
  e^{-\phi^{2}\left(\,n+1-\big\langle\,\Theta(T_{z}^{\beta
    0}\,X_{z}^{(\beta)})\,\Theta(\,h_{z}-T_{z}^{\beta 0}\,X_{z}^{(\beta)})\big\rangle- \sum_{\alpha=1}^{n}\,
        \big\langle\,\Theta\,(T_{z}^{\beta
    \alpha}\,X_{z}^{(\alpha)})\,\Theta(\,\lambda_{z}h_{z}-T_{z}^{\beta\alpha}\,X_{z}^{(\beta)})\big\rangle\right)}\\ 
  \label{eq:a0potensiaalsoften}
    e^{-\langle
    \mathbb{A}_{0}\rangle}&\longrightarrow& e^{+\phi^{2}\left(\,1-\big\langle\Theta\,(\,T_{z}^{00}\,X_{z}^{(0)})\,\Theta\,(h_{z}-T_{z}^{0 0}\,X_{z}^{(0)})\,\big\rangle\,\right) } \,,
\end{eqnarray}}
where $\langle \ldots \rangle$ denotes averaging with respect to the
Green's functions listed in Section \ref{sec:greensfunca}.

Next, we employ the fact that the integration limits, for the
``centre-of-mass'' coordinate of all the replicas $X_{z}^{(0)}$,  keep the
argument of the theta-function positive and in the allowed region:
\begin{eqnarray}
\label{eq:easypota}
\big\langle\Theta\,(\overbrace{X_{z}^{(0)}+n^{\scriptscriptstyle\half}\lambda
      X_{z}^{(1)}}^{\sqrt{1+n\lambda_{z}^{2}}\,R_{z}^{(0)}})\,\Theta\,(\,\overbrace{\sqrt{1+n\lambda_{z}^{2}} h_{z}-X_{z}^{(0)}-n^{\scriptscriptstyle
        \half}\lambda
      X_{z}^{(1)}}^{(1+n\lambda_{z}^{2})h_{z}-R_{z}^{(0)}})\big\rangle
    &=& 1 \, \,(\text{for }\mathbb{A}\text{--term}) \\ \label{eq:easypotb}
\text{and}\quad \big\langle\,\Theta(X_{z}^{(0)})\,\Theta\,(\sqrt{1+n\lambda_{z}^{2}}\, h_{z}-X_{z}^{(0)})\big\rangle &=& 1\, \, (\text{for }\mathbb{A}_{0}\text{--term}).
\end{eqnarray}
Substituting the above results for the $\beta=0$ and $\alpha=0$
terms in \eqref{eq:apotensiaalsoften} and \eqref{eq:a0potensiaalsoften}, we
obtain 
{\setlength\arraycolsep{1pt}
\begin{eqnarray}
  \label{eq:apotensiaalsoftenb}
  e^{\langle\,\mathbb{A}\rangle} &\simeq&
  e^{-\phi^{2}\left(n-\sum_{\alpha=1}^{n}\big\langle\,\Theta\,(T_{z}^{\beta
    \alpha}\,X_{z}^{(\alpha)})\,\Theta\,(\lambda_{z}h_{z}-T_{z}^{\beta
    \alpha}\,X_{z}^{(\beta)})\big\rangle\,\right)}\\ 
  \label{eq:a0potensiaalsoftenb}
    e^{-\langle
    \mathbb{A}_{0}\rangle}&\simeq& 1\,.
\end{eqnarray}}
The $n$ remaining terms in the sum in \eqref{eq:apotensiaalsoftenb} can
be written in terms of 
inverse Fourier integrals, where the Fourier transform of a
stepfunction is 
\begin{equation}
\label{eq:fourier}
\mathfrak{F}\,[\Theta\,(Z)]=\frac{1}{\sqrt{2
    \pi}}\left(\frac{i}{\omega}+\pi\delta(\omega)\right)\,. 
\end{equation}
Since the coordinates  $X_{z}^{(\beta)}$, for $\beta>0$, are simply
fluctuations relative to the ``centre-of-mass-of-all-the-replicas''
coordinate $X_{z}^{(0)}$, we may expand the Fourier expressions in
powers of $X_{z}^{(\beta)}$:
{\setlength\arraycolsep{1pt}
\begin{eqnarray}
  \label{eq:expanding}
  \lefteqn{\big\langle \Theta\,(T^{\beta
  \alpha}X^{(\beta)})\,\Theta\,(\lambda_{z}h_{z}-T^{\beta
  \alpha}X^{(\beta)})\, \big\rangle } \\
&=& \label{eq:expandingb}
\left(1+\frac{1}{2\pi}\,\int\,\frac{i\,d\omega}{\omega}\,\big\langle\,e^{-i\omega
    T_{z}^{\beta \alpha}X_{z}^{(\beta)}(s)}
  \big\rangle\,\right)\,\left(1+\frac{1}{2\pi}\,\int\frac{i\,d\omega'}{\omega'}\,\big\langle\,e^{-i\omega'
    (\lambda_{z}h_{z}-T_{z}^{\beta \alpha}X_{z}^{(\beta)}(s))}
  \big\rangle\,\right)\\ \label{eq:expandingc}
&=& 1+\frac{1}{q}\,\left[\int d\omega
    \text{ terms} \right] + \mathcal{O}\left(\frac{1}{q_{z}^{2m-1}}\right)\left[\int d\omega
    \text{ terms} \right]\,,\quad m = 1, 2, \ldots
\end{eqnarray}}
In the above average only the \emph{even} terms $[
X_{z}^{(\beta)}(s)]^{2 m}$ survive the averaging over the Gaussian
Green's functions [see Section \ref{sec:greensfunca}], and each of these
terms contribute a $q^{-(2m-1)}$ factor, where $m$ is a  positive
integer.

We thus have the following result:
\begin{equation}
  \label{eq:rsul}
  \langle\mathbb{A}-\mathbb{A}_{0}\rangle \simeq
  \phi^{2}\frac{n}{q_{z}}\,\left(\,q\text{-{\small independent
        terms}}\,\right)+\phi^{2}\mathcal{O}\left(n,
    \frac{1}{q_{z}^{2m-1}}\right)\text{-- {\small terms}}
\end{equation}
In this thesis, we assume a high crosslink density. This
implies that the localization parameter $q$
is large, so that terms of the order $q^{-1}$ are negligable relative
to  other terms, $\sim q$ and $\sim \ln q$, which play the dominant
roles in minimizing the free energy in \eqref{eq:partitionfb}. On the
other hand, the scalar $\phi$ is also large in order to decently
approximate the hard potential $\mathbb{A}$, and will tend to make the
$1/q$ terms important. However, since these path integrals are not
tractable, we approximate $\mathbb{A}$ by a \emph{soft} enough
potential so as to make the $\phi$-influence insignificant.  This
assumption is thus inherent in the Ansatz of \eqref{eq:ansatzAA0}.
The centre-of-mass coordinate $X_{z}^{(0)}$
is the \emph{only} transformed coordinate that represents physical
position of the polymer chains; the other coordinates are simply
relative to it. In the variational calculation, the $\beta>0$ (or
$\phi$-terms) are therefore only expected to play a relatively insignificant role in
localizing the network. 




\chapter{The First Variational Calculation}\label{chap:appaverages}
We set out here to evaluate the average $\langle\,
  \mathbb{Q}+\mu_{\text{c}}\mathbb{X}_{\text{c}}+\mu_{w}\mathbb{X}_{w}\,\rangle$ of
  Section \ref{sec:averages} by
  using the Green's functions listed in Section \ref{sec:greensfunca}. 

\section{The Bulk Cross-links   $\langle \mu_{\text{c}} \mathbb{X}_{\text{c}}\rangle$}
\label{sec:bulkappend}
The $\beta=0$ term for the constrained $\hat{z}$-dimension is given
by:
{\small
{\setlength\arraycolsep{1pt}\begin{eqnarray}
  \label{eq:avedeltazbulk0a}
  &\langle& \,\delta (X_{z}^{(0)}(s)-X_{z}^{(0)}(s'))
  \rangle_{\mathcal{G}_{0}}\\ &=& \frac{\int\,dX_{z 0}^{(0)}
    dX_{z s'}^{(0)} dX_{z s}^{(0)}dX_{z \mathcal{L}}^{(0)}\,
\mathcal{G}_{0}(X_{z s'}^{(0)},X_{z 0}^{(0)},{\scriptstyle s'})
\mathcal{G}_{0}(X_{z s}^{(0)},X_{z s'}^{(0)},{\scriptstyle |s-s'|}) \,\delta
(X_{z s}^{(0)}-X_{z s'}^{(0)}) 
\mathcal{G}_{0}(X_{z \mathcal{L}}^{(0)},X_{z
  s}^{(0)},{\scriptstyle |\mathcal{L}-s|})}{\int\,dX_{z 0}^{(0)}dX_{z \mathcal{L}}^{(0)}\,
\mathcal{G}_{0}(X_{z}^{(0)}(\mathcal{L}),X_{z}^{(0)}(0),\mathcal{L})}\nonumber
\\ 
&=&  \left(\frac{8 \sqrt{1+n\lambda^{2}}
    h_{z}}{\pi^{2}}\,\sum_{p=1,3,\dots}^{\infty}\frac{1}{p^{2}}\,\exp\left\{-\frac{\ell \pi^{2} p^{2}\mathcal{L}}{6\,(1+n\lambda^{2})h_{z}^{2}}\right\}
\right)^{-1}\, \nonumber \\ & &  { } \times 
{\displaystyle\left(\frac{2}{\sqrt{1+n\lambda^{2}}h_{z}}\right)^{3}\,
  \sum_{r=1}^{\infty}\,\sum_{p,\rho=1,3,\dots}^{\infty}\frac{4(1+n\lambda^{2})h_{z}^{2}}{\pi^{2} p \,\rho}
  \,e^{-\frac{\ell\pi^{2}p^{2}}{6\,(1+n\lambda^{2})h_{z}^{2}}|s'| }
\,e^{-\frac{\ell\pi^{2}r^{2}}{6\,(1+n\lambda^{2})h_{z}^{2}}|s-s'| } \,
e^{-\frac{\ell\pi^{2}\rho^{2}}{6\,(1+n\lambda^{2})h_{z}^{2}}|\mathcal{L}-s|}}\,\nonumber
\\ & & { }\times\,
{\displaystyle \int_{0}^{\sqrt{1+n\lambda^{2}}h_{z}}\, dX_{z}^{(0)}(s') 
  \,\sin\,\left[\frac{\pi
      pX_{z}^{(0)}(s')}{\sqrt{1+n\lambda^{2}}h_{z}}\right]\,\sin^{2}\,\left[\frac{\pi rX_{z}^{(0)}(s')}{\sqrt{1+n\lambda^{2}}h_{z}}\right]\,
  \sin\,\left[\frac{\pi \rho X_{z}^{(0)}(s')}{\sqrt{1+n\lambda^{2}}h_{z}}\right]
}\label{eq:avedeltazbulk0b}
\\ &\simeq & \frac{3}{2\sqrt{1+n\lambda^{2}}h_{z}}\,,\qquad \textrm{when}\qquad  p = r =\rho.\label{eq:avedeltazbulk0c}
\end{eqnarray}}}
The solution to the integral in \eqref{eq:avedeltazbulk0b} follows
from the following result: 
\begin{eqnarray}
  \label{eq:sinintegral}
  \lefteqn{\int\,dX\,\sin aX \sin^{2} bX \sin cX = \frac{\sin
      (a-c)X}{4(a-c)}-  \frac{\sin
    (a-2b-c)X}{8(a-2b-c)}}\nonumber \\ & &\, - \frac{\sin (a+2b-c)X}{8(a+2b-c)}+ 
  \frac{\sin (a-2b+c)X}{8(a-2b+c)}- \frac{\sin (a+c)X}{4(a+c)}+
  \frac{\sin (a+2b+c)X}{8(a+2b+c)}\,,
\end{eqnarray}
where $a \equiv \frac{\pi p}{\sqrt{1+n\lambda^{2}}\,h_{z}}$,  $b \equiv \frac{\pi r}{\sqrt{1+n\lambda^{2}}\,h_{z}}$,
$a \equiv \frac{\pi \rho}{\sqrt{1+n\lambda^{2}}\,h_{z}}$ and $X\equiv X_{z}^{(0)}$.
Each of the above terms can only contribute to an integration of $X=0$
to $X=\sqrt{1+n\lambda^{2}}\,h_{z}$ if the quantities in
round brackets go to zero. This immediately rules out the last two
terms given in \eqref{eq:sinintegral} since $p,r,\rho\geq 1$. The
most straightforward combination to choose, is when $a=b=c$ for each
term, which implies that $p=q=\rho$, so that the three exponential
functions reduce to one, with exponent $\frac{\ell\pi^{2}p^{2}}{6\,a^{2}\,h_{z}^{2}}L$. Also, in the limit $\sqrt{L\ell} \gg h_{z}$, the exponential function
\eqref{eq:avedeltazbulk0b} is
dominated by the first term in the sum, that is,
$p=q=\rho=1$. Consequently, for $\epsilon \ll h_{z}$, $\sin\big(\frac{\pi
  p\epsilon}{h_{z}}\big)\cos\big(\frac{\pi \rho
    \epsilon}{h_{z}}\big) \simeq \frac{\pi \epsilon}{h_{z}}$, which 
    leads to the result in \eqref{eq:avedeltazbulk0b}.

The $\beta=0$ terms for the $\hat{x}$ and $\hat{y}$
coordinates are given by:
{\setlength\arraycolsep{1pt}\begin{eqnarray}
  \label{eq:avedeltaxbulk0a}
  &\langle& \,\delta (X_{i}^{(0)}(s)-X_{i}^{(0)}(s'))
  \rangle_{\mathcal{G}_{0}} \\ &=&\frac{\int\,dX_{0}
    dX_{s'} dX_{s} dX_{\mathcal{L}}\,
\mathcal{G}_{0}(X_{s'},X_{0},s')
\mathcal{G}_{0}(X_{s},X_{s'},|s-s'|) \,\delta
(X_{s}-X_{s'}) 
\mathcal{G}_{0}(X_{\mathcal{L}},X_{
  s},|\mathcal{L}-s|)}{\int\,dX_{0}dX_{\mathcal{L}}\,
\mathcal{G}_{0}(X_{i}(\mathcal{L}),X_{i}(0),\mathcal{L})}\nonumber\\ \label{eq:avedeltaxbulk0b}
&=& \frac{1}{V_{i}\left(1+n\lambda_{i}^{2}\right)^{\frac{1}{2}}}+\left(\frac{3}{2 \pi \ell |s-s'|}\right)^{\frac{1}{2}}\,,
\end{eqnarray}}
where for example $X_{s}$ in the above is shortcut notation that stands for
$X_{i}^{(0)}(s)$.

The $\beta \in\{1,2,\ldots,n\}$ terms are identical, apart from
different $q$-values. In particular, for $X_{i}^{(1)}$ the average is given by:
{\setlength\arraycolsep{1pt}\begin{eqnarray}
  \label{eq:avedeltazbulk1a}
  &\langle& \,\delta (X_{i}^{(1)}(s)-X_{i}^{(1)}(s'))
  \rangle_{\mathcal{G}_{1}} \\ &=&\frac{\int\,dX_{0}\,
    dX_{s'}\, dX_{s}\, dX_{\mathcal{L}}\,
\mathcal{G}_{1}(X_{s'},X_{0},s')\,
\mathcal{G}_{1}(X_{s},X_{s'},|s-s'|) \,\delta
(X_{s}-X_{s'}) 
\mathcal{G}_{1}(X_{\mathcal{L}},X_{
  s},|\mathcal{L}-s|)}{\int\,dX_{ 0}dX_{\mathcal{L}}\,
\mathcal{G}_{1}(X_{i}(\mathcal{L}),X_{i}(0),\mathcal{L})}\nonumber
\\ \label{eq:avedeltazbulk1long}
&=&{\sqrt{\frac{q\,e^{-\frac{\ell\,q}{6}|2\mathcal{L}-s+s'|}\,\left(e^{\frac{2\ell\,\mathcal{L}\,q}{3}}-1 \right)}
       {2\pi\,
         \left(e^{\frac{\ell\,q}{3}|s- s'|}-1 \right) \,
         \left( \sinh \frac{\ell\,q\,}{6}|{\scriptstyle 2\mathcal{L} - s - 3\,s'}| - 
           \sinh \frac{\ell\,q\,}{6}|{\scriptstyle 2\mathcal{L} - 3\,s - s'}| + 
           2\,\sinh \frac{\ell\,q\,}{6}|{\scriptstyle 2\mathcal{L} - s + s'}| \right) }}}
\end{eqnarray}}
Each Green's function in \eqref{eq:avedeltazbulk1a} can be written as a sum of exponential functions
multiplied by a product of eigenfunctions \cite{Feynman}, as
follows:
\begin{eqnarray}
  \label{eq:feynmanexp1}
  \mathcal{G}_{1}(X_{s},X_{s'},|s-s'|)&=&\sum_{k=0}^{\infty}\,e^{-\frac{q}{3}(k+\half)\,|s-s'|}\,\phi_{k}\,[X_{s}]\phi_{k}\,[X_{s'}]
\\ \label{eq:feynmanexp2} &\simeq&  \left( \frac{q_{z}}{\pi}\right)^{\frac{1}{2}}
 e^{-\frac{q_{z}}{2}\Big[X_{s}^{(1)\,2}+X_{s'}^{(1)\,2}+\frac{\ell}{3}|s-s'|\,\Big]} \equiv \widetilde{\mathcal{G}}_{1}\,.
\end{eqnarray}
where the last step follows from the assumption that the lowest eigenfunction,
\begin{equation}
  \label{eq:groundstate}
  \phi_{0}\,(X_{s})=\left(\frac{q}{\pi}
  \right)^{\frac{1}{4}}\,e^{-\frac{q_{z}}{2}\,X_{i}^{(1)\,2}(s)}\,,
\end{equation}
dominates the sum in \eqref{eq:feynmanexp1}. This is the case when $\ell
q_{i}|s-s'|$ in \eqref{eq:greens1x} is large, so that \mbox{$\cosh \frac{1}{3} \ell
        q_{i}|s-s'| \to \frac{1}{2}\exp\Big\{\frac{\ell
          q_{i}}{3}|s-s'|\,\Big\}$}.
In this limit, the integration in \eqref{eq:avedeltazbulk1a} can be
repeated with the approximate Green's function
$\widetilde{\mathcal{G}}_{1}$ in \eqref{eq:feynmanexp2}, such
that the average becomes
\begin{equation}
   \label{eq:avedeltazbulk1b}
 \langle \,\delta (X_{i}^{(1)}(s)-X_{i}^{(1)}(s'))
  \rangle_{\mathcal{G}_{1}}\approxeq \left(
    \frac{q_{i}}{2\pi}\right)^{\frac{1}{2}}\,,\quad q_{i} \in \{q_{x},q_{z}\}\,.
\end{equation}
Since the Green's functions 
$\mathcal{G}_{1}(X_{i\,s}^{(1)},X_{i\,s'}^{(1)},|s-s'|)$ of
\eqref{eq:greens1x} and
$\mathcal{G}_{m}(Y_{i\,s}^{(m)},Y_{i\,s'}^{(m)},|s-s'|)$  of \eqref{eq:greensmx} for\mbox{ $i \in
\{x,y,z\}$}, all have the same form, the
remaining averages are:
\begin{eqnarray}
  \label{eq:avedeltazbulkma}
  \langle \,\delta (Y_{z}^{(m)}(s)-Y_{z}^{(m)}(s'))
  \rangle_{\mathcal{G}_{m}}
&\simeq& \Big( \frac{q_{z}}{2\pi}\Big)^{\frac{1}{2}} \\ 
  \label{eq:avedeltaxbulkma}
  \langle \,\delta (Y_{i}^{(m)}(s)-Y_{i}^{(m)}(s'))\,
  \rangle_{\mathcal{G}_{m}} &\simeq&\Big( \frac{q_{x}}{2\pi}\Big)^{\frac{1}{2}}.
\end{eqnarray}
Putting all the averages together, the bulk crosslink contribution
\eqref{eq:bulkdef} becomes
\begin{eqnarray}
\label{eq:completeXca}
\therefore\,\langle\, \mu_{\text{c}}\mathbb{X}_{\text{c}}\,\rangle &=&
  \mu_{\text{c}}\,\langle\mathbb{X}_{{\rm c} x} \mathbb{X}_{{\rm c}
    y}\mathbb{X}_{{\rm c}
    z}\rangle \\  \label{eq:completeXcb}
&=& \left \langle \mu_{\text{c}}\, \int_{0}^{\mathcal{L}}ds\int_{0}^{\mathcal{L}}ds'\,\,
       \prod_{i=x,y,z}\delta\big(\,X_{i}^{(0)}(s)-X_{i}^{(0)}(s')\big) \,
       \delta\big(\,X_{i}^{(1)}(s)-X_{i}^{(1)}(s')\big)\,\right.
      \nonumber \\ & & \left.\qquad\times\,\prod_{m=1}^{n-1}\,\delta\big(\,Y_{i}^{(m)}(s)-Y_{i}^{(m}(s')\big) \right\rangle \\ 
      & = &
      \frac{3\mu_{\text{c}}}{2\sqrt{1+n\lambda^{2}}\, h_{z}}
     \left(\frac{q_{z}q_{x}^{2}}{8 \pi^{3}}\right)^{\frac{n}{2}} \int\,ds\,ds'
      \left[\frac{1}{V_{x}V_{y}\left(1+n/\lambda\right)}+\left(\frac{3}{2 \pi \ell |s-s'|}\right)\right]\,
      \label{eq:completeXccapp}
 \\ & = &
     \frac{3\mu_{\text{c}}}{2\sqrt{1+n\lambda^{2}}\, h_{z}}
     \,\left(\frac{q_{z}}{2 \pi}\right)^{n/2}\,\left(\frac{q_{i}}{2 \pi}\right)^{n}\left[
    \frac{\mathcal{L}^{2}}{V_{x}V_{y}\left(1+n/\lambda\right)}\,+\,
    \frac{3\mathcal{L}}{2\pi \ell}\,\int_{\ell_{\text{c}}}^{\mathcal{L}}\frac{du}{u}\,\right]
\label{eq:completeXcdapp} \\ &=& \frac{3 \mu_{\text{c}}}{2
  \sqrt{1+n\lambda^{2}}\,h_{z}} \,
\left(\frac{q_{z}\,q_{i}^{2}}{8 \pi^{3}}\right)^{n/2}\,
    \left[
    \frac{\mathcal{L}^{2}}{V_{x}V_{y}\left(1+n/\lambda\right)}\,+
    \frac{3\mathcal{L}}{2 \pi \ell}\,\ln\left(\frac{\mathcal{L}}{\ell_{\text{c}}}\right)\right]\,. \label{eq:completeXceapp}
\end{eqnarray}
\section{The Wall Crosslinks   $\langle \mu_{\text{w}}
  \mathbb{X}_{\text{w}}\rangle$}
\label{sec:appwall}
The wall cross-links are completely specified by the vectors $\nu^{(0)}(x, y,\epsilon)$
and $\nu^{(0)}(x,y,h_{z}-\epsilon)$, for cross-links situated an
infinitesimal distance $\epsilon$ from the wall surface.
Letting $a \equiv \sqrt{1+n\lambda^{2}}$, the centre-of-mass average
is given by:
{\setlength\arraycolsep{1pt}
\begin{eqnarray}
  \label{eq:avedeltazwall0a}
  &\langle& \delta (X_{z}^{(0)}(s)-\nu^{(0)}(x,y,\epsilon))
  \rangle_{\mathcal{G}_{0}}\nonumber \\ &=& 
\frac{\int\,dX_{z 0}^{(0)}
    dX_{z s}^{(0)}dX_{z \mathcal{L}}^{(0)}\,
\mathcal{G}_{0}(X_{z s}^{(0)},X_{z 0}^{(0)},s)\, \delta(X_{z s}^{(0)}-\eta(x,y,\epsilon)) \,
\mathcal{G}_{0}(X_{z \mathcal{L}}^{(0)},X_{z
  s}^{(0)},|\mathcal{L}-s|)}{\int\,dX_{z 0}^{(0)}dX_{z \mathcal{L}}^{(0)}\,
\mathcal{G}_{0}(X_{z}^{(0)}(\mathcal{L}),X_{z}^{(0)}(0),\mathcal{L}) }
\\ \label{eq:avedeltazwall0b}
&=& \frac{{\displaystyle
  \left(\frac{2}{a h_{z}}\right)^{2}\,
  \sum_{p,r=1,3,\ldots}^{\infty} \, \left(\frac{a h_{z}}{\pi
      p}\right)\, \sin {\small\frac{2\pi p
    \epsilon}{h_{z}}}\,e^{-\frac{\ell^{2} \pi^{2} p^{2}s} {6\,(a h_{z})^2}
  } 
 \left(\frac{a h_{z}}{\pi r}\right)\,
 \sin{\small\frac{2\pi r \epsilon}{h_{z}}}
\,e^{-\frac{\ell^{2} \pi^{2}r^{2}|\mathcal{L}-s|}{6 a^{2}h_{z}^2}}} } 
{{\displaystyle \frac{8
      a  h_{z}}{\pi^{2}}\,\sum_{p=1,3,\dots}^{\infty}\frac{1}{p^{2}}\,\cos^{2}\left( \frac{\pi p
    \epsilon}{ h_{z}}\right)\,
      \exp\left\{-\frac{\ell \pi^{2}
          p^{2}\mathcal{L}}{6\,a^{2} h_{z}^{2}}\right\} }} \\\label{eq:avedeltazwall0c}
&\simeq& \frac{{\displaystyle
  \left(\frac{2}{a h_{z}}\right)\,\left(\frac{\pi \epsilon}{a h_{z}}\right)^{2}\,
  \sum_{p,r=1,3,\ldots}^{\infty} 
e^{-\frac{\ell^{2} \pi^{2} p^{2}s} {6\,h_{z}^2} } 
\,e^{-\frac{\ell^{2} \pi^{2}p^{2}|\mathcal{L}-s|}{6\,h_{z}^2}}} } 
{{\displaystyle \sum_{p=1,3,\dots}^{\infty}\frac{1}{p^{2}}\,
      \exp\left\{-\frac{\ell \pi^{2}
          p^{2}\mathcal{L}}{6\,h_{z}^{2}}\right\} }} \quad \underbrace{\textrm{if }
  {\small \left(\frac{\pi p \epsilon}{h_{z}} < 1\right) \,\textrm{and}\, \left(\frac{\pi r
    \epsilon}{h_{z}} < 1 \right)}}_{\therefore\, \,p\,=\,1\,=\,r}.\\ &=& \frac{2\pi^{2} \epsilon^{2}}{\sqrt{1+n\lambda^{2}}h_{z}^{3}}.
\label{eq:avedeltazwall0d} 
\end{eqnarray}}
The $\beta=0$ terms for the $\hat{x}$ and $\hat{y}$
coordinates are given by:
\begin{equation}
  \label{eq:avedeltaxwall0a}
  \langle \,\delta (X_{i}^{(0)}(s)-\nu^{(0)}\,(x,y,\epsilon))
  \rangle_{\mathcal{G}_{0}}\nonumber =\frac{2}{V_{i}
    \left(1+n\lambda_{i}^{2}\right)^{\frac{1}{2}}}\, , \qquad
  i={x,y}\,\,,
  \lambda_{i}= \frac{1}{\sqrt{\lambda}}.
\end{equation}

\begin{figure}[h!]
\centering
\psfrag{Int}[][l]{${\scriptstyle\mathcal{I}}$}\psfrag{Integrand}[][c]{\framebox[4.52cm]{${\scriptstyle\mathcal{I}=\left(\frac{\sinh \frac{1}{3} \ell q\mathcal{L}}{2\pi\cosh \frac{1}{3} \ell
        qs\,\,\cosh \frac{1}{3} \ell
        q|\mathcal{L}-s|}\right)^{\half}}$}}\psfrag{s}[][l]{$s$}\psfrag{25}[][l]{${\scriptstyle L/2}$}
\psfrag{50}[][l]{${\scriptstyle L}$}
{\includegraphics[width=0.4\textwidth]{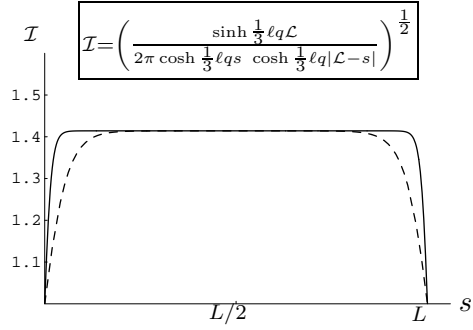}}
\caption[Illustrating the behaviour of the integrand in \eqref{eq:completeXwcapp}.]{\label{fig:wallintegral}Plot illustrating the constant behaviour of the
  integrand in \eqref{eq:completeXwcapp} and \eqref{eq:wallintexp}. The
  solid line corresponds to $\frac{1}{3}\ell q \geq 1$ (high crosslink
  density) and the dashed
  line to when the linking density decreases to $\frac{1}{3}\ell q \ll 1$.}
\end{figure}

Originally, in terms of the $\bR^{(\alpha)}$ coordinates, there was a
clear cut distinction between the undeformed ($ \alpha = 0 $) and deformed
($\alpha ={1,2,\ldots n}$) replica s, and these were different by the
deformation tensor $\mathbf{\Lambda}$ \eqref{eq:isomatrix}. The wall constraint vectors were given by  $\eta\,(x, y,h_{z}-\epsilon)$ for
the $\alpha=0$ replica, and  $\eta\,(\frac{x}{\sqrt{\lambda}},
\frac{y}{\sqrt{\lambda}},\lambda h_{z}-\epsilon)$ for the  $\alpha>0$
deformed replica systems. Note that after the coordinate
transformation $\mathbf{\textsf{T}}$, the only vectors that do not vanish are
$\mathbf\eta^{(0)}$ \eqref{eq:defcoornu0} for the top and bottom walls. 
When the network system undergoes an isovolumetric, affine deformation
the averages are given by
\begin{eqnarray}
 \label{eq:avedeltazwall1a}
   \langle \,\delta (X_{z}^{(1)}(s))
  \rangle_{\mathcal{G}_{1}}&=&
   \bigg[\frac{q_{z} \sinh \frac{1}{3} \ell q_{z}\mathcal{L}}{2 \pi\cosh \frac{1}{3} \ell
        q_{z}|s|\,\cosh \frac{1}{3} \ell q_{z}|\mathcal{L}-s|}
        \bigg]^{\frac{1}{2}}\\
  \label{eq:avedeltaxwall1a}
   \langle \,\delta (X_{i}^{(1)}(s))
  \rangle_{\mathcal{G}_{1}} &=&
   \bigg[\frac{q_{x} \sinh \frac{1}{3} \ell q_{x}\mathcal{L}}{2 \pi\cosh \frac{1}{3} \ell
        q_{x}|s|\,\cosh \frac{1}{3} \ell q_{x}|\mathcal{L}-s|}
        \bigg]^{\frac{1}{2}}\\ \label{eq:avedeltazwallma}
   \langle \,\delta (Y_{z}^{(m)}(s))
  \rangle_{\mathcal{G}_{m}} &=& 
   \bigg[\frac{q_{z} \sinh \frac{1}{3} \ell q_{z}\mathcal{L}}{2 \pi\cosh \frac{1}{3} \ell
        q_{z}|s|\,\cosh \frac{1}{3} \ell q_{z}|\mathcal{L}-s|}
        \bigg]^{\frac{1}{2}}\\
  \label{eq:avedeltaxwallma}
   \langle \,\delta (Y_{i}^{(m)}(s))
  \rangle_{\mathcal{G}_{m}} &=& 
   \bigg[\frac{q_{x} \sinh \frac{1}{3} \ell q_{x}\mathcal{L}}{2 \pi\cosh \frac{1}{3} \ell
        q_{x}|s|\,\cosh \frac{1}{3} \ell q_{x}|\mathcal{L}-s|}
        \bigg]^{\frac{1}{2}}\,.
\end{eqnarray}
Putting all the contributions together as dictated by
\eqref{eq:completeXwb},  results in the following:
\begin{eqnarray}
\label{eq:completeXwaapp}
\therefore \langle \,\mu_{\text{w}}\mathbb{X}_{\text{w}}\,\rangle &=&
  \mu_{\text{w}}\,\langle\mathbb{X}_{{\rm w} x} \mathbb{X}_{{\rm w}
    y}\mathbb{X}_{{\rm w}
    z}\rangle \\  \label{eq:completeXwbbapp}
&=&\left\langle 2\mu_{\text{w}}\,\int_{0}^{\mathcal{L}}ds\int_{-\infty}^{+\infty}dx\,dy\,
       \, \delta\big(\,X_{x}^{(0)}(s)-\nu_{x}^{(0)}\big)
       \,\delta\big(\,X_{y}^{(0)}(s)-\nu_{y}^{(0)}\big) \,
       \delta\big(\,X_{z}^{(0)}(s)-\nu_{z}^{(0)}\big) \right.\nonumber
     \\ & & { } \left. \times 
       \delta\big(\,X_{x}^{(1)}\big)\,
       \delta\big(\,X_{y}^{(1)}\big)\,
       \delta\big(\,X_{z}^{(1)}\big)
           \prod_{m=1}^{n-1}\,\delta\big(\,Y_{x}^{(m)}\big)\,\delta\big(\,Y_{y}^{(m)}\big)\,\delta\big(\,Y_{z}^{(m)}\big) 
       \right\rangle \\
&=& \label{eq:completeXwbapp} 2\mu_{\text{w}}\,\frac{2\pi^{2}
  \epsilon^{2}}{h_{z}^{3}\,\left(1+n\lambda^{2}\right)^{1/2}}\,\frac{4}{V_{x}V_{y}
  \left(1+n/\lambda\right)}\,
   \nonumber \\& & { }\times \,\int_{0}^{\mathcal{L}}ds\int_{-\infty}^{+\infty}dx\int_{-\infty}^{+\infty}dy\,
  \bigg(\frac{q\,\sinh \frac{1}{3} \ell q\mathcal{L}}{2\pi\cosh \frac{1}{3} \ell
        qs\,\,\cosh \frac{1}{3} \ell q|\mathcal{L}-s|}
        \bigg)^{n} \nonumber \\ & & { } \times
        \,\bigg(\frac{q_{z}\sinh \frac{1}{3} \ell
          q_{z}\mathcal{L}}{2\pi \cosh \frac{1}{3} \ell
        q_{z}s\,\,\cosh \frac{1}{3} \ell q_{z}|\mathcal{L}-s|}
        \bigg)^{\frac{n}{2}} \\
&=& \label{eq:completeXwcapp} 8\mu_{\text{w}}\,\frac{2\pi^{2}
  \epsilon^{2}}{h_{z}^{3}\,\left(1+n\lambda^{2}\right)^{1/2}}
    \,\int_{0}^{\mathcal{L}}ds\,
  \bigg(\frac{q\,\sinh \frac{1}{3} \ell q\mathcal{L}}{2\pi\cosh \frac{1}{3} \ell
        qs\,\,\cosh \frac{1}{3} \ell q|\mathcal{L}-s|}
        \bigg)^{n} \nonumber \\ & & { } \times
        \,\bigg(\frac{q_{z}\sinh \frac{1}{3} \ell
          q_{z}\mathcal{L}}{2\pi \cosh \frac{1}{3} \ell
        q_{z}s\,\,\cosh \frac{1}{3} \ell q_{z}|\mathcal{L}-s|}
        \bigg)^{\frac{n}{2}}
        \\  
   \label{eq:completeXwdapp} & \simeq &
   \frac{16 \mu_{\text{w}}\pi^{2}\epsilon^{2}\mathcal{L}}{h_{z}^{3}\,\left(1+n\lambda^{2}\right)^{1/2}}\, 
 \left(\frac{q_{z}}{\pi}\right)^{n/2}\,\left(\frac{q}{\pi}\right)^{n}\,\quad\text{for}\quad \frac{\ell q}{3} \ge 1\,.
\end{eqnarray}
The integral in \eqref{eq:completeXwcapp} cannot be evaluated
analytically. However, in the limit of  $\mathcal{L}\geq\frac{3}{\ell
  q}$, the
integrand is constant, except at the boundaries near $s=0$
and $s=\mathcal{L}$, as illustrated in Figure
\ref{fig:wallintegral}.
This limit contains the
  localization factor $q$, which is found from \eqref{eq:qwaardes} to be
  $q_{i} =  \frac{6\,(N_{\text{w}}+N_{\text{c}})}{\ell\,\mathcal{L}}$ for
  $i=\{x,y,z\}$. The magnitude of the non-constant contribution $\mathcal{I}_{\text{err}}$ (near $s=0$ and
  $s=\mathcal{L}$) is then of the order of
  $\mathcal{I}_{\text{err}}\thicksim\frac{\mathcal{L}}{2\,(N_{\text{c}}+N_{\text{w}})}$. This means that the
    assumption of a densely linked network ($\mathcal{I}_{\text{err}}<1$) ensures that the integrand
    in \eqref{eq:wallintexp} can be safely approximated by a
    constant. The approximate solution to the integral 
\begin{equation}
  \label{eq:wallintexp}
  \int\mathcal{I}^{n}\,ds =\left(\frac{\sinh \frac{1}{3} \ell q\mathcal{L}}{2\pi\cosh \frac{1}{3} \ell
        qs\,\,\cosh \frac{1}{3} \ell
        q|\mathcal{L}-s|}\right)^{\frac{n}{2}}\approx 2^{\frac{n}{2}}\,\mathcal{L}\, 
    \end{equation}
then leads to the result in \eqref{eq:completeXwdapp}. It 
can also be obtained by  rewriting the hyperbolic functions as exponentials and
investigating the limit of large exponents.

\section{The harmonic trial potential $\langle \,\mathbb{Q}\,\rangle$ } 
\label{sec:qappend}
Implementing only Green's functions $\mathcal{G}_{1}$ and
$\mathcal{G}_{m}$ which have the same form for all the coordinates,
results in $n$ identical terms. These solutions differ only in terms
of the parameters $q_{x}$ and $q_{z}$:
\begin{eqnarray}
\label{harmonicQz1}
\langle X_{z}^{(1)\,2}(s) \rangle_{\mathcal{G}_1} &=&
\frac{\cosh \frac{1}{3} \ell
  q_{z}|\mathcal{L}-s|\,\cosh \frac{1}{3} \ell q_{z}|s|}{q_{z}\sinh
  \frac{1}{3} \ell q_{z}\mathcal{L}}
 = \langle\, |Y_{z}^{(m)}(s)|^{2} \,\rangle_{\mathcal{G}_m}
\\
\label{harmonicQx1}
\langle X_{i}^{(1)\,2}(s) \rangle_{\mathcal{G}_1} &=&
\frac{\cosh \frac{1}{3} \ell
  q_{x}|\mathcal{L}-s|\,\cosh \frac{1}{3} \ell q_{x}|s|}{q_{x}\sinh
  \frac{1}{3} \ell q_{x}\mathcal{L}}
=\langle\, |Y_{i}^{(m)}(s)|^{2} \,\rangle_{\mathcal{G}_m}\,,
\end{eqnarray}
where $i$ collectively denotes $x$ and $y$ coordinates.

The complete average \eqref{eq:qdef} is found by integrating over the sum of the above
terms:
\begin{eqnarray}
  \label{eq:completeQapp}
   \therefore\langle \,\mathbb{Q} \, \rangle &=& \left\langle
   \sum_{i={x,y,z}}\,\frac{q_{i}^{2}\ell}{6}\int_{0}^{\mathcal{L}}\,ds\,\bigg[\,X_{i}^{(1)\,2}+\sum_{m=1}^{n-1}|Y_{i}^{(m)}|^{2}\,
   \bigg] \right\rangle\\ \label{eq:completeQaapp}&=& \frac{n\ell}{6}\int_{0}^{\mathcal{L}}ds\left[
   \frac{2q_{i}\cosh \frac{1}{3} \ell
  q_{i}|\mathcal{L}-s|\,\cosh \frac{1}{3} \ell
  q_{i}|s|}{\sinh\frac{1}{3} \ell q_{i}\mathcal{L}}\,\nonumber \right. 
\\    & & \left. \qquad\qquad+\quad \frac{q_{z}\cosh \frac{1}{3} \ell
  q_{z}|\mathcal{L}-s|\,\cosh \frac{1}{3} \ell
  q_{z}|s|}{\sinh\frac{1}{3} \ell q_{z}\mathcal{L}}\right] \\ \label{eq:completeQbapp}
&=&  \frac{n}{4}\,\left( 3+ \frac{\ell}{3}
  q_{z}\mathcal{L}\coth\frac{\ell}{3} q_{z}\mathcal{L}
  +\frac{2\ell}{3} q_{i}\mathcal{L}\coth\frac{\ell}{3}
  q_{i}\mathcal{L} \right)\\
&\simeq& \frac{n}{4}\left(3+\frac{\ell \mathcal{L}}{3}(q_{z}+2q_{x})
\right)\quad \textrm{for}\quad \frac{q\ell\mathcal{L}}{3} > 1.
\end{eqnarray}






\chapter{The Second Variational Calculation}\label{chap:appendc}
We set out here to evaluate the average $\langle\,
  \mathbb{Q}+\mu_{\rm{c}}\mathbb{X}_{\rm{c}}+\mu_{\rm{w}}\mathbb{X}_{\rm{w}}\,\rangle$ of
  Section \ref{sec:averagingb} by
  using the Green's functions listed in Section
  \ref{sec:greensfunca}.
\section{The Bulk Cross-links   $\langle \mu_{\rm{c}} \mathbb{X}_{\rm{c}}\rangle$}
\label{sec:bulkappendb}
The $\beta=0$ term for the $\hat{z}$-dimension corresponds with the
average in the previous chapter \eqref{eq:avedeltazbulk0c}, and is given by {\small
\begin{equation}
  \langle \,\delta (X_{z}^{(0)}(s)-X_{z}^{(0)}(s'))
  \rangle_{\mathcal{G}_{0}}
\simeq \frac{3}{2\sqrt{1+n\lambda^{2}}\,h_{z}}\,,\qquad \textrm{when}\qquad  p = r =\rho.\label{eq:avedeltazbulk0cb}
\end{equation}
The solution to the integral in \eqref{eq:avedeltazbulk0cb} follows
from the result \eqref{eq:sinintegral} and assumptions, previously
presented in \ref{sec:bulkappend}.

The $\alpha=0$ terms for the $\hat{x}$ and $\hat{y}$ 
coordinates are given by (where $i$ and $j$ are chain indices):
{\setlength\arraycolsep{1pt}\begin{eqnarray}
  \label{eq:avedeltaxbulk0a}
  &\langle& \,\delta (X_{xi}^{(0)}(s)-X_{xi}^{(0)}(s'))
  \rangle_{\mathcal{G}_{0}} \\  &=&\frac{\int\,dX_{0}
    dX_{i s'} dX_{i s} dX_{i L}\,
\mathcal{G}_{0}(X_{i s'},X_{i 0},s')
\mathcal{G}_{0}(X_{i s},X_{i s'},|s-s'|) \,\delta
(X_{i s}-X_{i s'}) 
\mathcal{G}_{0}(X_{i L},X_{i
  s},|\mathcal{L}-s|)}{\int\,dX_{i 0}dX_{i L}\,
\mathcal{G}_{0}(X_{i L},X_{i 0},L)}\nonumber\\ \label{eq:avedeltaxbulk0b}
&=& \frac{2}{V_{x}\left(1+n\lambda_{i}^{2}\right)^{\frac{1}{2}}}+\left(\frac{3}{2 \pi \ell |s-s'|}\right)^{\frac{1}{2}}\,,
\end{eqnarray}}
where $V_{x}$ denotes the length in the $x$-dimension  of the original
box volume, $V=V_{x}V_{y}h_{z}$, containing
the sample. 
The above result is for the special case when the polymer links with
itself ($i=j$) to form a loop of length $|s-s'|$. When $i\neq j$, two
\emph{different} chains are crosslinked, and the average is
$s$--independent:
{\setlength\arraycolsep{1pt}\begin{eqnarray}
  \label{eq:avedeltaxbulk0a}
  &\langle& \,\delta (X_{xi}^{(0)}(s)-X_{xj}^{(0)}(s'))
  \rangle_{\mathcal{G}_{0}} \\ &=& \frac{\int\,
    dX_{i 0} dX_{j 0}dX_{j s'}dX_{i L}dX_{j L}\,
\mathcal{G}_{0}(X_{j s'};X_{i 0},{\scriptstyle s})
\mathcal{G}_{0}(X_{i L},X_{j s'};{\scriptstyle |L-s|}) 
\mathcal{G}_{0}(X_{j s'},X_{j 0};{\scriptstyle
  s'})\mathcal{G}_{0}(X_{j L},X_{j s'};{\scriptstyle
  |L-s'|}}{\int\,dX_{i 0}
dX_{j 0} dX_{i L} dX_{j L}
\mathcal{G}_{0}(X_{i L},X_{i 0};\,L)\mathcal{G}_{0}(X_{j L},X_{j 0};\,L)}\nonumber\\ 
  \label{eq:avedeltaxbulk0a} 
  &=& \frac{2}{V_{x}\left(1+n\lambda_{i}^{2}\right)^{\frac{1}{2}}}\,.
\end{eqnarray}
The above result occurs $N_{\rm{w}}-1$ times. 

The rest of the bulk cross-link averages are performed in terms of
Green's functions which are either $q_0$ or $q_1$ dependent, and
therefore we have to employ the scheme in
\eqref{eq:avetwoparametersa}. For large $\ell
q|s-s'|$, we approximate the Green's function by
$\widetilde{\mathcal{G}}_{1}$ given in \eqref{eq:feynmanexp2}.
There are $3\,n$ remaining averages, given by
\eqref{eq:avebulkkk} -- \eqref{eq:avebulkgg}, where $X^{(\alpha)}$ 
stands for any cartesian component of $\{\mathbf{X}_{i}^{(1)},\mathbf{Y}_{i}^{(m)}|_{m=1\ldots\,n-1}\}$:

\begin{minipage}[t]{15.5cm}
 \parbox[]{3cm}{
     \includegraphics[width=0.4\textwidth]{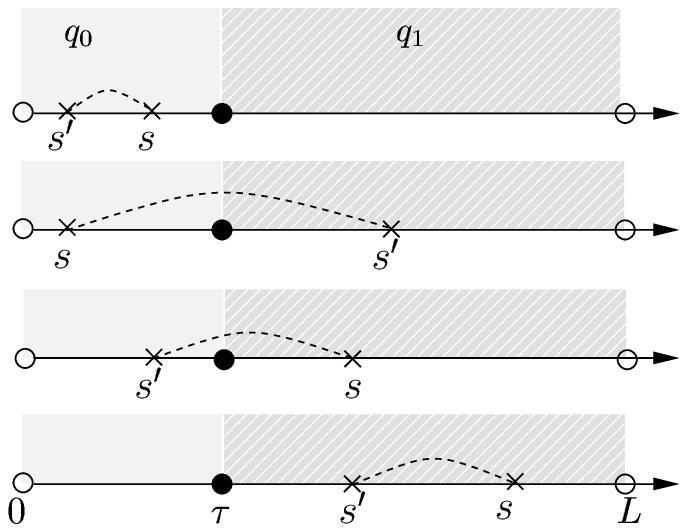}
  } \hfill
\parbox[]{10cm}{
\begin{eqnarray}
  \label{eq:avebulkkk}
   \left\langle \,\delta \big(X_{s}^{(\alpha)}-X_{s'}^{(\alpha)}\big)\,\right\rangle_{s<\tau,\,s'<\tau}
   &=&\sqrt{\frac{q_{0}}{2 \pi}}\\
  \label{eq:avebulkkg}
  \left \langle \,\delta \big(X_{s}^{(\alpha)}-X_{s'}^{(\alpha)}\big)\,\right\rangle_{s<\tau,\,s'>\tau}
   &=&\sqrt{\frac{q_{0}\,q_{1}}{\pi\,(q_{0}+q_{1})}}\\
  \label{eq:avebulkgk}
   \left\langle \,\delta \big(X_{s}^{(\alpha)}-X_{s'}^{(\alpha)}\big)\,\right\rangle_{s>\tau,\,s'<\tau}
    &=&\sqrt{\frac{q_{0}\,q_{1}}{\pi\,(q_{0}+q_{1})}}\\
  \label{eq:avebulkgg}
   \left\langle \,\delta \big(X_{s}^{(\alpha)}-X_{s'}^{(\alpha)}\big)\,\right\rangle_{s>\tau,\,s'>\tau}
   &=&\sqrt{\frac{q_{1}}{2 \pi}}
\end{eqnarray}}
\end{minipage}

Putting all the averages together, the complete bulk crosslink contribution
\eqref{eq:completeXcabrush} becomes
{\setlength\arraycolsep{1pt}
\begin{eqnarray}
\label{eq:completeXca}
\therefore\,\langle\, \mu_{\rm{c}}\mathbb{X}_{\rm{c}}\,\rangle &=&
  \mu_{\rm{c}}\,\langle\mathbb{X}_{{\rm c} x} \mathbb{X}_{{\rm c}
    y}\mathbb{X}_{{\rm c}
    z}\rangle \\  \label{eq:completeXcb}
&=&
\frac{3\mu_{\rm{c}}}{2\,h_{z}\,\sqrt{1+n\lambda^{2}}}\,\int_{0}^{\tau}ds\int_{0}^{\tau}ds'\left(\frac{2
    N_{\rm{w}} h_{z}}{V\,(1+n/\lambda)}+\frac{3}
  {2\pi\ell|s-s'|}\right)\,\left(\frac{q_{0}}{2\pi}
\right)^{\frac{3n}{2}}\nonumber \\& & +
2\,\int_{0}^{\tau}ds\int_{\tau}^{L}ds'\left(\frac{2 N_{\rm{w}} h_{z}}{V\,(1+n/\lambda)}+\frac{3}
  {2\pi\ell|s-s'|}\right)\,\left(\frac{q_{0}\,q_{1}}{\pi\,(q_{0}+q_{1})}
\right)^{\frac{3n}{2}}\nonumber\\& &
+\int_{\tau}^{L}ds\int_{\tau}^{L}ds'\left(\frac{2 N_{\rm{w}} h_{z}}{V\,(1+n/\lambda)}+\frac{3}
  {2\pi\ell|s-s'|}\right)\,\left(\frac{q_{1}}{2\pi}
\right)^{\frac{3n}{2}}
\end{eqnarray}
The above integrals should be evaluated carefully, 
since some of the terms diverge when $s=0$. However, the Gaussian
chain model and continuous chain coordinates $s$, are only valid when
we look at length scales larger than a certain length, say
$\ell_{\rm{c}}$. When $s<\ell_{\rm{c}}$, the molecule is no longer a
flexible, continuous chain: it consists of monomers with rigid
bonds. We therefore make the substitution: \mbox{$|s-s'|^{-1}\longrightarrow
\lim_{\ell_{\rm{c}}\to 0}\,[\,(s-s')^{2}+\ell_{\rm{c}}^{2}\,]^{-1/2}$}. The
integrals are now straightforward to calculate and lead to the result
shown in ~\eqref{eq:completeXce}.
\section{The Wall Crosslinks   $\langle \mu_{\rm{w}}
  \mathbb{X}_{\rm{w}}\rangle$}
\label{sec:appwallb}
The wall cross-links are completely specified by the vectors
$\nu^{(0)}(x, y,\epsilon)$ in \eqref{eq:nudefeps}
and $\nu^{(0)}(x,y,h_{z}-\epsilon)$ in \eqref{eq:nudefheps}, for cross-links situated an
infinitesimal distance $\epsilon$ from the wall surface. 
The wall crosslinks and endpoints $(s=0)$ of each chain will deform affinely
together with the walls. Letting $a \equiv \sqrt{1+n\lambda^{2}}$, the centre-of-mass average
is given by:
{\setlength\arraycolsep{1pt}
\begin{eqnarray}
  \label{eq:avedeltazwall0abrush}
  &\langle& \,\delta (X_{z}^{(0)}(0)-\nu^{(0)}(x,y,\epsilon))
  \rangle_{\mathcal{G}_{0}}\nonumber \\ &=& \frac{\int\,
    dX_{z}^{(0)}(0)\,dX_{z}^{(0)}(L)\,
\mathcal{G}_{0}(X_{z}^{(0)}(0),a\,\epsilon;\,0)\,\delta(X_{z}^{(0)}(0)-a\,\epsilon) \,
\mathcal{G}_{0}(X_{z}^{(0)}(L),X_{z}^{(0)}(0);\,L) }{\int\,dX_{z}^{(0)}(L)\,
\mathcal{G}_{0}(X_{z}^{(0)}(L),\,a\,\epsilon;\,L)}
\nonumber 
\\ &=& \frac{2\pi^{2} \epsilon^{2}}{\sqrt{1+n\lambda^{2}}h_{z}^{3}}\,,
\label{eq:avedeltazwall0dbrush} 
\end{eqnarray}
since $X_{z}^{(0)}(s=0)=a \epsilon=\sqrt{1+n\lambda^{2}}$. 

The $\hat{x}$ and $\hat{y}$ average for the
``centre-of-mass-of-replicas'' term, is 
\begin{equation}
  \label{eq:avedeltaxwall0a}
  \langle \,\delta (X_{i}^{(0)}(s)-\nu^{(0)}\,(x,y,\epsilon))
  \rangle_{\mathcal{G}_{0}} =\frac{h_{z}}{V
    \sqrt{1+n\lambda_{i}^{2}}}\, , \qquad
  i={x,y}\,\,,
  \lambda_{i}= \frac{1}{\sqrt{\lambda}}\,,
\end{equation}
where $V$ is the volume of the original undeformed system.

The remaining averages are the identity
because \begin{eqnarray}
  \label{eq:wallstuff}
  \langle\delta\big(\,\mathbf{X}^{(1)}(0)\big)\rangle &=& \langle\,\prod_{m=1}^{n-1}\,\delta\big(\,\mathbf{Y}^{(m)}(0)\big)\rangle=\mathbf{1},\\
\nonumber
  \text{because}\quad& &  \,\mathbf{X}^{(\alpha)}(s=0)=\mathbf{Y}^{(m)}(s=0)=\nu^{(\alpha)}(x,y,h_{z}-\epsilon)=\nu^{(\alpha)}(x,y,\epsilon)=0\,,
 \quad \text{for}\, \alpha>0 
\end{eqnarray}
as shown
in Appendix \ref{chap:appenda}. 
\section{The harmonic trial potential $\langle \,\mathbb{Q}\,\rangle$ } 
\label{sec:qappendb}
Implementing only Green's functions $\mathcal{G}_{1}$ and
$\mathcal{G}_{m}$ which have the same form for all the coordinates,
results in $n$ identical terms for each of the $s$ domains, for
example for the $X^{(1)}$ coordinate:
\begin{eqnarray}
\label{harmonicQz1brushk}
\lefteqn{\langle X_{z}^{(1)\,2}(s) \rangle_{s<\tau}} \\ &=&
\frac{3q_{0}\cosh \frac{\ell
  q_{0}}{3}s\,\left(q_{0}\cosh \frac{\ell q_{0}}{3}(\tau-s)\cosh
    \frac{1}{3} \ell q_{1}(L-\tau)+
    q_{1}\sinh \frac{\ell q_{0}}{3}(\tau-s)\sinh \frac{\ell
    q_{1}}{3}(L-\tau)\right)}{q_{0}\sinh \frac{\ell q_{0}}{3}\tau\,
  \cosh\frac{\ell q_{1}}{3}(L-\tau)+q_{1}\sinh \frac{\ell q_{1}}{3}(L-\tau)\,
  \cosh\frac{\ell q_{0}}{3}\tau}\nonumber
\end{eqnarray}
\begin{eqnarray}
\label{harmonicQz1brushg}
\lefteqn{\langle X_{z}^{(1)\,2}(s) \rangle_{s>\tau}} \\ &=&
\frac{3q_{0}\cosh \frac{\ell
  q_{1}}{3}(L-s)\,\left(q_{0}\sinh \frac{\ell q_{0}}{3}\tau\sinh
    \frac{1}{3} \ell q_{1}(s-\tau)+
    q_{1}\cosh \frac{\ell q_{1}}{3}(s-\tau)\cosh \frac{\ell
    q_{0}}{3}\tau\right)}{q_{0}\sinh \frac{\ell q_{0}}{3}\tau\,
  \cosh\frac{\ell q_{1}}{3}(L-\tau)+q_{1}\sinh \frac{\ell q_{1}}{3}(L-\tau)\,
  \cosh\frac{\ell q_{0}}{3}\tau}\nonumber
\end{eqnarray}

The complete average \eqref{eq:qdefb} is found by integrating over the sum of the above
terms, which is shown in the main text \eqref{eq:completeQbrush}.





  \raggedright\sloppy       


\addcontentsline{toc}{chapter}{Bibliography}
\begin{thebibliography}{10}

\bibitem{Allegra}
ALLEGRA,~G. and RAOS,~G., ``Confined polymer networks: The harmonic approach.''
  {\em J. Chem. Phys.}, 2002, Vol.~116, pp.~3109--3118.

\bibitem{Ball}
BALL,~R.~C. and EDWARDS,~S.~F., ``Elasticity and Stability of a Dense Gel.''
  {\em Macromolecules}, 1980, Vol.~13, pp.~748--761.

\bibitem{BallEdwardsDoi}
BALL,~R.~C., EDWARDS,~S.~F., DOI,~M., and WARNER,~M., ``Elasticity of entangled
  networks.'' {\em Polymer}, 1981, Vol.~22, pp.~1010--1018.

\bibitem{BinderYoung}
BINDER,~K. and YOUNG,~A.~P., ``Spin glasses: Experimental facts, theoretical
  concepts, and open questions.'' {\em Rev. Mod. Phys.}, 1986, Vol.~58,
  pp.~801--976.

\bibitem{Carslaw}
CARSLAW,~H.~S. and JAEGER,~J.~C., {\em Conduction of Heat in Solids}\/.
\newblock Second~edition.
\newblock Clarendon Press, 1959.

\bibitem{Cavagna}
CAVAGNA,~A., GIARDINA,~I., PARISI,~G., and M\'EZARD,~M., ``On the formal
  equivalence of the TAP and thermodynamic methods in the SK model.'' {\em
  Preprint cond-mat /0210665}, 2002.

\bibitem{DeGennes80}
DE~GENNES,~P.~G., ``Conformations of Polymers Attached to an Interface.'' {\em
  Macromolecules}, 1980, Vol.~13, pp.~1069--1075.

\bibitem{DeGennesScaling}
DE~GENNES,~P.~G., {\em Scaling Concepts in Polymer Physics}\/.
\newblock Ithaca: Cornell University Press, 1985.

\bibitem{DeamEdwards}
DEAM,~R.~T. and EDWARDS,~S.~F., ``The Theory of Rubber Elasticity.'' {\em Phil.
  Trans. R. Soc. London A. Math. Phys. Sciences}, 1976, Vol.~280, pp.~317--353.

\bibitem{Derrida1}
DERRIDA,~B., ``Random-Energy Model: Limit of a Family of Disordered Models.''
  {\em Physical Review Letters}, 1980, Vol.~45, pp.~79--82.

\bibitem{Derrida2}
DERRIDA,~B., ``Random-Energy model: An exactly solvable model of disordered
  systems.'' {\em Physical Review B}, 1981, Vol.~24, pp.~2613--2623.

\bibitem{DeCloiz}
DES~CLOIZEAUX,~J. and JANNINK,~G., {\em Polymers in Solution: their modelling
  and structure}\/.
\newblock Oxford: Oxford University Press, 1989.

\bibitem{Doi}
DOI,~M., {\em Introduction to Polymer Physics}\/.
\newblock Oxford: Clarendon Press, 1996.

\bibitem{Chompff}
EDWARDS,~S.~F., ``The statistical mechanics of rubbers.'' in {\em Polymer
  Networks} (CHOMPFF,~A. J. E.~A. (Ed.)), New York: Plenum Press, 1972.

\bibitem{EdwardsAnderson}
EDWARDS,~S.~F. and ANDERSON,~P.~W., ``Theory of Spin Glasses.'' {\em J. Phys.},
  1975, pp.~965--974.

\bibitem{DoiEdwards}
EDWARDS,~S.~F. and DOI,~M., {\em The Theory of Polymer Dynamics}\/.
\newblock Oxford: Clarendon Press, 1986.

\bibitem{MuthuEdwards}
EDWARDS,~S.~F. and MUTHUKUMAR,~M., ``The size of a polymer in random media.''
  {\em J. Chem. Phys.}, November~1988, Vol.~89, pp.~2435--2441.

\bibitem{VilgisEdwards}
EDWARDS,~S.~F. and VILGIS,~T.~A., ``The tube model theory of rubber
  elasticity.'' {\em Rep. Prog. Phys.}, 1988, Vol.~51, pp.~243--297.

\bibitem{Feynman}
FEYNMAN,~R.~P. and HIBBS,~A.~R., {\em Quantum mechanics and Path Integrals}\/.
\newblock New York: McGraw-Hill, 1965.

\bibitem{Fischer}
FISCHER,~K.~H. and HERTZ,~J.~A., {\em Spin Glass Theory}\/.
\newblock Cambridge: Cambridge University Press, 1991.

\bibitem{Flory}
FLORY,~P.~J., {\em Statistical mechanics of chain molecules}\/.
\newblock New York: Interscience, 1969.

\bibitem{FreedEdwards}
FREED,~K.~F. and EDWARDS,~S.~F., ``The entropy of a confined polymer chain.''
  {\em J. Phys. A (Gen. Phys.)}, November~1969, Vol.~2, pp.~145--150.

\bibitem{GoldbartI}
GOLDBART,~P. and GOLDENFELD,~N., ``Microscopic theory for cross-linked
  macromolecules. I. Broken symmetry, rigidity, and topology.'' {\em Phys. Rev.
  A}, 1989, Vol.~39, pp.~1402--1411.

\bibitem{GoldbartII}
GOLDBART,~P. and GOLDENFELD,~N., ``Microscopic theory for cross-linked
  macromolecules. II. Replica theory of the transition to the solid state.''
  {\em Phys. Rev. A}, 1989, Vol.~39, pp.~1412--1419.

\bibitem{Gradsteyn}
GRADSHTEYN,~I.~S. and RYZHIK,~I., {\em Table of Integrals, Series, and
  Products}\/.
\newblock Fifth~edition.
\newblock London: Academic Press, Inc., 1994.

\bibitem{GrosbergKhok}
GROSBERG,~A.~Y. and KHOKLOV,~A.~R., {\em Statistical Physics of
  Macromolecules}\/.
\newblock New York: American Institute of Physics, 1994.

\bibitem{GrossMezard}
GROSS,~D.~J. and M\'EZARD,~M., ``The Simplest Spin Glass.'' {\em Nucl. Phys.},
  1984, pp.~431--452.

\bibitem{Hassani}
HASSANI,~S., {\em Foundations of Mathematical Physics}\/.
\newblock Boston, Mass.: Allyn and Bacon, 1991.

\bibitem{Heinrich}
HEINRICH,~G., KL\"UPPEL,~M., and VILGIS,~T.~A., ``Reinforcement of
  Elastomers.'' {\em Current Opinion in Solid State and Materials Science},
  2002, Vol.~6, pp.~195--203.

\bibitem{imtek}
IMTEK,~U. O.~F., ``Chemistry and Physics of Interfaces: Research Projects.''
  \underline{\tt {http://www.imtek.de/cpi/projects-list.htm}}. 2001.

\bibitem{JamesGuth}
JAMES,~H.~M. and GUTH,~E., ``Theory of the Elastic Properties of Rubber.'' {\em
  J. Chem. Phys.}, 1943, Vol.~11, pp.~455--480.

\bibitem{VilgisKhol}
KHOLODENKO,~A.~L. and VILGIS,~T.~A., ``Some geometrical and topological
  problems in polymer physics.'' {\em Physics Reports}, 1998, Vol.~298,
  pp.~251--370.

\bibitem{Kuhn}
KUHN,~W. {\em Kolloid Z.}, 1934, Vol.~68, p.~2.

\bibitem{Parisi}
M\'EZARD,~P.~G., M. and VIRASARO,~M.~V., {\em Spinglass Theory And Beyond}\/.
\newblock First~edition.
\newblock Singapore: World Scientific, 1987.

\bibitem{Morse}
MORSE,~P.~M. and FESHBACH,~H., {\em Methods of Theoretical Physics}\/.
\newblock New York: McGraw-Hill, 1953.

\bibitem{KKMNEdwards}
M\"ULLER-NEDEBOCK,~K.~K. and EDWARDS,~S.~F., ``Entanglements in polymers: II.
  Networks.'' {\em J. Phys. A: Math. Gen.}, 1999, Vol.~32, pp.~3301--3320.

\bibitem{KKMNMc}
M\"ULLER-NEDEBOCK,~K.~K., EDWARDS,~S.~F., and M{c}LEISH,~T. C.~B., ``Scattering
  from deformed polymer networks.'' {\em J. Chem. Phys.}, 1999, Vol.~111,
  pp.~8196--8208.

\bibitem{Panyukov}
PANYUKOV,~S.~V., ``Inhomogeneities as consequences of a stretching of polymer
  networks.'' {\em JETP Lett.}, 1993, Vol.~58, p.~118.

\bibitem{Paul}
PAUL,~S., {\em Surface Coatings: Science and Technology}\/.
\newblock Second~edition.
\newblock J. Wiley and Sons, 1986.

\bibitem{Prucker}
PRUCKER,~O., M\"ULLER,~K., and R\"UHE,~J., ``Surface-attached Polymer
  Networks.'' in {\em Interfaces, Adhesion and Processing in Polymer Systems}
  (S.H.~ANASTASIADIS,~G.~F., A.~KARIM (Ed.)), vol.~629, Materials Research
  Society, April~2000.

\bibitem{Ratner}
RATNER,~B.~D., ``Biomedical Applications of Synthetic Polymers.'' in {\em The
  Synthesis, Characterization, Reactions and Applications of Polymers}, vol.~7,
  New York: Pergamon Press, 1989.

\bibitem{ReadMc}
READ,~D.~J. and MCLEISH,~T. C.~B., ``"Lozenge" Contour Plots in Scattering from
  Polymer Networks.'' {\em Phys. Rev. Lett.}, 1997, Vol.~79, pp.~87--90.

\bibitem{Read}
READ,~D.~J. and MCLEISH,~T. C.~B., ``Microscopic Theory for the "Lozenge"
  Contour Plots in Scattering from Stretched Polymer Networks.'' {\em
  Macromolecules}, 1997, Vol.~30, pp.~6376--6385.

\bibitem{SK}
SHERRINGTON,~D. and KIRKPATRICK,~S., ``Solvable Model of a Spin-Glass.'' {\em
  Phys. Rev. Lett.}, 1975, Vol.~35, pp.~1792--1796.

\bibitem{SolfVilgis}
SOLF,~M.~P. and VILGIS,~T.~A., ``Statistical mechanics of macromolecular
  networks without replicas.'' {\em J. Phys. A}, 1995, Vol.~28, p.~6655.

\bibitem{TreloarGee}
TRELOAR, GEE, and STERN {\em Trans. Faraday Soc.}, 1950, Vol.~46, p.~1101.

\bibitem{TreloarPolSci}
TRELOAR,~L. R.~G., {\em Introduction to Polymer Science}\/.
\newblock London: Wykeham Publications, 1974.

\bibitem{Treloar}
TRELOAR,~L. R.~G., {\em The Physics of Rubber Elasticity}\/.
\newblock Oxford: Clarendon Press, 1975.

\bibitem{Urayama}
URAYAMA,~K. E.~A., ``Experimental Tests of Molecular Entanglement Models of
  Rubber Elasticity.'' {\em Macromolecules}, 2001, Vol.~34, pp.~8261--8269.

\bibitem{VonHemmen}
VAN~HEMMEN,~J.~L. and PALMER,~R.~G., ``The replica method and a solvable spin
  glass.'' {\em J. Phys.}, 1979, Vol.~A 12, pp.~563--580.

\bibitem{Erman}
VILGIS,~T.~A., ``Rubber Elasticity and Network Defects: Inhomogeneities in
  Crosslink Density and Entanglements.'' in {\em Elastomeric Polymer Networks}
  (MARK,~J.~E. and ERMAN,~B. (Eds)), New Jersey: Prentice-Hall, Inc., 1992.

\bibitem{Warner}
WARNER,~M. and EDWARDS,~S.~F., ``Neutron scattering from strained polymer
  networks.'' {\em J. Phys. A}, 1978, Vol.~11, pp.~1649--1655.

\bibitem{Wiegel}
WIEGEL,~F.~W., {\em Introduction to Path Integral Methods in Physics and
  Polymer Science}\/.
\newblock First~edition.
\newblock Philadelphia: World Scientific, 1986.

\bibitem{Young}
YOUNG,~R.~J. and LOVELL,~P.~A., {\em Introduction to Polymers}\/.
\newblock Second~edition.
\newblock Cheltenham, Great Britain: Stanley Thornes, 1991.

\end{thebibliography}
 \addcontentsline{toc}{chapter}{\listfigurename}
  \listoffigures    
\end{document}